\documentclass[botnum, fleqn]{unmeethesis}
\usepackage{graphicx}
\usepackage{amssymb}
\usepackage{amsmath}
\usepackage{rotating}
\usepackage{cite}

\usepackage{braket, amsfonts, mathrsfs,bm,bbm}
\usepackage{bbold, dsfont, color}
\usepackage{float}
\usepackage{algorithm,algpseudocode}
\usepackage{array,ragged2e}

\DeclareMathAlphabet{\mathpzc}{OT1}{pzc}{m}{it}

\newcommand{\comment}[1]{}
\usepackage{hyperref,doi}
\hypersetup{colorlinks=true}

\usepackage[utf8x]{inputenc}
\usepackage[english]{babel}
\newtheorem{definition}{Definition}[chapter]
\newtheorem{corollary}{Corollary}[chapter]
\newtheorem{proposition}{Proposition}[chapter]
\newcommand{\ii}{{\rm i}}

\begin{document}

\frontmatter

\title{Efficient and Robust Methods for Quantum Tomography}

\author{Charles Heber Baldwin}

\degreesubject{Ph.D., Physics}

\degree{Doctor of Philosophy \\ Physics}

\documenttype{Dissertation}

\previousdegrees{B.S., Physics, Denison University, 2009 \\
                 M.S., Physics, Miami University, 2011}

\date{December, 2016}

\maketitle

\makecopyright

\setcounter{page}{2}

\begin{acknowledgments}
   \vspace{1in}
I would like to thank Amir Kalev who was my main collaborator for the research presented in this dissertation and who contributed equally to the results. He additionally gave me valuable feedback in writing this dissertation and encouragement along the way. I would also like to thank my advisor, Prof. Ivan Deutsch, for not only teaching me about physics, but also how to approach research and present ideas. I have greatly valued his mentorship during my time in graduate school. Much of the work in this dissertation is a result of collaboration between the experimental group of Prof. Poul Jessen with his students Hector Sosa-Martinez and Nathan Lysne. I have appreciated our discussions of different aspects of quantum tomography and control in an experimental setting and the opportunity to collaborate with them.

I would also like to thank my fellow CQuIC members from whom I have learned so much in the last five years. Especially current and former members of Prof. Deutsch's group: Carlos Riofr\'{i}o, Ezad Shojaee, Leigh Norris, Ben Baragiola, Rob Cook, and Bob Keating. Outside of Prof. Deutsch's group, Jacob Miller and Andy Ferdinand have provided me with valuable feedback for presentations and enlightening conversations about quantum information. I would also like to recognize the contributions of Profs. Carl Caves, Elohim Becerra, and Akimasa Miyake. Each of these professors has taught me a great deal in both the courses they have taught and in more informal discussions.

I would finally like to thank all of the people outside of CQuIC who have helped me complete the program. Fellow UNM graduate students, Mark Gorski and Ken Obenberger have been great friends for the last 5 years. Also, my parents, Frank and Barbara, who have encouraged and supported me through all levels of my academic journey. Finally, and most importantly, I would like to thank my girlfriend Sarah for here encouragement and support in the last 2$+$ years, and especially in the last few months while I wrote my dissertation.
\end{acknowledgments}

\thispagestyle{plain}

\maketitleabstract

\begin{abstract}
The development of large-scale platforms that implement quantum information processing protocols requires new methods for verification and validation of quantum behavior. Quantum tomography (QT) is the standard tool for diagnosing quantum states, process, and readout devices by providing complete information about each. However, QT is limited since it is expensive to not only implement experimentally, but also requires heavy classical post-processing of experimental data. In this dissertation, we introduce new methods for QT that are more efficient to implement and robust to noise and errors, thereby making QT a more widely practical tool for current quantum information experiments. 

The crucial detail that makes these new, efficient, and robust methods possible is prior information about the quantum system. This prior information is prompted by the goals of most experiments in quantum information. Most quantum information processing protocols require pure states, unitary processes, and rank-1 POVM operators. Therefore, most experiments are designed to operate near this ideal regime, and have been tested by other methods to verify this objective. We show that when this is the case, QT can be accomplished with significantly fewer resources, and produce a robust estimate of the state, process, or readout device in the presence of noise and errors. Moreover, the estimate is robust even if the state is not exactly pure, the process is not exactly unitary, or the POVM is not exactly rank-1. Such compelling methods are only made possible by the positivity constraint on quantum states, processes, and POVMs. This requirement is an inherent feature of quantum mechanics, but has powerful consequences to QT. 

Since QT is necessarily an experimental tool for diagnosing quantum systems, we discuss a test of these new methods in an experimental setting. The physical system is an ensemble of laser-cooled cesium atoms in the laboratory of Prof. Poul Jessen. The atoms are prepared in the hyperfine ground manifold, which provides a large, 16-dimensional Hilbert space to test QT protocols. Experiments were conducted by Hector Sosa-Martinez {\em et al.}~\cite{SosaMartinez2016} to demonstrate different QT protocols. We compare the results, and conclude that the new methods are effective for QT.

\clearpage
\end{abstract}

\bgroup
\hypersetup{linkcolor = blue}
\tableofcontents
\listoffigures
\listoftables
\egroup


\mainmatter

\chapter{Introduction} \label{ch:intro}
Quantum information processors hold the promise to carry out powerful new protocols in communication, computation, and sensing~\cite{Nielsen2002}. However, quantum information processing devices are notoriously delicate, and sources of errors, such as decoherence or inexact controls can easily diminish the advantage of quantum information protocols. Therefore, it is essential to characterize quantum systems in order to assure they are performing as expected, and to diagnose sources of errors. 

Quantum tomography (QT) is the standard method for diagnosing a quantum information processor, and is the focus of this dissertation. Originally, QT was proposed as a method to characterize a quantum state of light~\cite{Vogel1989}, and then generalized to estimate quantum states of arbitrary systems, in a protocol now called quantum state tomography (QST). Later, the methodology was extended to estimate quantum processes, or dynamical maps on quantum systems, as quantum process tomography (QPT)~\cite{Chuang1997}, and quantum readout devices as quantum detector tomography (QDT)~\cite{Luis1999}. Therefore, QT can be used to produce estimates of the three major components that make up a quantum informational processor: state preparation, evolution, and readout. QT protocols have two steps: measurement and estimation. Through measurements, one probes the quantum system to produce data that characterizes the component in question while estimation is the procedure to use this data to build a characterization of the state, process, or readout device. 

Most theoretical proposals for measurements and estimation procedures have been developed in the context of QST. There have been several different measurements proposed in this context. The original work used homodyne detection to reconstruct the Wigner function that describes a quantum state in a continuous variable representation~\cite{Vogel1989}. Other work proposed measurement schemes for finite dimensional systems such as single-qubits~\cite{Hradil2000}, two-qubits~\cite{James2001}, arbitrary spins~\cite{Newton1968}, and general qudits~\cite{Thew2002}. More recently, constructions based on symmetric mathematical properties have been proven optimal for QST when the estimate is limited only by finite sampling~\cite{Scott2006}, such as the symmetric informationally complete (SIC) POVM~\cite{Renes2004a} and a set of mutually unbiased bases (MUB)~\cite{Wootters1989}.

Estimation techniques for QST have benefited from developments in numerical optimization. The first proposals for quantum state estimation used classical methods to estimate the Wigner function~\cite{Vogel1989,Smithey1993}, or elements of the density matrix~\cite{DAriano1994}. However, these techniques did not produce a ``physical'' quantum state, that is the estimated state was not positive and/or unit trace. Therefore, the estimate cannot be used to predict future outcomes of the experiment, since it will, by definition, predict unphysical results for some measurements. With the advance of numerical methods, the physicality constraints can now be incorporated into estimation protocols. The first such protocol was maximum likelihood (ML)~\cite{Hradil1997}, which made use of the classical likelihood principle to determine the most likely state that produced the data within the set of quantum states. Later, a simplification of the ML estimator was proposed, called least-squares~\cite{James2001}, which approximates the likelihood function when there is Gaussian distributed noise.

QT has also been implemented in a variety of different physical systems. The original proposal of using homodyne detection to reconstruct the Wigner function of a single mode of light was implemented in Ref.~\cite{Smithey1993}. Since then, QST has been demonstrated in many different experimental platforms, for example, atomic ions~\cite{Roos2004,Home2006}, atomic spins of neutral atoms~\cite{Klose2001,Smith2013a,SosaMartinez2016}, orbital angular momentum modes of light~\cite{Giovannini2013,Bent2015}, and superconducting qubits~\cite{Steffen2006}. QPT has also been used to characterize the processes in many different systems, such as entangling gates with trapped atomic ions~\cite{Riebe2006} and optical systems~\cite{OBrien2004,Mitchell2003}, the motion of atoms in an optical lattice~\cite{Myrskog2005}, and three qubits in NMR~\cite{Weinstein2004}. Applications of QDT are more recent and primarily focus on characterizing detectors in optical systems~\cite{Lundeen2008,Zhang2012,Cooper2014,Brida2012,Humphreys2015}

Despite the promising theoretical and experimental work in QT, as well as the tremendous potential for QT as a diagnostic tool, it still faces two major difficulties that limit its future practicality. First, in any experiments there are sources of noise and errors in the implementation that can make the estimate inaccurate. Most importantly, to accomplish QT with high reliability, we must assume some parts of the quantum system are working perfectly. For example, with QPT, we assume that we can perfectly prepare quantum states and measurements to probe the unknown quantum process. However, in practice this will never be the case and errors in state preparation and measurement (commonly referred to as SPAM errors~\cite{Gambetta2012}) will limit the performance. Second, it is expensive both to perform the necessary measurements and to produce an estimate with classical post-processing of the data. This is especially true for large systems (e.g. of order 10 qubits), which are more typical of modern day experiments. In order for QT to be a useful strategy for the future of quantum information processing, these two issues must be addressed.

Recent work on QT has focused on both of these challenges. In order to deal with errors in the implementation, new techniques have been proposed that do not require one to assume some parts of the system are working perfectly~\cite{Merkel2012,Blume-Kohout2013,Greenbaum2015,Jackson2015}. These techniques can be understood as a combination of all three types of QT: state, process, and detector. For example, one method, called gate-set tomography (GST)~\cite{Blume-Kohout2013,Greenbaum2015}, only requires a finite set of quantum processes, or gates, that can be repeated consistently. Then, in GST the experimenter implements the set of gates in different orders and collects data from the outcomes. The data is used to reconstruct a description of each gate, thus accomplishing QPT for each gate. The procedure does not require a known set of quantum states or detectors, and therefore is not susceptible to SPAM errors. However, these types of methods require more measurements of the quantum system. Therefore, while these methods do not suffer from SPAM errors, they are even more limited by the size of the system than standard methods.

There has also been a considerable number of new proposals for specialized diagnostic schemes, unrelated to QT, that are independent of SPAM errors. Most notably is randomized benchmarking (RB), which is a protocol for measuring the average performance of a set of quantum processes~\cite{Knill2008,Magesan2012}. RB has also been expanded to measure the performance of a particular process~\cite{Magesan2012a} and to estimate other parameters that describe a particular quantum process~\cite{Kimmel2014,Kimmel2015}. RB is now a common procedure for verifying the performance of quantum systems in many laboratories~\cite{Gaebler2012,Anderson2015, Smith2012,Olmschenk2010,Chow2009}. There exist other techniques, such as phase estimation, which measures a few components that define a quantum readout device~\cite{Kimmel2015a}. However, all specialized diagnostic techniques  only characterize a few parameters that describe the quantum system, such as the average fidelity of a process, and most do not give information about particular errors that occurred. Consequently, there is still a need for QT in an experimental setting. 

In order to make QT a more practical tool, new methods have been proposed that take advantage of prior information about the quantum system to reduce the resources required. For example, most quantum information processing protocols require pure states, so theoretical methods have been designed to reconstruct pure states that require less resources than standard techniques~\cite{Carmeli2014,Carmeli2015,Chen2013,Flammia2005,Finkelstein2004,Goyeneche2015,Heinosaari2013,Kech2015a,Ma2016}. However, these methods have no guarantees on performance in the presence of noise and errors, and in an experiment such circumstances will necessarily exist. Another closely related technique for efficiently estimating pure quantum states is called quantum compressed sensing~\cite{Gross2010,Flammia2012,Liu2011,Kalev2015}. Quantum compressed sensing is based on the classical technique of compressed sensing, which allows for the estimation of low-rank matrices or sparse signals more efficiently than classical limits~\cite{Candes2006,Candes2008}. Quantum compressed sensing estimates low-rank quantum states, such as pure states, more efficiently than standard techniques by using a set of special measurements~\cite{Liu2011} and a specific optimization program~\cite{Gross2010, Flammia2012}. This protocol offers the advantage that the estimate produced is provably robust to noise and errors~\cite{Gross2010,Flammia2012}. Quantum compressed sensing has been experimentally demonstrated for QST~\cite{Smith2013a,Tonolini2014,Liu2012} and QPT~\cite{Shabani2011,Rodionov2014}. However, the technique has limited practicality since it requires special measurements and optimization programs. One goal of this dissertation is to show how these two techniques for efficient QT with prior information fit into a general, more flexible framework.

Even with new proposals, like GST~\cite{Blume-Kohout2013,Greenbaum2015} and quantum compressed sensing~\cite{Gross2010,Flammia2012}, QT is unsuitable for many experiments. GST and related methods are independent of SPAM errors, but only feasible for small systems (e.g. at most two-qubits). Efficient methods can work for larger systems, but may not perform well in the presence of noise and errors, or require special types of measurements and estimation. Experimental efforts have pushed the size of typical quantum systems beyond a few qubits, and therefore there is a growing demand for QT protocols that are flexible to a particular implementation while still being robust to noise and errors in these regimes.

In this dissertation, we develop measurements and estimation techniques that are more efficient to implement for larger quantum systems, robust to any type of noise and errors, and are flexible to suit a given experiment. The fundamental aspect that allows for the creation of such measurements is prior information about the physical system. In any experimental implementation of a quantum information protocol, there is a wealth of prior information. Most quantum information protocols require pure states, coherent evolution, and projective measurements. Therefore, most experiments try to engineer systems that operate near these requirements. Through a variety of separate calibrations and experiments, such as RB~\cite{Johnson2015,Smith2013,Anderson2015} or phase estimation, one often has confidence that the experiment is operating near the desired regime before QT is performed. We show how to include this type of prior information into QST, QDT and QPT.

We begin with a discussion of standard methods for QT in Chapter~\ref{ch:background}. We review previously proposed measurements and estimation techniques as well as formalize the effect of noise and errors on the three types of QT. In Chapter~\ref{ch:new_IC}, we consider QST with the prior information that the state is pure, or, more generally, close to pure. We define two types of measurements called complete and strictly-complete, that rely on different prior information. We further prove that the estimates derived from strictly-complete measurements are robust to all noise and errors.  In Chapter~\ref{ch:constructions}, we present methods to construct both complete and strictly-complete measurements for QST, and provide examples of such measurements. We also show that strictly-complete measurements require roughly the same amount of resources as complete measurements. In Chapter~\ref{ch:PT}, we study how the complete and strictly-complete measurements can be generalized to QPT when there is prior information that the process is unitary.  We simulate these methods for unitary QPT in the presence of errors and show how comparing different estimators can be used as a diagnostic tool.

In Chapter~\ref{ch:experiment}, we consider an experimental implementation of QST and QPT in order to demonstrate the power of techniques described in Chapters~\ref{ch:new_IC}--\ref{ch:PT}. The platform involves ensembles of laser-cooled cesium atoms in which quantum information is encoded in the spin of each atom.  The large nuclear spin of cesium, together with the electron spin, leads to a large dimensional Hilbert space and thus provides a rich test-bed in which to explore QT protocols. The spins are controlled by four separate magnetic fields which allow for a variety of different evolutions and measurements. We discuss how different methods for measurement and estimation perform in this system, and draw conclusions on the best ways to implement QT. Finally in Chapter~\ref{ch:conclusions}, we offer conclusions on the methods discussed as well as an outlook to future work in QT.

The dissertation follows the following published articles and manuscripts in preparation:
\begin{table}[ht]
\centering
\def\arraystretch{1}
\begin{tabular}{lll}
\cline{1-3}
\multicolumn{1}{|c|}{Reference}      & \multicolumn{1}{c|}{Authors}  & \multicolumn{1}{c|}{Chapter}  \\ \hline
\multicolumn{1}{|l|}{PRA {\bf 93}, 052105 (2016)}  & \multicolumn{1}{c|}{CHB, I. H. Deutsch, and A. Kalev} & \multicolumn{1}{c|}{Ch.~\ref{ch:new_IC} and \ref{ch:constructions}}         \\ \hline
\multicolumn{1}{|l|}{PRA {\bf 90}, 012110 (2014)} & \multicolumn{1}{c|}{CHB, A. Kalev and I.H. Deutsch} & \multicolumn{1}{c|}{Ch.~\ref{ch:PT}}  \\ \hline
\multicolumn{1}{|l|}{\em in preparation} & \multicolumn{1}{c|}{\begin{tabular}[c]{@{}c@{}}H. Sosa-Martinez, N. Lysne,\\ CHB, A. Kalev, \\I. H. Deutsch, and P. S. Jessen \end{tabular}} & \multicolumn{1}{c|}{Sec.~\ref{sec:ST_exp}}  \\ \hline
\multicolumn{1}{|l|}{\em in preparation} & \multicolumn{1}{c|}{\begin{tabular}[c]{@{}c@{}}H. Sosa-Martinez, N. Lysne, \\CHB, A. Kalev,\\ I. H. Deutsch, and P. S. Jessen \end{tabular}} & \multicolumn{1}{c|}{Sec.~\ref{sec:PT_exp}} \\ \hline
\multicolumn{1}{|l|}{arXiv:1607.03169} & \multicolumn{1}{c|}{\begin{tabular}[c]{@{}c@{}} T. Keating, CHB, \\ Y.-Y. Jau, J. Lee, \\ G. W. Biedermann, and I. H. Deutsch \end{tabular}} & \multicolumn{1}{c|}{}  \\ \hline
\end{tabular}
\caption[Work published and in preparation, with reference to text]{{\bf Work published and in preparation, with reference to text.} }
\label{tbl:full_IC}
\end{table}

\chapter{Standard methods for quantum tomography} \label{ch:background}
Quantum tomography (QT) is a well established protocol for characterizing the three components of a quantum information processor: state preparation, evolution, and readout. In this chapter, we formally describe these components and then introduce the standard methods for QT. We divide the discussion into two regimes. First, an ideal setting where the states, evolutions, and readout devices are errorless and we have direct access to the probability of each measurement outcome. Second, the realistic setting where noise and errors exist in all components. Any experimental implementation will necessarily fall into the second regime, so the first regime serves as mathematical tool to establish the framework for QT.

\section{Quantum information processing devices} \label{sec:components}
A quantum information processing device can be broken into three components: state preparation, evolution, and readout. In an experiment, the quantum system is usually prepared in some fiducial state by cooling. The system is then evolved with external control fields, such as electromagnetic fields. After the evolution, the system is measured by coupling to an ancilla system and then reading out the values of the ancilla. In this section, we describe each component and present mathematical descriptions. 

In the following discussion, we denote the $d$-dimensional Hilbert space that describes the quantum system as $\mathcal{H}_d$. We will also describe linear operators on the Hilbert space that are elements of the Hilbert space $\mathcal{H}_{d^2}$, which we refer to as the operator space. In analogy to Dirac notation, we describe elements of $\mathcal{H}_{d^2}$ as ``rounded kets,'' $| \cdot )$. The procedure to take an operator, which is represented as a $d \times d$ matrix to a rounded ket, which is represented by a $d^2 \times 1$ vector, is called vectorization. This can be accomplished in many ways but the most common is a ``stacking'' of the matrix columns,
\begin{equation}
A = \begin{pmatrix}
a_{1,1} & a_{1,2} & \cdots & a_{1,d} \\
\vdots & \vdots &\ddots & \vdots \\
a_{d,1} & a_{d,2} & \cdots& a_{d,d} 
\end{pmatrix} \rightarrow
| A) = \begin{pmatrix}
a_{1,1} \\
\vdots \\
a_{d,1} \\
a_{1,2} \\
\vdots \\
a_{d,2} \\
\vdots  \\
a_{1,d} \\
\vdots \\
a_{d,d}
\end{pmatrix}.
\end{equation}
Vectorization preserves the trace inner product such that $\textrm{Tr}(A^{\dagger} B) = ( A | B)$. We denote $\{ \Upsilon_{\alpha} \}$ as an arbitrary orthonormal basis on operator space. One useful choice is the Hermitian basis, $\{ H_{\alpha}  \}$, where $H_{0} = \mathds{1}/\sqrt{d}$ and $H_{\alpha >0}$ are traceless Hermitian matrices. In the Hermitian basis, Hermitian operators are represented by real vectors, where $a_0 = \textrm{Tr}(\rho H_0) = \frac{1}{\sqrt{d}} \textrm{Tr}(\rho)$. We may apply a $d^2 \times d^2$ unitary map to change the basis of the rounded ket in the operator space. For example, the unitary, $V = \sum_{\alpha} |H_{\alpha})( \Upsilon_{\alpha} |$ maps a vectorized operator from a given basis $\{ \Upsilon_{\alpha} \}$ to the Hermitian basis. 

\subsection{State preparation} \label{ssec:states_intro}
State preparation is the first step in a quantum information processing. A quantum state is mathematically described by density operator, $\rho$, which is represented by a positive semi-definite (PSD), $\rho \geq 0$, and trace one, $\textrm{Tr}(\rho) = 1$, matrix. The set of all quantum states is the convex set $\mathcal{Q} = \Set{ \rho  | \rho \geq 0, \, \textrm{Tr}(\rho) = 1}$. We will make many definitions with respect to the set of all PSD matrices, labelled as $\mathcal{S} = \Set{S | S \geq 0}$, which is also convex, such that $\mathcal{Q} \subset \mathcal{S}$. We refer to the PSD constraint as ``positivity'' and it will play an important part in later chapters. We use the greek letters, $\rho, \, \sigma,$ and $\tau$ to represent quantum states and the capital letters $S$ or $X$ to represent an arbitrary PSD matrix. An arbitrary quantum state is specified by $d^2-1$ free parameters (real numbers) because quantum states are elements of the operator space but are constrained to have unit trace. 

\subsection{Quantum evolution} \label{ssec:process_intro}
Once the system is prepared in the desired quantum state, external control is applied to evolve the state. The external control produces a quantum process, $\mathcal{E}[\cdot]$, or dynamical map, on the quantum state. Mathematically, 
we consider such maps that satisfy two conditions, complete positivity (CP) and trace preserving (TP)~\cite{Nielsen2002}. To understand CP maps, first let us define a positive map. A positive map, which is applied to a quantum state $\rho^{\rm in}$, produces an output state, $\rho^{\rm out} = \mathcal{E}[\rho^{\rm in}]$ that is positive, i.e. if $\rho^{\rm in} \geq 0$ then $\rho^{\rm out} \geq 0$. A completely positive (CP) map satisfies the same definition as a positive map but additionally maintains the positivity of any bipartite state $\rho_{AB}$ when $\mathcal{E}$ acts on one subsystem, i.e. if $\rho_{AB} \geq 0$ then $(\mathcal{E}_A \otimes \mathcal{I}_B) [ \rho_{A,B} ]\geq 0$. A TP map is a map that preserves the trace of the quantum state, i.e. $ \textrm{Tr}\left( \rho \right) = \textrm{Tr} \left( \mathcal{E}[\rho] \right) $. 

There are many ways of representing a quantum process but we focus on two methods. First, we consider the Kraus representation, 
\begin{equation} \label{Kraus_rep}
\mathcal{E}[\rho] = \sum_{\mu} A_{\mu} \rho A_{\mu}^{\dagger}.
\end{equation}
where $A_{\mu}$ are called Kraus operators. By construction, the Kraus representation describes a CP map. If the Kraus operators resolve the identity,
\begin{equation} \label{TP_const}
\sum_{\mu} A_{\mu}^{\dagger} A_{\mu} = \mathds{1} ,
\end{equation}
then the map is also TP. The Kraus representation is not unique; a given map can be described by infinitely many different sets of Kraus operators. A special type of CPTP map is a unitary map. A unitary map preserves the eigenvalues of the density operator. Unitary maps have a single Kraus operator, $U$, which is represented by a unitary matrix, and therefore satisfies Eq.~\eqref{TP_const}. 

Another representation that will be important for QT is the process matrix, which is a $d^2 \times d^2$ matrix denoted $\chi$. We can relate the process matrix to the Kraus representation by expanding each Kraus operator in a basis on operator space, $\{ \Upsilon_{\alpha} \}$, where $\textrm{Tr}(\Upsilon_{\alpha} \Upsilon_{\beta})= \delta_{\alpha, \beta}$ and $\alpha, \beta = 1,\dots, d^2$. Then writing the Kraus operators in this basis gives $A_{\mu} = \sum_{\alpha} a_{\alpha, \mu} \Upsilon_{\alpha}$, where $a_{\alpha, \mu} = \textrm{Tr}(A_{\mu} \Upsilon_{\alpha})$ is a complex expansion coefficient. Applying this expansion to Eq.~\eqref{Kraus_rep} gives,
\begin{equation}
\mathcal{E} [ \rho ] = \sum_{\alpha, \beta} \chi_{\alpha, \beta} \Upsilon_{\alpha} \rho \Upsilon_{\beta}^{\dagger},
\end{equation}
where the expansion coefficients have been grouped to $\chi_{\alpha,\beta} = \sum_{\mu} a_{\mu, \alpha} a_{\mu,\beta}^*$, which defines the elements of the process matrix. By construction $\chi$ is Hermitian and when $\chi \geq 0$ it corresponds to a CP map. We apply the expansion to Eq.~\eqref{TP_const} to write the TP constraint in terms of $\chi$,
\begin{equation} \label{TP_const_chi}
\sum_{\alpha, \beta} \chi_{\alpha, \beta} \Upsilon_{\beta}^{\dagger} \Upsilon_{\alpha} = \mathds{1}.
\end{equation}
An arbitrary CP map is specified by $d^4$ real numbers, which is made clear in the process matrix representation since $\chi$ is a $d^2 \times d^2$ Hermitian matrix. If the map is TP, then there are an additional $d^2$ linear constraints on the process matrix given in Eq.~\eqref{TP_const_chi}. Therefore, the number of free parameters that describes an arbitrary CPTP map is $d^4-d^2$. 

\subsection{Information readout} \label{ssec:detector_intro}
After evolution, one needs to read out the desired information about the final state in order to determine the result of the quantum information protocol. Readout is typically accomplished by coupling the quantum system to an ancilla system and then measuring the ancilla~\cite{Nielsen2002}. The result is described mathematically by a POVM, which is a set of operators, called POVM elements. An ancilla may have several different orthonormal states that correspond to different outcomes of the measurement. We index the outcomes with $\mu$ and the probability of getting an outcome, $\mu$, is described by a POVM element, which is a  positive operators $E_{\mu} \geq 0$. The POVM elements are represented by PSD matrices that resolve the identity, $\sum_{\mu} E_{\mu} = \mathds{1}$. We focus on POVMs with a finite number of $N$ elements but mathematically a POVM may have infinite (even continuous) elements. The POVM can also be expressed by a $N \times d^2$ matrix, referred to as the POVM matrix, 
\begin{equation}
\Xi \triangleq \begin{pmatrix}
\leftarrow & ( E_{1} | &\rightarrow \\
& \vdots & \\
\leftarrow & ( E_{N} | &\rightarrow \\
\end{pmatrix},
\end{equation}
The POVM matrix, $\Xi$, maps elements of the operator space to an $N$-dimensional vector space. When $\Xi$ acts on positive operators, i.e. vectorized PSD matrices, the $N$-dimensional vector space is real. When $\Xi$ acts on a quantum state, i.e. vectorized PSD matrices with unit trace, the result is probabilities of the different possible measurement outcomes, $\Xi | \rho ) = \bm{p}$. A single POVM can be described by $(N-1)d^2$ free parameters. This corresponds to the $d^2$ real numbers that describes each one of the $N$ POVM elements. The identity resolution condition consists of a set of $d^2$ linear constraints that relate the $N$ POVM elements.

There are many different types of POVMs.  A particular example we will use throughout this dissertation is a ``basis'' measurement. A basis measurement is a POVM consisting of $d$, rank-1 orthonormal elements, $\textrm{Tr}(E_{\mu} E_{\nu}) = \delta_{\mu, \nu}$. This is the familiar case of measurement of a Hermitian observable, whose measurement outcomes correspond to its (non-degenerate) eigenvalues.  We denote a basis measurement by its eigenvectors,
\begin{equation}
\mathbbm{B} = \{ \ket{e_0}, \dots, \ket{e_{d-1}} \},
\end{equation}
which has corresponding POVM elements, $E_{\mu} = | e_{\mu} \rangle \langle e_{\mu} |$. 

In QT, we often require multiple readout devices. This corresponds to a collection of POVMs. We will use an additional subscript, $b$, to denote the POVM, and $v$ to denote the POVM element, $E_{b,v}$. A collection of $B$ POVMs is also a POVM, however we must normalize the POVM elements, $E_{b,v} \rightarrow \frac{1}{B} E_{b,v}$  so that they resolve the identity, $ \sum_{b,v} \frac{1}{B} E_{b,v} = \mathds{1}$. 

\subsection{The Born rule}
A quantum information processor makes use of the three different components in a given experiment. The combination produces an outcome with probability expressed mathematically by the Born rule,
\begin{equation} \label{Born_rule}
p_{\mu} = \textrm{Tr}\left( E_{\mu} \mathcal{E}[\rho] \right).
\end{equation}
The Born rule establishes a linear relationship between the probability of each outcome and the mathematical description of the state, process, or readout device. We organize these probabilities into a vector $\bm{p} = [ \textrm{Tr}(E_{1} \mathcal{E}[\rho]), \dots, \textrm{Tr}(E_{N} \mathcal{E}[\rho]) ]$, referred to as the probability vector. 

We previously constrained $\rho$, $\mathcal{E}$, and $E_{\mu}$ to be positive (or CP for the process) and have linear constraints related to the trace. These constraints were necessary to ensure that the Born rule return probabilities. The positivity constraints ensure that all $p_{\mu}$'s are positive while the trace and resolution of the identity constraints, assure that $\sum_{\mu} p_{\mu} = 1$. 

In any real experiment, it is impossible to determine the probabilities $\{ p_{\mu} \}$ exactly due to finite sampling limits and experimental sources of errors. We return to these issues in a later section but for now, we consider the unrealistic case where we have access to $\{ p_{\mu} \}$. This idealization allows us to define an important notion in QT, known as informational completeness.

\section{Ideal quantum tomography} \label{sec:full-IC}
There exists a QT protocol to reconstruct a mathematical description of each part of a quantum information processor (state preparation, evolution, and readout) respectively called state, process, and detector tomography. In each protocol, one of the components is unknown, but we assume complete knowledge of the other two. For example, in quantum state tomography, the quantum state is unknown but we assume knowledge of all quantum evolutions and readout devices. To accomplish standard QT, we probe the unknown component to determine a measurement vector. In this section, we describe an ideal version of QT, where the measurement vector is the probability vector defined by the Born rule in Eq.~\eqref{Born_rule}. With the prior knowledge of the other components, the Born rule establishes a linear relationship between the measurement vector and the unknown component. However, not all methods of probing the unknown component are sufficient to completely characterize the unknown component. In order to reconstruct a description of a quantum state, process, or readout device, we need to fully characterize all free parameters. When the probabilities provide information about all the free parameters, we call them {\em fully informationally complete} (full-IC). In this section, we describe full-IC methods for state, process, and detector tomography in the ideal setting.

\subsection{Ideal quantum state tomography} \label{ssec:noiseless_ST}
In quantum state tomography (QST), we measure an unknown quantum state with a POVM. The probability of each outcome is found from the Born rule, $p_{\mu} = \textrm{Tr}[E_{\mu} \rho]$. For convenience, we sometimes notate the POVM as a map from density matrix space to probabilities, $\mathcal{M}[\rho] = (\textrm{Tr}[E_{\mu} \rho], \dots, \textrm{Tr}[E_{\mu} \rho] )= \bm{p}$. A full-IC POVM uniquely identifies the $d^2-1$ free parameters that describe an arbitrary quantum state. A mathematical definition of a full-IC POVM for QST is given below.
\begin{definition}{\bf (Fully informationally complete, QST)} \label{def:full-IC}
Let  $\mathcal{Q}=\{ \rho: \rho\geq0,\, {\rm Tr}(\rho) = 1 \}$ be the set of all quantum states.  A POVM is said to be fully informationally complete if  
\begin{equation} 
\forall \, \rho_1, \rho_2 \in \mathcal{Q}, \, \rho_1\neq \rho_2 \textrm{ iff }\mathcal{M}[ \rho_1] \neq \mathcal{M}[\rho_2],
\end{equation}
\end{definition}
\noindent We can determine when a POVM is full-IC by the POVM matrix, $\Xi$. In vectorized form, the Born rule is,
\begin{equation} \label{ST_lin}
\bm{p} = \Xi | \rho ).
\end{equation}
When $\Xi$ is invertible, that is $\Xi^+ = (\Xi^{\dagger} \Xi)^{-1} \Xi^{\dagger}$\footnote{The superscript ``$^+$'' denotes the left inverse since, in principle, there may be more than $d^2$ POVM elements so the POVM matrix is not square, and the standard inverse does not apply.} exists, all elements of $\rho$ are uniquely determined. This is only possible if there are $d^2$ linearly independent POVM elements.\footnote{One might then think that in order for a POVM to be full-IC, $\textrm{rank}(\Xi) = d^2-1$, since this is the number of free parameters that describe an arbitrary quantum state. However, due to the identity resolution constraint, $\sum_{\mu} p_{\mu} = \sum_{\mu} \textrm{Tr}(E_{\mu} \rho) = \textrm{Tr}(\rho)$, i.e. the sum of all probabilities is equal to the trace of the quantum state for all POVMs. Therefore, all POVMs measure the trace of a quantum state. This overlaps with the trace constraint, and therefore the trace constraint does not reduce the number of POVM elements required.}

Two important examples of full-IC POVMs are the symmetric informationally complete (SIC) POVM~\cite{Renes2004a} and the set of mutually unbiased bases (MUB)~\cite{Wootters1989}. The SIC POVM is a single POVM with $d^2$, rank-1 POVM elements. The POVM elements have a constant inner product such that they are symmetrically separated in operator space. The inner product is defined as 
\begin{equation} \label{SIC_def}
\textrm{Tr}[E_{\mu} E_{\nu} ] = \frac{1}{d^2(d+1)}, \, \, \mu \neq \nu,
\end{equation}
and $\textrm{Tr}[E_{\mu}^2] = \frac{1}{d^2}$. The MUB consist of $B = d+1$ basis measurements. A measurement of one of the bases projects the quantum state into an unbiased state with respect to the other bases. For example, in $d = 2$, the bases that make up the MUB consist of the eigenvectors of the well known set of Pauli matrices. If we measure the basis corresponding to  $\sigma_z$ the resulting state is either $\ket{\uparrow_z}$ or $\ket{\downarrow_z}$. Therefore, if we measure this state with the corresponding basis to $\sigma_x$ (or to $\sigma_y$), we have equal probability of getting each of the possible outcomes. The unbiased nature of the measurement outcomes are defined by the inner product relation (where each POVM element is normalized such that $\sum_{b,v} E_{b,v} = \mathds{1}$),
\begin{equation} \label{MUBs_def}
\textrm{Tr}(E_{b,v} E_{b',v'}) = 
\begin{cases}
\frac{\delta_{v,v'}}{(d+1)^2} & \textrm{if } b = b' \\
\frac{1}{d (d+1)^2} & \textrm{if } b\neq b'
\end{cases}
\end{equation}

\subsection{Ideal quantum detector tomography} \label{ssec:noiseless_DT}
The goal of quantum detector tomography (QDT) is to determine the unknown POVM that describes the readout device.  To accomplish this, we probe the POVM element with a set of $M$, known quantum states, which we organize into a $d^2 \times M$ matrix $\Theta$, defined as,
\begin{equation} \label{Theta}
\Theta = \begin{pmatrix}
\uparrow &  &\uparrow \\
|\rho_1)& \cdots & | \rho_M) \\
\downarrow & &\downarrow \\
\end{pmatrix}.
\end{equation}
Then the Born rule can be written as the linear matrix relation, $P = \Xi \Theta$, where the elements of the matrix $P_{\mu, \nu} = \textrm{Tr}[E_{\mu} \rho_{\nu}]$ are the conditional probability of getting outcome $\mu$ given the $\nu$ state. A mathematical definition of full-IC for QDT is similar to Definition~\ref{def:full-IC} but applies to the set of probing states. The collection of states is full-IC when the matrix $P$ uniquely identifies every POVM element. This occurs when $\Theta$ is invertible, i.e. $\Theta^+ = \Theta^{\dagger} (\Theta^{\dagger} \Theta)^{-1}$ exists. For $\Theta$ to be invertible the POVM must be probed with $d^2$ linearly independent quantum states. For example, the of set $d^2$ pure states,
\begin{align} \label{standard_QPT_states}
&\ket{k},\textrm{ for } k=1,\ldots,d, \nonumber \\
&\frac{1}{\sqrt{2}}(\ket{k}+\ket{n}),\textrm{ for } k=1,\ldots,d-1,\textrm{ and }n=k+1,\ldots,d,\nonumber \\
&\frac{1}{\sqrt{2}}(\ket{k}+ {\rm i} \ket{n}),\textrm{ for } k=1,\ldots,d-1,\textrm{ and }n=k+1,\ldots,d,
\end{align}
are linearly independent~\cite{Chuang1997}. No matter how many elements there are in a given POVM, ideal QDT only requires $d^2$ linearly independent states to be full-IC. This is because applying the unknown POVM matrix to a single state produces an $N \times 1$ probability vector. Each element in the probability vector relates to one free parameter in each of the $N$ POVM elements. 

We could also accomplish QDT by characterizing each POVM element independently,
\begin{equation} \label{DT_lin}
\bm{p}^{\top} = (E_{\mu} | \Theta.
\end{equation}
Similar to Eq.~\eqref{ST_lin}, we can solve for $(E_{\mu}|$ when $\Theta$ is invertible. This technique is advantageous when there are many POVM elements, which may make it computationally expensive to store the matrix $\Xi$. 

\subsection{Ideal quantum process tomography} \label{ssec:noiseless_PT}
The goal of quantum process tomography (QPT) is to determine the unknown quantum process. To accomplish this, we prepare a set of known quantum states and evolve them with the unknown process. The output states from the unknown process are then determined by a full-IC POVM. By Eq.~\eqref{Born_rule},
\begin{align} \label{PT_relation}
p_{\mu, \nu} &= \textrm{Tr} \left[ E_{\mu} \sum_{\alpha, \beta} \chi_{\alpha, \beta} \Upsilon_{\alpha} \rho_{\nu} \Upsilon_{\beta}^{\dagger} \right], \nonumber \\
		   &= \sum_{\alpha, \beta} \chi_{\alpha, \beta} \textrm{Tr} \left[ E_{\mu} \Upsilon_{\alpha} \rho_{\nu} \Upsilon_{\beta}^{\dagger} \right], \nonumber \\
		   &= \textrm{Tr} \left[ \mathpzc{D}_{\mu, \nu} \chi \right],
\end{align}
where $(\mathpzc{D}_{\mu, \nu})_{\beta, \alpha} \triangleq \textrm{Tr} \left[ E_{\mu} \Upsilon_{\alpha} \rho_{\nu} \Upsilon_{\beta}^{\dagger} \right] $ are elements of a four dimensional array~\cite{Chuang1997}. We can also express the elements, $(\mathpzc{D}_{\mu, \nu})_{\beta, \alpha}$ in vectorized form,
\begin{equation} \label{D_matrix}
(\mathpzc{D}_{\mu, \nu})_{\beta, \alpha} = ( \Upsilon_{\beta} | E_{\mu} \Upsilon_{\alpha} \rho_{\nu} ) = ( \Upsilon_{\beta} |\rho_{\nu}^{\top}  \otimes E_{\mu}  | \Upsilon_{\alpha}),
\end{equation}
yielding $\mathpzc{D}_{\mu, \nu} \triangleq  \rho_{\nu}^{\top} \otimes E_{\mu}$, which is an operator. The relation $| A X B) = B^{\top} \otimes A | X)$ is a property of the Kronecker product, ``$\otimes$''. This is similar to the relation found by the Choi-Jamio\l{}kowski isomorphism, which is another representation of a quantum process~\cite{Watrous2008}. If the probabilities, $\{ p_{\mu,\nu} \}$, uniquely identifies an arbitrary process matrix then the states and POVMs are full-IC for QPT. The mathematical definition of full-IC for QPT is similar to Definition~\ref{def:full-IC}, but applies to the set of probing states and POVM elements.

The result of Eq.~\eqref{PT_relation} is a linear relationship between the process matrix and the probabilities, similar to QST and QDT. We can also express this relationship in vectorized form where the process matrix is transformed to a $d^4 \times 1$ vector, Eq.~\eqref{ST_lin}, the 4-dimensional array, $\mathpzc{D}$ becomes a $MN \times d^2$ matrix that operates on $| \chi )$ (we use the rounded bra-ket notation for simplicity even though $\chi$ is not an element of the operator space) and the probabilities form a $MN \times 1$ vector,
\begin{equation} \label{PT_lin}
\bm{p} = \mathpzc{D} | \chi ).
\end{equation}
If $\mathpzc{D}$ is invertible, then the solution $| \chi ) = \mathpzc{D}^+ \bm{p}$ is unique, so the states and POVM that determine $\mathpzc{D}$ are full-IC. In order for $\mathpzc{D}$ to be invertible there must be $d^2$ linearly independent states, such as the ones introduces in Eq.~\eqref{standard_QPT_states}, and $d^2$ linearly-independent POVM elements, such as the SIC POVM or MUB. 

\subsection{General QT}
There are clearly many parallels between QST, QDT, and QPT. In each procedure, we look to reconstruct one component in the quantum system, either the state, evolution, or readout device. Each component is represented by a positive semidefinite (PSD) matrix. This property, simply referred to as positivity, is a powerful constraint that will have important implications in future chapters. Another commonality between all three methods is the linear relationship between the probabilities and the PSD matrix that represents the component. We can generalize this relationship as follows,
\begin{equation}
\bm{p} = \mathcal{M}[X],
\end{equation}
where $X$ is the PSD matrix that represents the given component and $\mathcal{M}$ is referred to as the ``sensing map.'' The sensing map is a linear mapping between the PSD matrix that represents a given component and the probability of each outcome. For example, in QST the sensing map is proportional to the POVM. While the sensing map always provides a linear relation, in practice its form is dependent on the type of QT. For example, in QPT, the sensing map is the matrix $\mathpzc{D}$, which has elements dependent on the input states and the POVM that is applied to the output states. So, while the linear relation is an inherent feature of QT, the form is dependent on the type of QT being implemented.

Another difference between the three types of QT comes from the trace constraint. For QST, we saw the quantum state is constrained to be unit trace, while for QDT we saw that the POVM elements are constrained to resolve the identity. For QPT, the trace constraint is more complicated, and contains $d^2$ linear constraints on the process matrix. Therefore, the three different types of QT are differentiated by the sensing map and the trace constraint. However, we will see that the positivity constraint is a very important feature of QT, and since all three methods share this constraint, many results we present in future chapters in terms of one type of QT can be generalized to the other two.

\section{Noise and errors in quantum tomography} \label{sec:noisy_QT}
In any experimental implementation of QT there necessarily exists sources of noise and errors. One fundamental source of noise is due to a finite number of copies of the system, referred to as ``projection noise.''  There may also be other sources of noise within the experimental setup. Errors correspond to inexact characterizations of the other parts of the quantum information processor. For example, in QST the readout device may be not be described by the expected POVM. Despite the fact that many current systems have a very high level of control, there will always be physical mechanisms that are not known. We take here a frequentist perspective, that the probabilities are inherent to the system and the measurement vector returned in the experiment is a perturbation from the probabilities based on the the noise and errors. In a real application of QT, we only have access to the measurement vector and not the probabilities.

\subsection{Noise in quantum tomography} \label{ssec:noise}
In any quantum system there exists some level of noise due to finite sampling. Additionally, there may be noise in the readout device, such as shot noise. For a noisy system, the experiment produces a measurement vector, $\bm{f}$, which we relate to the probability vector $\bm{p}$, discussed in the previous section, by the noise vector, $\bm{e}$,
\begin{equation} \label{noise}
\bm{f} = \bm{p} + \bm{e}.
\end{equation}
The elements of the noise vector are random variables with zero mean and distribution dependent on the type of noise. The magnitude, $\| \bm{e} \|_2$, which we take as the $\ell_2$-norm of the vector but in general could be any norm, depends on the distribution that defines the random variable. In general, the expected noise magnitude is proportional to the variance of the distribution, $\mathbb{E} \left[ \| \bm{e} \|_2^2 \right] = \sum_{\mu} \mathbb{E}[ e_{\mu}^2 ]$.

\subsubsection{Projection noise}
In any realization there will be projection noise due to finite sampling of the system. For example, in QST we may have access to a finite number of copies, $m$, of the quantum state. Therefore, each POVM outcome occurs a finite number of times. The random variable that describes the noise then follows a multinomial distribution,
\begin{align}
\mathbb{E}[ e_{\mu}] &= 0, \nonumber \\
\mathbb{E}[ e_{\mu} e_{\nu} ] &= \frac{p_{\mu} ( \delta_{\mu,\nu} - p_{\nu})}{m}.
\end{align}
The expected magnitude of projection noise is $\mathbb{E}[ \| \bm{e} \|_2^2] =  \frac{1}{m}\sum_{\mu} p_{\mu} (1- p_{\mu})$. One can easily show that the expected magnitude is bounded by $p_{\mu} = 1/N$ for all $\mu$, i.e., the probabilities associated with the maximally mixed state. Then,
\begin{equation}
\mathbb{E}[\| \bm{e} \|^2_2] \leq \frac{1-1/N}{m} = \xi^2.
\end{equation}
The expression can be generalized for multiple POVMs, or for QDT and QPT with multiple states being measured.

\subsubsection{Shot noise}
When the noise in the measurement vector is caused by shot noise from the readout device, we treat the random variable, $e_{\mu}$, as being normally distributed, with mean zero, $\mathbb{E}[e_{\mu}] = 0$ and constant variance, $\mathbb{E}[e_{\mu}e_{\nu}] = \sigma^2 \delta_{\mu, \nu}$. This assumption may also apply to the case of finite sampling when the number of samples is very large if the probabilities are not too small. Then the expected magnitude of the noise is bounded by $\mathbb{E}[\| \bm{e} \|^2_2] \leq \sigma^2 N = \xi^2$. It is important to keep in mind that for both the types of noise discussed, and perhaps other examples, the bound $\mathbb{E}[\| \bm{e} \|_2^2] \leq \xi^2$ is only approximate. In a given experiment it may be violated.

\subsection{Errors in quantum tomography} \label{ssec:errors}
In each type of QT, we require perfect knowledge of other parts of the quantum system. For example, in QST we must know the POVM that describes the readout device exactly. This will never be possible in real experiments. Therefore, we must study how QT performs when this assumption breaks down. In this section we consider the effect of these errors on the measurement vector for the three different types of QT.

\subsubsection{Errors in QST} \label{ssec:errors}
For QST, we assume the readout device is described by the POVM, $\{ E_{\mu} \}$, called the target POVM. However, due to unknown errors such as imperfect control, or technical noise in the detector, the device is actually described by a different POVM, $\{E'_{\mu}\}$. These errors are commonly referred to as ``measurement errors.'' The actual POVM can be written in terms of the target POVM,
\begin{equation}
E_{\mu}' = E_{\mu} + X_{\mu}.
\end{equation}
where the matrix $X_{\mu}$ describes the error in the readout device. The error matrix can have any form such that both $E_{\mu}$ and  $E_{\mu}' $ are POVMs. In the vectorized form we can write the actual POVM as a sum of the two POVM matrices,
\begin{equation} \label{meas_errors}
\Xi' = \Xi + \mathcal{X},
\end{equation}
where $\mathcal{X}$ is the matrix form of $\{ X_{\mu} \}$. Then acting the implemented POVM matrix on a quantum state gives
\begin{equation} \label{ST_errors}
\bm{p}' = \Xi' | \rho ) = \Xi | \rho ) + \mathcal{X} |\rho) = \bm{p} + \bm{x},
\end{equation}
where $\bm{x} = \mathcal{X} | \rho )$ is the error vector, similar to the noise vector discussed in the previous section. The elements of the error vector, $x_{\mu}$, have a distribution dependent on the physical process that causes errors in the readout device. Some processes may also cause the elements of the error vectors to have mean not equal to zero. We call these systematic errors, as they correspond to a systematic offsets in the experimental setting as opposed to random fluctuations. Similar to the noise vector, the error vector may be bounded  such that $\| \bm{x} \|_2 \leq \eta$. This bound will depend on the physical process that produces the error and requires some prior knowledge about the performance of the readout device. 

\subsubsection{Errors in QDT} 
For QDT, we require the perfect preparation of many quantum states in order to characterize a detector. However, the state preparation procedure will never be perfect due to decoherence and/or imperfections in the control fields. In this case, there is a set of target states, $\{ \rho_{\nu} \}$, but due to errors in the state preparation procedure, the set of states actually prepared are $\{ \rho'_{\nu} \}$. These errors are commonly referred to as ``preparation errors.'' We can express the prepared density matrices in terms of the target density matrices,
\begin{equation}
\rho_{\nu}' = \rho_{\nu} + Y_{\nu},
\end{equation}
where the matrix $Y_{\nu}$ describes the errors in the state preparation. As with QST, the exact form of $Y_{\nu}$ is dependent on the physical process that is causing the prepared state not to match the target state. In vectorized form,
\begin{equation} \label{prep_errors}
\Theta' = \Theta + \mathcal{Y},
\end{equation}
where $\mathcal{Y}$ is the matrix form of $\{ Y_{\mu} \}$. Then acting the prepared state matrix on the POVM matrix gives
\begin{equation}
\bm{p}' = \Xi \Theta' = \Xi \Theta + \Xi \mathcal{Y} = P + {\bf Y},
\end{equation}
where ${\bf Y}$ is the error matrix corresponding to preparation errors. Systematic errors occur when the elements of ${\bf Y}$ do not have zero mean. We can similarly bound the magnitude of the error matrix, $\| {\bf Y} \|_2 \leq \upsilon$. This bound will depend on the physical process that produces the error and requires some prior knowledge about the state preparation procedure. 

\subsubsection{Errors in QPT}
QPT can suffer from both state preparation and measurement errors (commonly known as SPAM~\cite{Gambetta2012}). Taking the linear relation in Eq.~\eqref{D_matrix} and applying Eq.~\eqref{meas_errors} and~\eqref{prep_errors} gives,
\begin{equation}
D_{\mu, \nu}'=  \rho_{\nu}'^{\top} \otimes E_{\mu}' = \rho_{\nu}^{\top} \otimes E_{\mu}+  Y_{\nu}^{\top} \otimes E_{\mu} +  \rho_{\nu}^{\top} \otimes X_{\mu} +  Y_{\nu}^{\top} \otimes X_{\mu} .
\end{equation}
Or in vector form,
\begin{equation}
\mathpzc{D}' = \mathpzc{D} + \mathpzc{Z},
\end{equation}
where $ \mathpzc{Z}_{\mu, \nu} =Y_{\nu}^{\top} \otimes E_{\mu}  + \rho_{\nu}^{\top} \otimes X_{\mu} + Y_{\nu}^{\top} \otimes X_{\mu}$. Then the probability of getting the outcome for the prepared states and implemented measurements is related to the outcome of getting the assumed states and measurements,
\begin{equation}
\bm{p}' = \mathpzc{D}' | \chi ) = \mathpzc{D} | \chi ) + \mathpzc{Z} | \chi ) = \bm{p} + \bm{z},
\end{equation}
where $\bm{z}$ is the error vector for SPAM errors. As with the state preparation and measurement errors independently, we can bound the magnitude of the error vector, $\| \bm{z} \|_2 \leq \zeta$. This bound will depend on the physical process that produces the error and requires some prior knowledge about the performance of the POVM and state preparation. 

\subsection{Additivity of noise and errors}
We make the assumption that the noise and errors are additive. That is, increasing the magnitude of statistical noise does not affect the preparation or measurement errors and vice versa. This assumption allows us to easily incorporate noise and errors together. For example, in QST the measurement vector is equal to the probability of getting each outcome plus a noise vector, given in Eq.~\eqref{noise}, plus the error vector. We combine these two expressions to give,
\begin{equation} \label{add_mag}
\bm{f} = \bm{p}' + \bm{e} = \bm{p} + \bm{x} + \bm{e}.
\end{equation}
The total noise plus error vector is $\bm{x} + \bm{e}$ and has magnitude bounded by the two independent vectors, $\| \bm{x} + \bm{e} \|_2 \leq \| \bm{x} \|_2 + \|\bm{e} \|_2 \leq \eta + \xi$. Therefore, in this assumption the magnitude of the error vectors are additive. The same procedure can be applied for QDT and QPT.

The additivity assumption breaks down for certain sources of noise, such as projection noise. With projection noise, as shown in Sec.~\ref{ssec:noise}, the magnitude of the noise is proportional to the probability of the outcome. This probability is dependent on the measured quantum state, and therefore proportional to the state preparation errors. However, in most cases we can choose a bound for the noise magnitude that is independent of the state, as was done for projection noise in Sec.~\ref{ssec:noise}

\section{Numerical estimation methods} \label{sec:numerical_methods}
Since noise and errors are inherent to any application of QT, we need methods that produce reasonable estimates of quantum states, processes, and readout devices in this case. The most basic approach is to determine the matrix that best represents the measurement vector, $\bm{f}$. For example, in QST this can be found by minimizing the least-squares function between the measurement vector and a model in the following program:
\begin{equation} \label{LI_prog}
\underset{R}{\textrm{minimize:}}  \quad  \| \Xi | R ) - \bm{f} \|_2 .
\end{equation}
When the POVM is full-IC, there is a unique $R$ that minimizes this function called the ``linear-inversion estimate'', with analytic form,
\begin{equation} \label{LI}
| \hat{R} ) = \Xi^{+} \bm{f},
\end{equation}
which is related to the method we discussed in Sec.~\ref{sec:full-IC}. However, due to noise and errors, the linear-inversion estimate, $\hat{R}$, is not necessarily a ``physical'' quantum state, i.e. a PSD matrix with unit trace. This posses a problem for many reasons. For one, many quantities, such as fidelity, purity, entanglement measures, etc., are defined for PSD matrices. Another problem is the estimate may produce nonphysical predictions for outcomes of future experiments. For example, if the density matrix estimated is not positive it will predict a ``negative'' probability for certain outcomes. Therefore, we need a method to produce an estimate that is constrained to be a physical quantum state.

In general, to find a physical estimate for QT, we use numerical optimization techniques that are constrained over the physical set. The physical set contains the positivity constraint and a trace constraint, which is dependent on the type of QT. These two constraints define a convex set. When there are noise/errors we wish to find an estimate that is ``close'' to the measurement vector but still within the physical set. The closeness of the estimate is defined by some function. If the function is convex then, since we are searching over a convex set, this fits the standard convex optimization paradigm. 

\subsection{Convex optimization}
Convex optimization is advantageous for several reasons. First, it has been proven that for convex optimization only global minima exist, giving guaranteed convergence of numerical programs. Second, there exists efficient algorithms to solve convex programs that are freely available. The goal of convex optimization is to determine the minimum of a convex function $f(x)$ over a convex set. The defining property of a convex function is,
\begin{equation}
f(a x_1 + b x_2) \leq a f(x_1) + b f(x_2)
\end{equation}
such that $a + b =1$ and $a,b \geq 0$. A convex optimization problem has the following general form,
\begin{align}
\underset{x}{\textrm{minimize:}} \quad& f(x) & \textrm{(convex function)}, \nonumber \\
\textrm{subject to:}\quad  & g_i(x) \geq 0 & \textrm{(convex functions), }\nonumber \\
				& h_j(x) = 0 & \textrm{(affine functions), }
\end{align}
where $\{ g_i(x) \}$ are called the convex inequality constraints and $\{ h_i(x) \}$ are called the affine (a linear function plus a constant) equality constraints. We denote $\hat{x}$ as the value of $x$ that produces the minimum value of $f$ while still satisfying the constraints. There are many different types of convex programs, but the standard version of QT falls into semidefinite programs (SDP). SDPs have an inequality constraint that the variable is a PSD matrix. See Ref.~\cite{Boyd2004} for further information on convex optimization.

\subsection{Convex constraints for QT}
In QT, the variable is the matrix that describes the unknown component, which we constrain to be physical. We derived the physical constraints for each component in Sec.~\ref{sec:components}. We also introduce an additional inequality constraint that the any estimate for QT should have a probability vector that is close to the measurement vector, which we call this the measurement constraint. The variable and constraints for each type QT is defined in Table~\ref{tbl:cvx_constraints}.
\begin{table}[ht] 
\centering
\def\arraystretch{2}
\begin{tabular}{l|c|c|c|}
\cline{2-4}
                                  & \multicolumn{1}{|c}{QST} & \multicolumn{1}{|c}{QDT} & \multicolumn{1}{|c|}{QPT} \\ \hline\multicolumn{1}{|l|}{Variable}    &   $\rho$  &  $E_{\mu}$  & $\chi$     \\ \hline
\multicolumn{1}{|l|}{Positivity}  &  $\rho \geq 0$   &  $E_{\mu} \geq 0$   &  $\chi \geq 0$   \\ \hline
\multicolumn{1}{|l|}{Trace}       &  $\textrm{Tr}[\rho] = 1$   & $\sum_{\mu} E_{\mu} = \mathds{1}$    &   $\sum_{\alpha,\beta} \chi_{\alpha,\beta} \Upsilon_{\beta}^{\dagger} \Upsilon_{\alpha} = \mathds{1}$  \\ \hline
\multicolumn{1}{|l|}{Measurement} &  $\| \Xi | \rho ) - \bm{f} \|_2 \leq \varepsilon $   &  $\| \Xi \Theta - \bm{f} \|_2 \leq \varepsilon$    &   $\| \mathpzc{D} | \chi ) - \bm{f} \|_2 \leq \varepsilon$   \\ \hline
\end{tabular}
\caption[Convex constraints for QT]{{\bf Convex constraints for QT}}
\label{tbl:cvx_constraints}
\end{table}

From the table, we see the parallels between the different constraints in QT. The positivity constraint is a convex inequality constraint that is shared in all three versions of QT. The trace constraint is an affine equality constraint that is different for each type of QT. The measurement constraint is also a convex inequality constraint. In the measurement constraint, the value of $\varepsilon$ is dependent on prior information about the noise and errors present in the experiment. We define $\varepsilon$ as the sum of the magnitude of random noise, $\xi$, plus the magnitude of errors for each version of QT, (QST: $\eta$, QDT: $\upsilon$, and QPT: $\zeta$), discussed in Sec.~\ref{ssec:errors}.

\subsection{Estimation programs for QST} \label{sec:cvx_QT}
In principle we can choose the optimization function as any convex function. There are, however, some preferred choices. These functions also determine which constraints to apply in the convex optimization program. We will discuss each in terms of QST, since it has the simplest form, but generalizations can be made for QDT and QPT. 

\subsubsection{Least-squares}
The first program we consider for QST is constrained least-squares (LS). This is similar to the linear-inversion program considered previously, except we include the constraint that $\rho$ is a quantum state, i.e. it is positive and has unit trace. The corresponding convex optimization program is,
\begin{align} \label{LS}
\underset{\rho}{\textrm{minimize:}} \quad & \| \Xi | \rho ) - \bm{f} \|_2 \nonumber \\
\textrm{subject to:} \quad & \rho \geq 0, \nonumber \\
& \textrm{Tr}(\rho) = 1.
\end{align}
LS returns the quantum state that matches the measurement vector as closely as possible measured by the $\ell_2$-norm. 

\subsubsection{Maximum-likelihood}
The second convex estimator we consider is called maximum-likelihood (ML), originally proposed for QST in Ref.~\cite{Hradil1997}. The program is based on the classical maximum-likelihood technique, which returns the estimate that maximizes the likelihood function, $\mathcal{L}(\rho | \bm{f} ) = \prod_{\mu} \textrm{Tr}(E_{\mu} \rho)^{m f_{\mu}}$, for a finite sample of $m$ quantum states. The state that maximizes the likelihood function also minimizes the negative log-likelihood function,
\begin{equation}
-\textrm{log} \left[\mathcal{L}(\rho |\bm{f} ) \right] = -m\sum_{\mu} f_{\mu} \textrm{logTr}(E_{\mu} \rho),
\end{equation}
which is a convex function. Therefore, we can determine which quantum state minimizes the negative log-likelihood function with convex optimization. The ML program for QST is,
\begin{align}\label{ML}
\underset{\rho}{\textrm{minimize: }} \quad & -\textrm{log} \left[\mathcal{L}(\rho |\bm{f} ) \right]  \nonumber \\
\textrm{subject to:} \quad  & \rho \geq 0, \nonumber \\
& \textrm{Tr}(\rho) = 1,
\end{align}
where the factor of $m$ is dropped since it does not effect the optimization. ML returns the most likely quantum state to have produced the measurement vector. In the limit that the noise in QST is Gaussian distributed, then the likelihood function is well approximated by a Gaussian. Therefore, the negative log-likelihood function is $-\textrm{log}\left[ \mathcal{L}(\rho| \bm{f} ) \right] = \left( \sum_{\mu} \left| \textrm{Tr}(\rho E_{\mu}) - f_{\mu} \right|^2 \right)^{1/2}$, which is the LS function and the ML program is the same as LS.

\subsubsection{Tr-norm minimization}
The third estimator that we will consider is Tr-norm minimization, which was used in the context of quantum compressed sensing~\cite{Gross2010,Flammia2012}. Quantum compressed sensing is inspired by the classical protocol of compressed sensing, which is a technique to reconstruct an unknown matrix without sampling every element in the matrix~\cite{Candes2008,Recht2009,Candes2011}. Compressed sensing is made possible by the fact that many matrices we are interested in estimating have low rank. Low-rank matrices are specified by fewer free parameters than an arbitrary matrix. Given this prior information, it was shown that a set of measurements that satisfy a property called, the restricted isometry property (RIP), are sufficient to perfectly reconstruct a low-rank matrix without noise~\cite{Candes2008,Candes2011}. Classical compressed sensing requires the convex optimization program,
\begin{alignat}{2} \label{nuclear-norm}
\underset{X}{\textrm{minimize:}}\quad & \| X \|_* \nonumber \\
\textrm{subject to:} \quad  & \| \mathcal{M}[X ] - \bm{f} \|_2 \leq \varepsilon, 
\end{alignat}
where $\| X \|_* = \textrm{Tr}[\sqrt{X^{\dagger} X}]$ is called the nuclear-norm (also known as the trace-norm) and $\mathcal{M}[\cdot]$ represents the sensing map of the measurements that satisfy the RIP condition. It was also proven that in the presence of noise or errors, the RIP measurements and convex program in Eq.~\eqref{nuclear-norm} produce a robust estimate.

Liu~\cite{Liu2011} proved that a random set of $\mathcal{O}(d \,{\rm polylog} \, d)$ expectation values of Pauli matrices satisfy the RIP condition and Gross {\em et al.}~\cite{Gross2010} translated the compressed sensing results to QST, where there is the additional constraint that $X \geq 0$. For QST, the compressed sensing estimation program is,
\begin{alignat}{2} \label{Tr-min}
\underset{\rho}{\textrm{minimize:}} \quad & \textrm{Tr}(X) \nonumber \\
\textrm{subject to:} \quad  & \| \Xi |X ) - \bm{f} \|_2 \leq \varepsilon, \nonumber \\
&X \geq 0, 
\end{alignat}
where the nuclear-norm becomes the trace due to the positivity constraint, $X \geq 0$, and the trace constraint is dropped in order for $\textrm{Tr}(X)$ to be the free parameter. The program in Eq.~\eqref{Tr-min} estimates a PSD matrix, $\hat{X}_{\rm Tr}$, that must be renormalized to produce an estimated quantum state, $\hat{\rho}_{\rm Tr} = \hat{X}_{\rm Tr}/ \textrm{Tr}(\hat{X}_{\rm Tr})$. By relation to classical on compressed sensing, it was proven that $\hat{\rho}_{\rm Tr}$ is a robust estimate~\cite{Gross2010,Flammia2012} even though it only requires $\mathcal{O}(d \,{\rm polylog} \, d)$ expectation values. It was recently shown by Kalev {\em et al.}~\cite{Kalev2015} that the program in Eq.~\eqref{Tr-min} is not required to produce such an estimate. We will discuss this result in the next chapter.

\subsection{Robustness bound on estimation} \label{ssec:full_robustness}
The estimate returned by any of the convex programs described above are robust to the noise and errors, when the measurement vector comes from a full-IC POVM. Here, robustness means that the quality of the estimation is only linearly proportional to the magnitude of the noise and errors. To see this is true for QST, we first consider two arbitrary quantum states $\rho_a$ and $\rho_b$, which have probability vectors $\bm{p}_a = \Xi |\rho_a )$ and $\bm{p}_b = \Xi |\rho_b )$. Then, the square of the distance between the two probability vectors is,
\begin{align} \label{dist_expansion}
\| \bm{p}_a - \bm{p}_b \|_2^2 = \| \Xi | \rho_a -\rho_b) \|_2^2 = (\rho_a - \rho_b| \Xi^{\dagger} \Xi | \rho_a - \rho_b).
\end{align}
We can bound $\Xi^{\dagger} \Xi$ by the identity, $\mathds{I}$, times its smallest and largest eigenvalues, $\lambda_{\textrm{min}}(\Xi^{\dagger} \Xi) \mathds{I} \leq \Xi^{\dagger} \Xi \leq \lambda_{\textrm{max}}(\Xi^{\dagger} \Xi) \mathds{I}$. We apply this relation to Eq.~\eqref{dist_expansion}
\begin{equation} \label{meas_bound}
\sqrt{\lambda_{\textrm{min}}} \| \rho_a - \rho_b \|_2 \leq \| \bm{p}_a - \bm{p}_b\|_2 \leq \sqrt{\lambda_{\textrm{max}}} \| \rho_a - \rho_b \|_2,
\end{equation}
where $\| \rho_a - \rho_b \|_2 = \textrm{Tr}\left[ (\rho_a - \rho_b)^2 \right]^{1/2}$ is the Hilbert-Schmidt (HS) distance between the two matrices $\rho_a$ and $\rho_b$. The HS-distance is equivalent to the $\ell_2$-distance between the vectorized density matrices, $\| \rho_a - \rho_b \|_2 = \| | \rho_a ) - | \rho_b ) \|_2$.

Now, let us choose $\rho_a= \hat{\rho}$, the estimate returned by one of the convex programs, and $\rho_b = \rho$ the actual state that was measured in the presence of noise and errors. Each state has a corresponding probability vector, $\hat{\bm{p}} = \Xi | \hat{\rho})$ and $\bm{p} = \Xi | \rho)$. By Eq.~\eqref{dist_expansion},
\begin{equation}
\| \hat{\rho} - \rho \|_2 \leq \frac{1}{\sqrt{\lambda_{\textrm{min}}} } \| \hat{\bm{p}} - \bm{p}\|_2 \leq \frac{1}{\sqrt{\lambda_{\textrm{min}}} } \left( \| \hat{\bm{p}} - \bm{f} \|_2 + \| \bm{p} - \bm{f} \| \right),
\end{equation}
where the second line is found by inserting $+\bm{f} - \bm{f}$ and applying the triangle inequality. The first term on the LHS is bounded by the measurement constraint, $\| \hat{\bm{p}} - \bm{f} \|_2 = \| \Xi | \hat{\rho} ) - \bm{f} \|_2 \leq \varepsilon$ or by the minimum value returned in the LS program. The second term on the LHS is constrained by the definition of the noise and error magnitude, $\| \bm{p} - \bm{f} \| = \| \bm{x} + \bm{e} \| \leq \eta + \xi = \varepsilon$. Therefore, the HS-distance between the estimated state and the actual state is bounded,
\begin{equation} \label{full-IC_robustness}
\| \hat{\rho} - \rho \|_2 \leq \frac{2 \varepsilon}{\sqrt{\lambda_{\textrm{min}}} },
\end{equation}
which is saturated when the noise/error bound is saturated and the the estimated state and differs from the actual state in the direction of operator space that corresponds to the largest eigenvalue of $\Xi^{\dagger} \Xi$. The bound shows that the HS-distance between the estimated state and the actual state is linearly proportional to the magnitude of the noise and errors present with proportionality constant dependent on the POVM. Therefore, Eq.~\eqref{full-IC_robustness} satisfies our definition of robustness, such that the estimate produces by standard QST does not ``blow up'' when there is noise and/or errors present. This makes standard QST feasible in most experimental settings. A similar analysis can be applied to QDT and QPT.

\section{Summary and conclusions}
We have presented standard methods for the three types of QT: state, process, and detector tomography. We also discussed how to apply QT in the ideal case, when we have direct access to the probabilities, and the realistic case, where noise and errors exist. In the realistic case, we proved that full-IC measurements are robust to noise and errors. However, these methods, while widely used in experimental settings, are limited to small quantum systems. For example, a full-IC POVM, such as the SIC or MUB, require at least $d^2$ elements. Even for systems consisting of only five qubits, QST requires POVMs with at least 1024 elements. Implementing such measurements is experimentally challenging. Moreover, if such measurements are possible, the classical estimation is still demanding even with convex optimization. Therefore, standard full-IC methods, while useful due to the robustness property, are not applicable to many modern day experiments. In order to feasibly perform QT with experiments of five or more qubits, we need new types of measurement techniques, which will be the subject of subsequent chapters.

\chapter{Informational completeness in bounded-rank quantum tomography} \label{ch:new_IC}
In general quantum tomography (QT) is an expensive task. For example, in the context of quantum state tomography (QST), we saw in Sec.~\ref{sec:components} that the reconstruction of an arbitrary quantum state requires a fully informationally complete (full-IC) POVM, which has at least $d^2$ elements. However, often when we wish to implement QT, we have prior information about the component. This prior information can be applied to reduce the resources required. 

We focus here on QST, and consider the prior information that the quantum state being measured is pure or, more generally, close to pure. Most quantum information tasks require pure states, and therefore most experiments work to engineer these states. In practice, we can use other techniques, such as randomized benchmarking~\cite{Knill2008,Magesan2012,Gaebler2012}, to ensure the experiment operates in this regime. As we shall see, the prior information can be applied to design measurements that uniquely identify pure states with less POVM elements than are required for full-IC measurements. In any practical application, we do not know the state is pure (and in fact it will never be {\em exactly} pure). Therefore, we construct POVMs that are robust to small imperfections in this prior knowledge. We also show that these types of measurements can be generalized to the prior information that the state has bounded-rank, i.e., the rank is less than or equal to some value, $r$. 

The inherent feature of QST that allows for the design of efficient and robust measurements is the positivity constraint on the density matrix. Therefore, the ideas and results presented in terms of QST, can easily be generalized to quantum detector tomography (QDT) and quantum process tomography (QPT) since POVM elements and process matrices are also constrained to be positive. We will return to the generalization at the end of this chapter.

\section{Prior information in QST} \label{sec:prior_info}
In order to reduce the number of resources required for QST, we employ the prior information about the measured quantum state. The goal in most experiments is to prepare pure states, since these are required for the best performance in any quantum information processing task. A pure state is a rank-1 density matrix, $\rho = | \psi \rangle \langle \psi |$, where, $\ket{\psi} = \sum_{k=1}^d c_k \ket{k}$, is fully specified by the $d$ complex state amplitudes $\{ c_k \}$ in a given basis. The state amplitudes are normalized by the trace constraint and the measurements are insensitive to the global phase of the state vector. Therefore, there are $2d-2$ free parameters that specify an arbitrary pure state. The probability of each outcome is quadratically proportional to the state amplitudes,
\begin{equation} \label{ST_quad}
p_{\mu} =  \bra{\psi} E_{\mu} \ket{\psi} = \sum_{j,k} c^*_j c_k (E_{\mu})_{j,k}.
\end{equation}
where $E_{\mu}$ is the $\mu$th POVM element. This quadratic relation is in contrast to the linear relation between the probabilities and the free parameters for full-IC POVMs, which we derived in Sec.~\ref{ssec:noiseless_ST}. Therefore, the number of POVM elements required for pure-state QST is not necessarily equal to the number of free parameters as was the case with standard QST.

Despite the difficulty of the quadratic relationship, POVMs that uniquely identify pure states have been constructed~\cite{Carmeli2014,Carmeli2015,Carmeli2016,Chen2013,Flammia2005,Finkelstein2004,Goyeneche2015,Heinosaari2013,Kech2015,Kech2015a,Ma2016}, and shown to require only $\mathcal{O}(d)$ POVM elements. In fact,
Flammia {\em et al.}~\cite{Flammia2005} proved that the minimum number of POVM elements to reconstruct a pure state is $2d$, not much larger than the number of free parameters. Another approach is based on the compressed sensing methodology~\cite{Gross2010,Flammia2012,Liu2011,Kalev2015}, where certain measurements guarantee a robust estimation of low-rank states with high probability, based on a particular convex optimization program. Compressed sensing techniques were shown to require $\mathcal{O}(d\, \textrm{polylog} \, d)$ measurements~\cite{Liu2011}. In this chapter, we connect these two independent methods by formalizing the notion of informational completeness for pure-state QST.

Since a pure state is represented by a rank-1 density matrix, then the prior information that a state is pure can be generalized to the notion that a state has bounded-rank. A bounded-rank state, $\rho$, has a bounded-number of nonnegative eigenvalues, $\textrm{rank}(\rho) \leq r$. The prior information that the state is pure is then a special case when $r = 1$. A bounded-rank state is in general described by $2dr - r^2$ free parameters~\cite{Kech2015a}, which can be seen in the eigendecomposition. We consider bounded-rank QST for two reasons. First, in many applications, even when the goal is to create a pure state, due to errors in the state preparation the actual state may more closely match a state with higher rank. No actual prepared state will be exactly bounded rank, but may be close to such a state. Second, the mathematical formalism that describes pure-state QST  is easily generalized to the bounded-rank case. We will show there exist POVMs for bounded-rank QST that are more efficient that full-IC POVMs and produce a robust estimate.

\section{Informational completeness in bounded-rank QST} \label{sec:bounded_rank_IC}
We commonly think of POVMs, which are the mathematical descriptions of the readout device, as maps from measured quantum states to probabilities. More generally, we can apply the POVM map to any positive semidefinite (PSD) matrix. In this work we discuss POVMs mapping PSD matrices to a vector of positive numbers (which are not necessarily probabilities), as it highlights the fact that our definitions and results are independent of the trace constraint of quantum states, and only depend on the positivity property. Therefore, we treat the quantum readout device represented by the POVM $\{ E_{\mu} \}$,  more generally as a map, ${\cal M}[\cdot]$, between the space of PSD matrices and the real vector space, $\mathbbm{R}^N$. Particularly, the action of this map on a PSD matrix, $X \geq 0$, is given as ${\cal M}[X] = \bm{s}$, where the elements of the vector $\bm{s}$ satisfy, $s_{\mu} \geq0$ and $\sum_{\mu=1}^N s_\mu=\textrm{Tr} (X)$. The later expression shows that since, by definition, the POVM elements sum to the identity, the POVM always ``measures'' the trace of the matrix $X$. If $X = \rho$, a density matrix, then the condition $s_{\mu} \geq 0$ and $\sum_{\mu} s_{\mu} = \textrm{Tr}(\rho) = 1$ implies that $\{ s_{\mu} \}$ is a probability distribution and thus consistent with the Born rule. It is also useful to define the kernel of the map, ${\rm Ker}({\mathcal M})\equiv \{X:{\cal M}[X]={\bf 0}\}$. Since the POVM elements sum to the identity matrix, we immediately obtain that every $X\in{\rm Ker}({\mathcal M})$ is traceless, $\textrm{Tr}(X)=0$. The converse is not true; a traceless matrix is not necessarily entirely contained in the Kernel of $\mathcal{M}$.

When considering bounded-rank QST, a natural notion of informational completeness emerges~\cite{Heinosaari2013, Carmeli2014, Kalev2015}, referred to as {\em rank-$r$ completeness}. A measurement is rank-$r$ complete if the outcome probabilities uniquely distinguish the PSD matrix $X$, with rank $\leq r$, from any other PSD matrix with rank $\leq r$, more formally:
\begin{definition}{\bf (Rank-$\bm{r}$ complete)} \label{def:rankr_comp}
Let  $\mathcal{S}_r=\{ X | X \geq0, {\rm rank}(X)\leq r \}$ be the set of PSD matrices with  rank $\leq r$.  A POVM is said to be rank-$r$ complete if  
\begin{equation} 
\forall \, X_1, X_2 \in \mathcal{S}_r,\,  X_1\neq X_2 \textrm{ iff } \mathcal{M}[X_1]\neq\mathcal{M}[X_2],
\end{equation}
except for possibly a set of rank-$r$ PSD matrices that are dense on a set of measure zero, called the ``failure set.''
\end{definition}
\noindent We can alternatively write the definition in terms of any norm, $\| \cdot \|$: a POVM is rank-$r$ complete when $\| X_1 - X_2 \| = 0$ if and only if $\| \mathcal{M}[X_1] - \mathcal{M}[X_2]\|=0$.

\begin{figure}[ht]
\centering
\includegraphics[width=\linewidth]{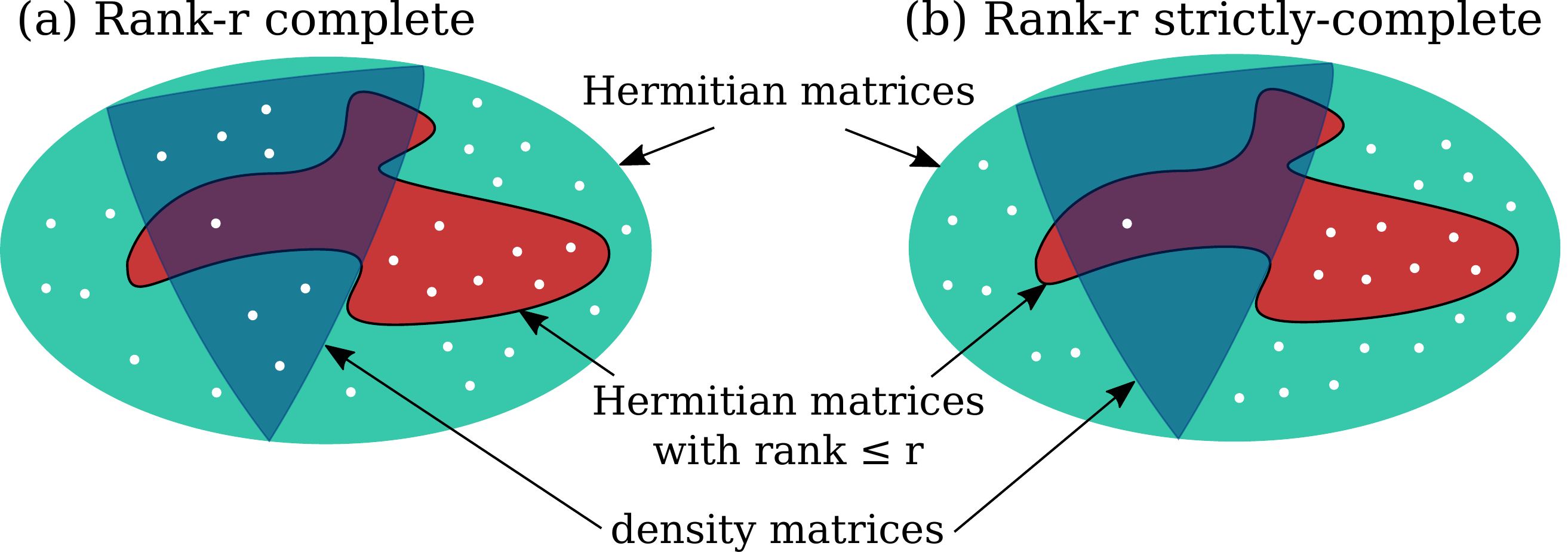}
\caption[Two notions of completeness in bounded-rank QST]{{\bf Two notions of completeness in bounded-rank QST.} The white dots represent Hermitian matrices, positive or not, that are consistent with the (noiseless) measurement record.  {\bf (a) Rank-$r$ completeness.} The measurement record, distinguishes the rank $\leq r$ state from any other rank $\leq r$ PSD matrix. However, there generally will be infinitely many other states,  with rank greater than $r$, that are consistent with the measurement record.  {\bf (b)  Rank-$r$ strict-completeness.} The measurement record  distinguishes the rank $\leq r$ state from any other PSD matrix. Thus it is unique in the convex set of PSD matrices.}
\label{fig:illustration}
\end{figure}

When applied to quantum states, the probabilities from a rank-$r$ complete POVM uniquely identify the rank $\leq r$ state from within the set of all PSD matrices with rank $\leq r$, $\mathcal{S}_r$, which includes all rank-$r$ density matrices. Fig.~\ref{fig:illustration}a illustrates the notion of rank-$r$ completeness. The measurement probabilities cannot uniquely identify states in this way if they lie in the failure set, as was considered in~\cite{Flammia2005,Goyeneche2015}. However, in the ideal case of no noise, the chances of randomly hitting a state in that set is vanishingly small. We comment on the implications and structure of the failure set in the next chapter.

In Ref.~\cite{Carmeli2014} an alternative, but equivalent, definition of rank-$r$ complete was proven using $\textrm{Ker}(\mathcal{M})$: A POVM is rank-$r$ complete if for all $X_1,X_2 \in \mathcal{S}_r$, with $\, X_1\neq X_2$, the difference $\Delta= X_1-X_2$ is not in $\textrm{Ker}(\mathcal{M})$, i.e., there exists an $E_{\mu}$, such that $\textrm{Tr}(E_\mu \Delta)\neq0$.  Carmeli~{\em et al.}~\cite{Carmeli2014} showed that a necessary and sufficient condition for a measurement to be rank-$r$ complete for QST is that every nonzero $A\in \textrm{Ker}(\mathcal{M})$ has $\textrm{max}(n_+[A], n_-[A]) \geq r+1$, where $n_{+}[\cdot]$ and $n_{-}[\cdot]$ are the number of strictly positive and strictly negative eigenvalues of a matrix, respectively. Carmeli~{\em et al.}~\cite{Carmeli2014} also showed a sufficient condition for rank-$r$ completeness is every nonzero $A\in \textrm{Ker}(\mathcal{M})$ has $\textrm{rank}(A)\geq 2r+1$. Using the sufficient condition alone, it was shown that the expectation values of particular $4r(d-r)$ observables corresponds to rank-$r$ complete measurement~\cite{Heinosaari2013}.

The notion of rank-$r$ completeness can also be applied to bounded-rank (not necessarily positive) Hermitian matrices. Let $\mathcal{A}$ be the set of all bounded-rank Hermitian matrices, $\mathcal{A} = \{ H | H = H^{\dagger}, \, \textrm{rank}(H) \leq r\}$. Then there exists POVMs whose measurement vector uniquely identifies an arbitrary Hermitian matrix within $\mathcal{H}$. We call these Hermitian rank-$r$ complete, and a formal definition can be made similar to Definition~\ref{def:rankr_comp}. In Ref.~\cite{Kech2015a}, it was shown that a set of of $4r\lceil\frac{d-r}{d-1}\rceil$ random orthonormal bases are Hermitian rank-$r$ complete. Hermitian rank-$r$ completeness is a sufficient condition for rank-$r$ completeness, since the set of bounded-rank PSD matrices is a subset of bounded-rank Hermitian matrices. In fact, it can be shown that the sufficient condition for rank-$r$ completeness given by Carmeli~{\em et al.}~\cite{Carmeli2014} is equivalent to this definition for Hermitian matrices.

The definition of rank-$r$ complete POVMs guarantees the uniqueness of the reconstructed state in the set ${\cal S}_r$, but it does not say anything about higher-rank states. There may be other density matrices, with rank greater than $r$ that are consistent with the measurement probabilities. Since ${\cal S}_r$ is a nonconvex set it may be difficult to differentiate between the unique rank-$r$ density matrix and these higher-rank states, particularly in the presence of noise or other experimental imperfections. To overcome this difficulty, we consider a ``stricter" type of POVM which excludes these higher-rank states.  This motivates the following definition~\cite{Chen2013,Carmeli2014,Kalev2015}:
\begin{definition}{\bf (Rank-$r$ strictly-complete)} \label{def:rankr_strictcomp}
Let  $\mathcal{S}=\{X | X\geq0\}$ be the set of PSD matrices. A measurement is said to be rank-$r$ strictly-complete if
\begin{equation}
 \forall \, X_1 \in \mathcal{S}_r, \textrm{ and } \forall \, X_2 \in \mathcal{S},\, X_1\neq X_2 \textrm{ iff } \mathcal{M}[X_1]\neq\mathcal{M}[X_2],
\end{equation}
except for possibly a set of rank-$r$ PSD matrices that are dense on a set of measure zero, called the ``failure set.''
\end{definition}
\noindent Alternatively, a POVM is rank-$r$ complete when $\| X_1 - X_2 \| = 0$ if and only if $\| \mathcal{M}[X_1] - \mathcal{M}[X_2]\|=0$. Clearly, a POVM that satisfies Definition~\ref{def:rankr_strictcomp} also satisfies Definition~\ref{def:rankr_comp}. For QST, when the rank of the state being measured is promised to be less than or equal to $r$, the probabilities from a rank-$r$ strictly-complete POVM distinguish this state from any other PSD matrix, of {\em any} rank (except on the failure set). Fig.~\ref{fig:illustration}b illustrates the notion of rank-$r$ strict-completeness. 

Carmeli~{\em et al.}~\cite{Carmeli2014} showed that a POVM is rank-$r$ strictly-complete if, and only if, every nonzero $A\in{\rm Ker}({\cal M})$ has $\min(n_{+}[A],n_{-}[A])\geq r+1$. This condition relies on the PSD property of the matrices. To date, there are only a few known POVMs that are proven to be rank-$r$ strictly-complete~\cite{Chen2013,Ma2016}. In Chapter~\ref{ch:constructions}, we present new strictly-complete POVMs with ${\cal O}(rd)$ elements, which is the same number of POVM elements as a rank-$r$ complete POVM.

The definition of strict-completeness does not have a related notion for Hermitian matrices, in contrast to rank-$r$ completeness. To see this, let us apply the definition of strict-completeness for bounded-rank Hermitian matrices in the context of QST, ignoring positivity. Let $A$ be a Hermitian matrix with $\textrm{rank}(A)\leq r$. To be (nontrivially) strictly-complete the POVM should be able to distinguish $A$ from any Hermitian matrix, of any rank, with less than $d^2$ linearly independent POVM elements. (If the POVM has $d^2$ linearly independent POVM elements, it is fully-IC and can distinguish any Hermitian matrix from any other.) However, for a POVM with less than $d^2$ linearly independent elements there are necessarily infinitely many Hermitian matrices with rank $> r$ that produce the same noiseless measurement vector as $A$. Therefore, positivity is the essential ingredient that allows us to define strict-completeness with less than $d^2$ linearly independent elements. The positivity condition, which appears in all three types of QT, is powerful constraint for efficient QT.

\section{Reconstruction with ideal bounded-rank QST } \label{sec:estimation_wout_noise}
The differences between rank-$r$ complete and rank-$r$ strictly-complete has implications for the way we reconstruct the unknown quantum state. For now, we assume that such measurements exist that satisfy the above definitions (we will construct examples of these measurements in the next chapter). The definitions above state that the measurements uniquely identify the bounded-rank PSD matrix within some set. In order to accomplish QST, we need methods to identify this unique PSD matrix. In this section, we consider the ideal situation that the probabilities are known exactly and there are no other errors in the system. The noiseless and errorless case does not correspond to any real application but is useful in establishing the fundamental properties of the different measurements. We return to the realistic case in the next section. 

In each definition we allowed a failure set where the probabilities do not uniquely identify the quantum state with respect to the set given in the definition. We also specified that this set must have zero volume. Therefore, if the measured state is random with respect to the measurement basis, it is vanishingly unlikely that it will be an element of this set.  Therefore, in the ideal limit for QST, it is vanishingly unlikely that the failure set will impact the reconstruction.

\subsection{Reconstruction with rank-$r$ complete POVMs} \label{sec:noiseless_rankr_comp}
Many rank-$r$ complete POVMs are constructed by deriving a set of quadratic equations that are solvable when the measured state is bounded-rank, e.g., the constructions provided in Refs.~\cite{Goyeneche2015, Flammia2005}. Therefore, when we consider the ideal case of QST, we can solve these quadratic equations to uniquely reconstruct the bounded-rank quantum state. Further details are provided in Sec.~\ref{sec:decomp_method}.

Some rank-$r$ complete measurements, however, do not provide a set of quadratic equations in their derivation, e.g., the POVM provided in Ref.~\cite{Carmeli2015}. Therefore, we must use numerical methods to reconstruct the quantum state. The numerical search must be constrained to the set of rank-$r$ states. One possible optimization program is based on minimizing the least-squares (LS) distance between the probabilities and the expected probabilities from a rank-$r$ quantum state,
\begin{align} \label{rankr_comp_program}
\underset{X}{\textrm{minimize:}} \quad & \| \mathcal{M}[X] - \bm{p} \|_2 \nonumber \\
\textrm{subject to:} \quad & X \in \mathcal{S}_r
\end{align}
However, the constraint $X \in \mathcal{S}_r$ is nonconvex, and therefore this constrained LS program cannot be solved with convex optimization techniques, like the ones discussed in Sec.~\ref{sec:numerical_methods}. Nonconvex optimization is, in general, difficult due to the existence of local minima.

One possible algorithm to solve the program in Eq.~\eqref{rankr_comp_program} is based on gradient-projection. The basic procedure for gradient-projection is to alternate a gradient descent approach with a projection onto the set $\mathcal{S}_r$~\cite{Meka2009, Calamai1987,Figueiredo2007}. We refer to this method throughout as ``rank-$r$-projection.'' We denote the LS optimization function as $g(X) = \| \mathcal{M}[X] - \bm{p} \|_2$, with gradient,
\begin{equation}
\vec{\bigtriangledown} g(X_r) = 2 \mathcal{M}^{\dagger} \left[ \mathcal{M}[X_r] - \bm{p} \right],
\end{equation}
where $\mathcal{M}^{\dagger}[\cdot]$ is the conjugate map defined by $\textrm{Tr}(A \mathcal{M}[B]) = \textrm{Tr}(\mathcal{M}^{\dagger}[A]B)$. The algorithm starts by generating a random rank-$r$ PSD matrix, $X_r^{(0)}$. We then evaluate $g(X_r^{(0)})$, and if $g(X_r^{(0)}) > \gamma_1$, which is some stopping threshold, then we also evaluate $\vec{\bigtriangledown}g(X_r^{(0)})$. From the gradient we produce a new estimate $X^{(1)} = X_r^{(0)} - a \vec{\bigtriangledown}g(X_r^{(0)})$, where $a$ is a small constant. The new estimate is not necessarily a rank-$r$ PSD matrix, and so we project $X^{(1)}$ onto the set $\mathcal{S}_r$ to give $X^{(1)}_r = \mathcal{P}[X^{(1)}]$. The projection, $\mathcal{P}[\cdot]$, is accomplished by diagonalizing $X^{(1)}$ and setting the smallest $d-r$ eigenvalues to zero (if there are greater than $d-r$ negative eigenvalues we must also set these to zero in order for the matrix to be PSD). We then repeat the procedure until either $g(X_r) \leq \gamma_1$ or $\| \vec{\bigtriangledown} g(X_r) \| \leq \gamma_2$ for some pre-specified $\gamma_1$ and $\gamma_2$ based on the implementation. If the algorithm stops due to the gradient threshold, then we have likely found a local minimum, which is not the desired result. In order to find the desired global minimum, we repeat the procedure with a different initial guess, $X_r^{(0)}$. This entire processes is repeated until the function threshold, $\gamma_1$, is reached. When such a solution is found, the result produces a rank-$r$ PSD matrix, $\hat{X}_r$ which has $\| \mathcal{M}[X_r] - \bm{p} \|_2 \leq \gamma_1$. For ideal QST, we take $\gamma_1$ and $\gamma_2$ near zero, approximately $10^{-5}$. 

Empirically, we find that the run time of this algorithm can be very long. The time is very dependent on the local minima that necessarily exist, since the set $\mathcal{S}_r$ is not a convex set. These minima act as traps for the gradient descent search and require that the algorithm restart with a new random seed. We do not know the number of minima and thus how likely it is to encounter one in the optimization. Therefore, this method is not generally a practical method for reconstruction.

\subsection{Reconstruction with rank-$r$ strictly-complete POVMs} \label{ssec:strict_comp_noiseless}
In the previous section, we saw that rank-$r$ complete POVMs are not compatible with convex optimization. However, this is not the case for rank-$r$ strictly-complete POVMs. The ideal measurement vector from a rank-$r$ strictly-complete POVM uniquely identifies the rank-$r$ PSD within the convex set of all PSD matrices. Therefore, we can design convex optimization programs for reconstruction of bounded-rank quantum states. This is formalized in the following corollary for the ideal measurement case:
\begin{corollary}{\bf (Uniqueness)} \label{cor:uniqueness}
Let $X_{\rm r}$ be a PSD matrix with rank $\leq r$, and let $\bm{s}= \mathcal{M}[X_r]$ be the corresponding measurement vector of a rank-$r$ strictly-complete POVM. Then, the estimate, $\hat{X}$, which produces the minimum of either,
\begin{align} \label{general_positive_CS}
\underset{X}{\rm minimize:} \quad &\mathcal{C}(X) \nonumber \\
{\rm subject\, to:}\quad &\mathcal{M}[X]=\bm{s}\nonumber \\
	& X \geq 0,
\end{align}
or,
\begin{align}
\label{general_norm_positive_CS}
\underset{X}{\rm minimize:} \quad & \Vert\mathcal{M}[X]-\bm{s}\Vert \nonumber \\ 
{\rm subject \, to:} \quad &X \geq 0,
\end{align}
where $\mathcal{C}(X)$ is a any convex function of $X$, and $\| \cdot \|$ is any norm function, is uniquely:  $\hat{X} = X_r$.
\end{corollary}
\noindent {\em Proof:} This is a direct corollary of the definition of strict-completeness, Definition~\ref{def:rankr_strictcomp}. Since, by definition, the probabilities of rank-$r$ strictly-complete POVM uniquely determine $X_r$ from within the set of all PSD matrices, its reconstruction becomes a feasibility problem over the convex set $\{ \mathcal{M}[X]=\bm{s},X\geq0\}$,
\begin{equation} \label{feasibility}
{\rm find}\; X\;\; {\rm s.t.}\; \mathcal{M}[X]=\bm{s}\, \,  \textrm{and} \, \, X \geq 0.
\end{equation}
The solution for this feasibility problem is $X_r$ uniquely. Therefore, any optimization program, and particularly an efficient convex optimization program that looks for the solution within the feasible set, is guaranteed to find $X_r$. $\square$

\noindent In Ref.~\cite{Kech2015} this was proven for the particular choice, $\mathcal{C}(X)=\textrm{Tr}(X)$, and also in the context of compressed sensing measurements in Ref.~\cite{Kalev2015}. 

The corollary implies that strictly-complete POVMs allow for the reconstruction of bounded-rank PSD matrices via convex optimization even though the set of bounded-rank PSD matrices is nonconvex. Moreover, all convex programs over the feasible solution set, i.e., of the form of Eqs.~\eqref{general_positive_CS} and~\eqref{general_norm_positive_CS}, are equivalent for this task. For example, this result applies to maximum-(log)likelihood estimation for QST~\cite{Hradil1997}, given in Eq.~\eqref{sec:cvx_QT}, where $\mathcal{C}(\rho) =-\log( \prod_{\mu}\textrm{Tr}(E_\mu\rho)^{p_\mu})$.  Corollary~\ref{cor:uniqueness} does not apply for PSD matrices in the measurements failure set, if such set exists. 

One can also include the trace constraint in Eqs.~\eqref{general_positive_CS} and~\eqref{general_norm_positive_CS}. For noiseless QST, this is redundant since any POVM ``measures'' the trace of a matrix.  Thus, if we have prior information that $\textrm{Tr}(X)=1$, then the feasible set in Eq.~\eqref{feasibility} is equal to the set $\{ X \, | \,\mathcal{M}[X]=\bm{p}, \, X \geq 0, \, \textrm{Tr}(X) = 1 \}$.

\section{Estimation in the presence of noise and errors} \label{sec:estimation_wnoise}
Any real implementation of QST will necessarily have sources of noise and errors, and therefore it is imperative that the QST protocol be robust to such effects. In order to produce an estimate for this realistic case, we use numerical optimization. In the previous section we saw that rank-$r$ complete POVMs require nonconvex programs. Due to the complicated nature of this type of program, we forgo a discussion of estimation with rank-$r$ complete POVMs and focus only on rank-$r$ strictly-complete POVMs. In this section, we use the formalism for describing noise and errors that was introduced in Sec.~\ref{ssec:errors}. We additionally model a new type of error that is inherent to rank-$r$ strictly-complete POVMs. The definition rank-$r$ strict-completeness assumes that the measured state has bounded rank. However, in any application the measured state will never be exactly bounded-rank due to unavoidable errors in the experimental apparatus. We call these preparation errors, since they cause the prepared quantum state to differ from the target bounded-rank quantum state.



We denote the state that is actually prepared, $\rho_{\rm a}$, which is, in general, full rank. However, since the goal was to prepare a bounded-rank state, the actual state is close to such a state, $\rho_r$. The ``closeness'' will depend on the magnitude of the preparation errors based on some measure. We can relate the two states with the error matrix, $Y$, such that $\rho_{\textrm{a}} \triangleq \rho_r + Y$. The matrix $Y$ is only constrained by the fact that $\rho_{\textrm{a}}$ and $\rho_r$ are both quantum states. The prior information that the state is close to a bounded-rank state then corresponds to $\| \rho_{\rm a} - \rho_r \|_2 \leq \upsilon$ where $\| \cdot \|_2$ is the Hilbert-Schmidt distance and $\upsilon$ is a small constant. 

In Sec.~\ref{sec:noisy_QT}, we derived expressions for the measurement vector when there exists errors in the POVM and noise in the measurement. We express the actual POVM as $\mathcal{M}' = \mathcal{M} + \mathcal{X}$, where $\mathcal{M}$ is the target POVM map and $\mathcal{X}$ represents the errors in the map. The noise in each outcome is expressed by the vector, $\bm{e}$. We can also include preparation errors in this expression,
\begin{align} \label{f_with_all}
\bm{f} &= \mathcal{M}'[\rho_{\rm a}]+ \bm{e}, \nonumber \\
	  &= \mathcal{M}[\rho_r] + \mathcal{X}[  \rho_{\rm a} ] + \mathcal{M}[Y] + \bm{e}, \nonumber \\
	  &= \bm{p} + \bm{x} + \bm{y} + \bm{e},
\end{align}
where $\bm{p} = \mathcal{M}[ \rho_r ] $ the probability of each outcome expected from a rank-$r$ state, $\bm{x} =  \mathcal{X} [ \rho_{\rm a} ]$ the contribution of the measurement errors, and $\bm{y} = \mathcal{M}[ Y]$ the contribution from the preparation errors. We assume that the contribution of measurement errors and noise is bounded for any quantum states, $\sigma$, $\| \mathcal{X} [ \sigma ] \|_2 \leq \eta$, and $\| \bm{e} \|_2 \leq \xi$. Then the total error and noise level can be bounded,
\begin{equation} \label{noise_bound}
\| \bm{f} - \bm{p} \|_2 = \| \bm{x} + \bm{y}  + \bm{e} \|_2 \leq \eta + \xi + \| \mathcal{M}[Y] \|_2 = \varepsilon + \| \mathcal{M}[Y] \|_2.
\end{equation}
where we define $\varepsilon = \eta + \xi$, for reasons that will be clear later. 

The value of $\| \mathcal{M}[Y] \|_2$ is related to the magnitude of the preparation errors, $\upsilon$. This is seen by separating the distance, $\| \rho_{\rm a} - \rho_r \|_2$ into two terms corresponding to the projection onto the Kernel ($\pi^{\perp}[\cdot]$) and Image ($\pi[\cdot]$) (the subspace of the operator space orthogonal to the Kernel),
\begin{equation} \label{proj_sep}
\| \rho_{\rm a} - \rho_r \|_2^2 = \| \pi[\rho_{\rm a} - \rho_r] \|_2^2 + \| \pi^{\perp}[\rho_{\rm a} - \rho_r] \|_2^2 \leq \upsilon^2.
\end{equation}
The first term can be bounded by an inequality similar to Eq.~\eqref{meas_bound}, $ \frac{1}{\lambda_{\rm max}} \| \mathcal{M}[ [\rho_{\rm a} - \rho_r]\|_2 \leq \| \pi[\rho_{\rm a} - \rho_r] \|_2^2 $, where $\lambda_{\rm max}$ is the maximum eigenvalue of $\Xi^{\dagger} \Xi$, the POVM matrix squared. Rearranging Eq.~\eqref{proj_sep} gives,
\begin{align} \label{prep_bound}
\| \pi[\rho_{\rm a} - \rho_r] \|_2^2 &\leq \upsilon^2 - \| \pi^{\perp}[ \rho_{\rm a} - \rho_r] \|_2^2, \nonumber \\
\| \mathcal{M}[ \rho_{\rm a} - \rho_r] \|_2^2 &\leq \lambda_{\rm max} \upsilon^2 -  \lambda_{\rm max} \| \rho_{\rm a} - \rho_r \|_2^2 \leq \lambda_{\rm max} \upsilon^2 .
\end{align}
This leads to the bound, $\| \mathcal{M}[Y]\|_2 \leq \sqrt{\lambda_{\rm max}} \upsilon$.

Noise and errors also cause the failure set to have an effect on bounded-rank QST. Finkelstein showed that in the presence of noise and errors the failure-set in fact has a finite measure~\cite{Finkelstein2004}. Therefore, there is a nonzero probability that the actual state lies within this failure set. In this case the measured outcomes from the rank-$r$ strictly-complete POVMs would fail to produce a robust estimate. In this section, we ignore the effects of the failure set but discuss it in the context of specific POVMs in the next chapter.

\subsection{Estimation with Rank-$r$ strictly-complete POVMs} \label{ssec:rankr_strictcomp_robustness}
The estimate produced from a strictly-complete POVM are provably robust to all sources of noise and errors, including all preparation errors. This is formalized in the following corollary:
\begin{corollary}{\bf (Robustness)} \label{cor:robustness}
Let $X_{\textrm{a}}$ be the actual prepared PSD and let $\bm{f}= \mathcal{M'}[X_{\textrm{a}}]+\bm{e}$ be the measurement vector (with noise and errors) of a rank-$r$ strictly-complete POVM, such that $\Vert \mathcal{M}[X_a] - \bm{f}\Vert \leq \varepsilon$ and $\| X_{\rm a} - X_r\| \leq \upsilon$, for some bounded-rank PSD matrix $X_r$. Then the PSD matrix, $\hat{X}$, that produces the minimum of,
\begin{align} \label{general_positive_CS_noisy}
\underset{X}{\rm minimize:} \quad &\mathcal{C}(X) \nonumber \\
{\rm subject\, to:}\quad &\Vert\mathcal{M}[X]-\bm{f}\Vert_2 \leq \varepsilon \nonumber \\
	& X \geq 0,
\end{align}
or,
\begin{align}
\label{general_norm_positive_CS_noisy}
\underset{X}{\rm minimize:} \quad & \Vert\mathcal{M}[X]-\bm{f}\Vert_2 \nonumber \\ 
{\rm subject \, to:} \quad &X \geq 0,
\end{align}
where $\mathcal{C}(X)$ is a any convex function of $X$, is robust:  $\Vert\hat{X} - X_r \Vert_2 \leq C_1\varepsilon + C_2 \upsilon$ and $\Vert\hat{X} - X_{\textrm{a}} \Vert_2\leq C_1 \varepsilon + 2 C_2 \upsilon$, where $\Vert\cdot\Vert_2$ is the Hilbert-Schmidt distance, and $C_1$ and $C_2$ are constants which depends only on the measurement.
\end{corollary}
\noindent {\em Proof:} The proof comes from Definition~\ref{def:rankr_strictcomp}, which states for a strictly-complete POVM, $X_r \in \mathcal{S}_r$ and $X \in \mathcal{S}$, $\| X_r - X\| = 0$ if and only if $\| \mathcal{M}[X_r] - \mathcal{M}[X] \|=0$. We can express this as an inequality relation,
\begin{equation} \label{strict_comp_ineq}
\alpha \| X_r - X\| \leq \| \mathcal{M}[X_r] - \mathcal{M}[X] \| \leq \beta \| X_r - X\|,
\end{equation}
where $\alpha$ and $\beta$ are real and depend on the POVM. The definition of rank-$r$ strict-completeness constrains the value of $\alpha$ to be strictly positive, $\alpha > 0$~\cite{Blumensath2014}. Otherwise, there may exist a case where $\| \mathcal{M}[X_r] - \mathcal{M}[X] \|= 0$ when $ \| X_r - X\| \neq 0$, which contradicts the definition. The RHS side can be derived from Eq.~\eqref{prep_bound}, such that $\beta = \sqrt{\lambda_{\rm max}}$.

Now, if we take $X = \hat{X}$ for Eq.~\eqref{strict_comp_ineq}, which is the estimated PSD matrix from either program in Eqs~\eqref{general_positive_CS_noisy} or~\eqref{general_norm_positive_CS_noisy}, then,
\begin{align} \label{strict_comp_bound_Xr}
\| X_r - \hat{X}\| &\leq  \frac{1}{\alpha} \| \mathcal{M}[X_r] - \mathcal{M}[\hat{X}] \|, \nonumber \\
 			&\leq  \frac{1}{\alpha} ( \| \mathcal{M}[X_r] - \bm{f} \| + \underbrace{\| \mathcal{M}[\hat{X}]  - \bm{f} \|}_{\leq \varepsilon} ), \nonumber \\
			&\leq  \frac{1}{\alpha} ( \underbrace{\| \mathcal{X}[X_{\rm a} ] + \mathcal{M}[Y] + \bm{e} \|}_{\leq \varepsilon + \beta \upsilon} + \varepsilon ), \nonumber \\
			&\leq \frac{2 (\varepsilon + \beta \upsilon/2)}{\alpha} = C_1 \varepsilon + C_2  \upsilon,
\end{align}
by expanding $\bm{f}$ and where $C_1 = 2/\alpha$ and $C_2 = \beta/\alpha$. The second term in the second line is from the the constraint in the convex optimization program in Eq.~\eqref{general_positive_CS_noisy} or the optimization function in Eq.~\eqref{general_norm_positive_CS_noisy}. The first term in the third line is from the bound on the noise and magnitudes in Eq.~\eqref{noise_bound} as well as the bound on preparation errors from Eq.~\eqref{prep_bound}. To get the second inequality of the corollary, which compares the prepared PSD matrix to the estimate, we apply the triangle inequality,
\begin{align}
\| X_{\textrm{a}} - \hat{X}\| &\leq \| X_{\rm a} - X_r \|+ \underbrace{\| \hat{X} - X_r \|}_{\leq \frac{2 \varepsilon}{\alpha} + \frac{\beta \upsilon}{\alpha}} , \nonumber \\
			&\leq  \frac{1}{\alpha}   \underbrace{\| \mathcal{M}[Y] \|}_{\leq \beta \upsilon}+ \frac{2 \varepsilon}{\alpha} + \frac{\beta \upsilon}{\alpha} , \nonumber \\
			&\leq \frac{2 (\varepsilon + \beta \upsilon)}{\alpha} = C_1 \varepsilon + 2 C_2 \upsilon.
\end{align}
The first line uses the result from Eq.~\eqref{strict_comp_bound_Xr} and the second line uses the bound on the preparation errors in Eq.~\eqref{prep_bound}. $\square$

\noindent In the context of QST, $X_{\rm a} = \rho_{\rm a}$ and $X_{\rm r} = \rho_r$, the actual density matrix prepared and a nearby bounded-rank density matrix, respectively. We do not have an analytic expression for the constant $\alpha$. In Ref.~\cite{Kech2015}, a similar proof was given for the particular choice $\mathcal{C}(X)=\textrm{Tr}(X)$ for QST. In this proof the constant $C_1$ is derived in more detail, but still has no known analytic form. 

In Ref.~\cite{Kalev2015}, Corollary~\ref{cor:robustness} was also studied in the context of compressed sensing measurements. As in the ideal case, the trace constraint is not necessary for Corollary~\ref{cor:robustness}, and in fact leaving it out allows us to make different choices for $\mathcal{C}(X)$, as was done in Ref.~\cite{Kech2015}. However, for a noisy measurement vector, the estimated matrix $\hat{X}$ is generally not normalized, $\textrm{Tr}(\hat{X})\neq1$. The final estimation of the state is then given by $\hat{\rho} = \hat{X}/\textrm{Tr}(\hat{X})$. In principle, we can consider a different version of Eqs.~\eqref{general_positive_CS_noisy} and~\eqref{general_norm_positive_CS_noisy} where we explicitly include the trace constraint.

The corollary assures that if the actual quantum state is close to bounded-rank and is measured with strictly-complete POVM, then it can be robustly estimated with any convex program, constrained to the set of PSD matrices. In particular, it implies that all convex estimators perform qualitatively the same for low-rank state estimation. This may be advantageous, especially when considering QT of high-dimensional systems. This also unifies previously proposed estimation programs for bounded-rank QST, such as trace-minimization~\cite{Gross2010}, maximum-likelihood, and maximum entropy~\cite{Liu2012,Teo2012}. While we cannot currently derive an analytic expression for the constant $\alpha$ for an arbitrary POVM, the scaling of the robustness bound in Corollary~\ref{cor:robustness} is linear, which is exactly the same as full-IC POVMs, derived in Sec.~\ref{ssec:full_robustness}. Therefore, strictly-complete POVMs perform very similar to full-IC POVMs in realistic applications.

\section{General bounded-rank quantum tomography}
The methodology we applied to bounded-rank QST can be generalized to both detector (QDT) and process tomography (QPT). The inherent feature that allows for this conversion is that, like quantum states, both detectors and processes are represented by PSD matrices. For detector tomography the PSD matrices are the POVM elements while for QPT the PSD matrix is the process matrix. Moreover, there often exists prior information that these PSD matrices are bounded-rank, or near bounded-rank. Therefore, QDT and QPT fit the framework outlined for bounded-rank QST. This means we can create ways to characterize bounded-rank readout devices and processes that are more efficient than the standard methods described in Sec.~\ref{sec:full-IC}.

Mathematically, the estimation problem for the three different types of QT differ by the trace constraint, as outlined in Sec.~\ref{sec:numerical_methods}. However, in Definitions~\ref{def:rankr_comp} and~\ref{def:rankr_strictcomp} as well as in Corollaries~\ref{cor:uniqueness} and~\ref{cor:robustness}, we ignored the trace constraint for QST. We comment on the effect of this constraint in bounded-rank QDT and QPT below.

\subsection{Bounded-rank QDT} \label{ssec:br_QDT}
Many quantum information protocols require quantum readout devices that are described by rank-1 POVM elements, for example, the SIC POVM introduced in Ref.~\cite{Renes2004a}. Rank-1 POVM elements can be expanded similar to pure states, $E_{\mu} = | \tilde{\phi}_{\mu} \rangle \langle \tilde{\phi}_{\mu} |$, except in this case $\ket{\tilde{\phi}_{\mu}}$ is an unnormalized vector. This differs from QST only in that the trace of the POVM elements are not constrained, since $\ket{\tilde{\phi}_{\mu}}$ is unnormalized. Therefore, if we perform the estimation for QDT on individual POVM elements, which was the second method discussed in Sec.~\ref{ssec:noiseless_DT}, then we can directly apply the definitions and corollaries from above to develop efficient methods for QDT.

The notion of rank-1 completeness and strict-completeness for QDT applies to the set of probing states used to characterize the POVM elements. We can construct sets of probing states that satisfy Definition~\ref{def:rankr_comp} and~\ref{def:rankr_strictcomp}, and we consider such sets in the next chapter. Therefore, these probing states are able to fully characterize rank-1 projectors with less than the $d^2$ states required for full-IC QDT, discussed in Sec.~\ref{ssec:noiseless_DT}. 

In most real applications, the POVM elements that describe the detector are not exactly rank-1. In this case, measurement errors in the physical apparatus cause the readout device to be described by a different POVM. This is equivalent to the preparation errors we discussed in Sec.~\ref{sec:estimation_wnoise} for QST. By analogy to the the robustness bounds derived above for preparation errors, a set of rank-1 strictly-complete probing states are robust to errors in the implementation of the POVMs, and also to errors in the preparation of the states and noise in the measurement.

We have so far discussed QDT with the second method from Sec.~\ref{ssec:noiseless_DT}, which is individually estimating the POVM elements. However, in Sec.~\ref{ssec:noiseless_DT}, we introduced another method for estimation in QDT, which performs the estimation collectively with all POVM elements. For this method, we are able to apply the trace constraint within the convex optimization program. While the definitions of rank-$r$ complete and strictly-complete are independent of this constraint, including it in the estimation may allow for the creation of sets of probing states with even less elements.

\subsection{Bounded-rank QPT} \label{ssec:br_QPT}
The prior information that a process is rank-1 corresponds to knowledge that it is a unitary process. Unitary processes are required in most quantum information protocols such as quantum computing. The process matrix that represents a unitary process is $\chi = | U)( U|$, where $| U)$ is the vectorized form of a unitary matrix $U$. While the process matrix is a $d^2 \times d^2$ matrix, we can still directly apply the definitions and corollaries from above to QPT.

The notion of rank-1 complete and strictly-complete for QPT applies to the combination of probing states and POVMs used to characterize the process matrix. We can construct a combination of states and POVMs that satisfy Definition~\ref{def:rankr_comp} and~\ref{def:rankr_strictcomp}. Therefore, rank-1 complete and strictly-complete measurements are able to fully characterize a unitary process with less than the $d^2$ probing states and full-IC POVM that is required for the standard method of QPT. We consider such methods in Chapter~\ref{ch:PT}.

In most real applications, the process is not exactly unitary due to sources of errors such as decoherence, inhomogeneity in the control, or imperfect calibrations. We call these process errors, and they cause the process matrix that describes the actual process to not match the target unitary process. This is equivalent to the preparation errors we discussed in terms of QST in Sec.~\ref{sec:estimation_wnoise}. By analogy to the robustness bounds derived above for QST, a set of rank-1 strictly-complete probing states and POVMs for QPT is robust to process errors. Moreover, by the same reasoning, such sets are also robust to to errors in the preparation of the states, implementation in the POVMs, and noise in the measurement.

We have so far discussed QPT without applying the TP constraint, which was derived in Sec.~\ref{ssec:process_intro}. This constraint can be used to create sets of probing states and POVMs with less elements than ones derived from Definitions~\ref{def:rankr_comp} and~\ref{def:rankr_strictcomp}. In Chapter~\ref{ch:PT}, we introduce such measurements and discuss how the TP constraint plays in a role in their construction.

\section{Summary and conclusions}
QST is a demanding experimental protocol, but in this chapter, we showed that certain types of POVMs, called rank-$r$ complete and rank-$r$ strictly-complete, can accomplish QST more efficiently when there is prior information that the prepared state has bounded-rank. This prior information corresponds to the goal of most quantum information processors, so it is reasonable in most applications. Moreover, we proved that even when the actual state is not exactly pure, strictly-complete POVMs still produce a robust estimate. This is very similar to the result for full-IC POVMs.  We also generalized these results to QDT and QPT where the same definitions and corollaries hold, since processes and readout devices are described by PSD matrices, and we often have prior information that they are bounded-rank. While strictly-complete POVMs are robust to preparation errors, we still have yet to show how many POVM elements are required. We answer this question in the next chapter.

\chapter{POVMs for bounded-rank quantum state tomography} \label{ch:constructions}
In this chapter, we construct rank-$r$ complete and strictly-complete POVMs for bounded-rank QST that have significantly less elements than fully informationally complete (full-IC) POVMs. We present three separate construction techniques. The advantage of having multiple construction techniques is that one can chose the method that is best suited for the experimental apparatus. Many experiments have so-called natural measurements, that are easier to implement. For example, some experiments can easily apply bases, introduced in Sec.~\ref{ssec:detector_intro}. Therefore, for these experiments, it is best to construct POVMs for bounded-rank QST that consist of bases. In each technique, we assume the ideal limit of QST, where there are no errors and the probabilities are known exactly as this defines informational completeness. From the previous chapter, we know that if we can prove a POVM to be rank-$r$ strictly-complete in the ideal limit, then it will be robust to noise and errors. We also present examples of each construction technique, though the methods are general and can be used to build new constructions based on the specific operation of a given experiment. With these construction we will also be able to determine which is more efficient, rank-$r$ complete or rank-$r$ strictly-complete.

\section{Decomposition methods} \label{sec:decomp_method}
The first method we consider applies to the construction of rank-1 complete POVMs. The method is based on the decomposition of a rank-1 density matrix into the state vector, $\ket{\psi}$. The state vector is described by $2d-2$ free parameters that make up the state amplitudes in some basis, $\ket{\psi} = \sum_k c_k \ket{k}$. If we take the ideal limit for QST, when the probability of each outcome is known exactly, we can relate the free parameters in $\{ c_k \}$ to the probabilities by the Born rule. If we can solve for each free parameter, then we can reconstruct $\ket{\psi}$ and thus $\rho$. Since the decomposition assumes that $\rho$ is rank-1 then this technique can show if the POVM is rank-1 complete. 

An examples of this technique was studied by Flammia~{\em et al.}~\cite{Flammia2005}, who introduced the following POVM,
\begin{align}\label{psi-complete}
&E_0=a\ket{0}\bra{0},\nonumber \\
&E_k=b(\mathds{1}+\ket{0}\bra{k}+\ket{k}\bra{0}),\; \;k=1,\ldots,d-1,\nonumber \\
&\widetilde{E}_k=b(\mathds{1}-\textrm{i} \ket{0}\bra{k}+\textrm{i} \ket{k}\bra{0}),\; \;k=1,\ldots,d-1,\nonumber \\
&E_{2d}=\mathds{1}-\left[E_0 +\sum_{n=1}^{d-1}(E_k+\widetilde{E}_k)\right],
\end{align}
with $a$ and $b$ chosen such that $E_{2d}\geq0$.  When $c_0>0$, we can chose  $c_0=\sqrt{p_0}/a$ (setting the phase of this amplitude to zero). The real and imaginary parts of $c_k$, $k=1,\ldots, d-1$, are related to the probabilities by $\textrm{Re}(c_k)=\frac1{2c_0}(\frac{p_k}{b}-1)$ and $\textrm{Im}(c_k)=\frac1{2c_0}(\frac{\tilde{p}_k}{b}-1)$, respectively when we assume $\textrm{Tr}(\rho) = 1$. There are then a set of $2d -1$ quadratic equations that we can use to uniquely solve for all amplitudes, $\{ c_k \}$. When $c_0 =0$, the set of equations are not solvable; however this is a set of zero volume corresponding to the failure set allowed in Definition~\ref{def:rankr_comp}. The POVM has a total of $2d$ POVM elements, and therefore it is efficient compared to standard QST, which requires at least $d^2$ POVM elements. Flammia~{\em et al.}~\cite{Flammia2005} also proved this to be the minimum number of POVM elements to be rank-1 complete. 

Goyeneche {\em et al.}~\cite{Goyeneche2015} constructed another POVM and proved it was rank-1 complete by this strategy. They proposed four orthogonal bases,
\begin{align}\label{4gmb}
\mathbbm{B}_{1} &=\left\{ \frac{\ket{0}\pm\ket{1}}{\sqrt{2}}, \frac{\ket{2}\pm\ket{3}}{\sqrt{2}}, \ldots, \frac{\ket{d-2}\pm\ket{d-1}}{\sqrt{2}}\right\}, \nonumber \\
\mathbbm{B}_{2} &=\left\{ \frac{\ket{1}\pm\ket{2}}{\sqrt{2}}, \frac{\ket{3}\pm\ket{4}}{\sqrt{2}}, \ldots, \frac{\ket{d-1}\pm\ket{0}}{\sqrt{2}}\right\}, \nonumber \\
\mathbbm{B}_{3} &=\left\{ \frac{\ket{0}\pm \textrm{i} \ket{1}}{\sqrt{2}}, \frac{\ket{2}\pm \textrm{i} \ket{3}}{\sqrt{2}}, \ldots, \frac{\ket{d-2}\pm \textrm{i} \ket{d-1}}{\sqrt{2}}\right\}, \nonumber \\
\mathbbm{B}_{4} &=\left\{ \frac{\ket{1}\pm \textrm{i} \ket{2}}{\sqrt{2}}, \frac{\ket{3}\pm \textrm{i} \ket{4}}{\sqrt{2}}, \ldots, \frac{\ket{d-1}\pm \textrm{i} \ket{0}}{\sqrt{2}}\right\}.
\end{align}
Denoting $p_{k}^{\pm}= |\frac{1}{2}( \bra{j}\pm\bra{k+1}) | \psi \rangle|^2$, and $p_{k}^{\pm \textrm{i}}= |\frac{1}{2}(\bra{k}\mp \textrm{i} \bra{k+1}) |\psi \rangle|^2$, we obtain, $c_k^* c_{k+1} {=}\frac{1}{2}[(p_{k}^{+}-p_{k}^{-})+\textrm{i} (p_{k}^{+\textrm{i}}-p_{k}^{-\textrm{i}})]$ for $k = 0,\ldots,d-1$, and addition of indices is taken modulo $d$. We then have a set of $d$ quadratic equations, which Goyeneche {\em et al.}~\cite{Goyeneche2015} showed has a unique solution when we include the trace constraint, $\sum_k |c_k |^2 = 1$; therefore, the construction is rank-1 complete. When $c_k = 0$ and $c_{k+l} = 0$, for $l >1$ the quadratic equations do not have have a unique solutions. This corresponds to the failure set of the POVM. Since the bases have a total of $4d$ POVM elements, this construction requires less resources than standard QST but more elements than the minimum POVM proposed by Flammia {\em et al.}~\cite{Flammia2005}.

While the method of reconstructing the state vector amplitudes is very intuitive, it is limited to rank-1 complete POVMs. To construct a rank-1 strictly-complete POVM, we cannot assume the pure-state structure of the measured state, as we did here. Moreover, the generalization to rank-$r$ complete constructions is not obvious. In this case, one needs to consider ensemble decompositions, $\rho = \sum_{i=0}^{r-1} \lambda_i | \psi_i \rangle \langle \psi_i |$, where $\langle \psi_i | \psi_j \rangle = \delta_{i,j}$, which require a greater number of quadratic equations.

\section{Element-probing POVMs} \label{sec:EP}
Another, more adaptable method to construct both rank-$r$ complete and rank-$r$ strictly complete POVMs applies to a class of POVMs we will define as element-probing (EP) POVMs. An EP-POVM allow for the reconstruction of matrix elements of $\rho$. More formally, there is a linear mapping between the probabilities from an EP-POVM and the elements of the density matrix, which is an inverse of the Born rule, $\{p_{\mu}\} \rightarrow \{ \rho_{i,j} \}$. If the POVM is not full-IC, then there necessarily exist a subset of elements reconstructed, called the measured elements.\footnote{An EP-POVM may give information about other parts of the density matrix besides the measured elements. For this case, we ignore this additional information and only study the measured elements.} We denote the remaining elements as the unmeasured elements. In this section, we will show that based on the structure of the measured elements, we can determine if a given EP-POVM is rank-$r$ complete or rank-$r$ strictly-complete for any value of $r$.

The POVMs considered in the previous section, given in Eq.~\eqref{psi-complete} and Eq.~\eqref{4gmb}, are in fact examples of EP-POVMs. For Eq.~\eqref{psi-complete}, the measured elements are the first row and column of the density matrix. The probability $p_0 = \textrm{Tr}(E_0 \rho)$ trivially determines $\rho_{0,0}=\langle 0 | \rho|0\rangle$, and the probabilities $p_n=\textrm{Tr}(E_n\rho)$ and $\tilde{p}_n=\textrm{Tr}(\widetilde{E}_n\rho)$ determines $\rho_{n,0}=\langle n |\rho |0\rangle$ and $\rho_{0,n}=\langle 0 |\rho | n \rangle$, respectively. For Eq.~\eqref{4gmb}, the probabilities, $\{ p_k^{\pm},  p_k^{\pm \textrm{i}}\}$ determine the density matrix elements $\rho_{k,k+1}{=}\frac{1}{2}[(p_{k}^{+}-p_{k}^{-})+\textrm{i} (p_{k}^{+\textrm{i}}-p_{k}^{-\textrm{i}})]$ for $k = 0,\ldots,d-1$, and addition of indices is taken modulo $d$. 

\subsection{Linear algebra relations for EP-POVMs}
We prove here whether an EP-POVM is rank-$r$ complete or strictly-complete based on the Schur complement and the Haynsworth matrix inertia~\cite{Haynsworth1968,Zhang2011}. Consider a block-partitioned $k \times k$ Hermitian matrix,
\begin{equation} \label{block_mat}
M =
\begin{pmatrix}
{A} & {B^{\dagger}} \\
{B} &{C} 
\end{pmatrix},
\end{equation}
where $A$ is a $r \times r$ Hermitian matrix, and the size of ${B^{\dagger}}$, $B$ and $C$ is determined accordingly.
The Schur complement of $M$ with respect to $A$, assuming $A$ is nonsingular, is defined by
\begin{equation}
M/A \equiv C - B A^{-1} B^{\dagger}.
\end{equation}
The inertia of a Hermitian matrix is the ordered triple of the number of negative, zero, and positive eigenvalues of the matrix, $\textrm{In}(M)=(n_-[M], n_0[M], n_+[M])$, respectively. 

We will use the Haynsworth inertia additivity formula, which relates the inertia of $M$ to that of $A$ and of $M/A$~\cite{Haynsworth1968},
\begin{equation} \label{Schur_iner}
\textrm{In}(M) = \textrm{In}(A)+\textrm{In}(M/A),
\end{equation}
A corollary of the inertia formula is the rank additivity property,
\begin{equation} \label{Schur_rank}
\textrm{rank}(M) = \textrm{rank}(A) + \textrm{rank}(M/A).
\end{equation}
With these relations, we can determine the informational completeness of any EP-POVM. A similar approach was taken for classical matrix completion in Ref.~\cite{Smith2008}.

\subsection{Application to rank-$r$ complete POVMs} \label{ssec:EP_comp}
As an instructive example, we use the above relations in an alternative proof that the POVM in Eq.~\eqref{psi-complete} is rank-1 complete without referring to the state amplitudes. The POVM in Eq.~\eqref{psi-complete} is an EP-POVM, where the measured elements are $\rho_{0,0}$, $\rho_{n,0}$ and $\rho_{0,n}$ for $n=1,\ldots,d-1$. Supposing that $\rho_{0,0}>0$ and labeling the unmeasured $(d-1)\times(d-1)$ block of the density matrix by $C$, we write
\begin{equation} \label{block_rho}
\rho=  \left(
    \begin{array}{cccc}
{\rho_{0,0}} &  {\rho_{0,1}}& \cdots &{\rho_{0,d-1}}\\
\cline{2-4}\multicolumn{1}{c|}{\rho_{1,0}}&
      {} &{}& \multicolumn{1}{c|}{} \\
      \multicolumn{1}{c|}{\vdots}&
      {} &{\;\;\Large\textit{C}}& \multicolumn{1}{c|}{}\\
\multicolumn{1}{c|}{\rho_{d-1,0}}&
      {} &{}& \multicolumn{1}{c|}{}\\\cline{2-4}
    \end{array}
    \right)
\end{equation}
Clearly, Eq.~\eqref{block_rho} has the same form as Eq.~\eqref{block_mat}, such that $M = \rho$, $A = \rho_{0,0}$, $B^{\dagger}=({\rho_{0,1}}\cdots {\rho_{0,d-1}})$, and $B = ({\rho_{0,1}}\cdots {\rho_{0,d-1}})^{\dagger}$. Assume $\rho$ is a pure state so $\textrm{rank}(\rho)=1$. By applying Eq.~\eqref{Schur_rank} and noting that $\textrm{rank}(A)=1$, we obtain $\textrm{rank}(\rho/A)=0$. This implies that $\rho/A=C - B A^{-1} B^{\dagger}=0$, or equivalently, that $C= B A^{-1} B^{\dagger}= \rho_{0,0}^{-1}B B^{\dagger}$. Therefore, by measuring every element of $A$, $B$ (and thus of ${B^{\dagger}}$), the rank additivity property allows us to algebraically reconstruct $C$ uniquely without measuring it directly. Thus, the entire density matrix is determined by measuring its first row and column. Since we used the assumption that $\textrm{rank}(\rho){=}1$, the reconstructed state is unique to the set ${\cal S}_1$, and the POVM is rank-1 complete. 

This algebraic reconstruction of the rank-$1$ density matrix works as long as $\rho_{0,0}\neq0$. When $\rho_{0,0}=0$, the Schur complement is not defined, and Eq.~\eqref{Schur_rank} does not apply. This, however, only happens on a set of states of measure zero (the failure set), i.e. the set of states where $\rho_{0,0} = 0$ exactly. It is exactly the same set found by Flammia~{\em et al.}~\cite{Flammia2005}.

The above technique can be generalized to determine if any EP-POVM is rank-$r$ complete for a state $\rho\in{\cal S}_r$. In general, the structure of the measured elements will not be as convenient as the example considered above. Our approach is to study $k \times k$ principle submatrices (square submatrices that are centered on the diagonal) of $\rho$ such that $k > r$. Since $\rho$ is a rank-$r$ matrix, it has at least one nonsingular $r \times r$ principal submatrix,
\begin{equation}
\rho = \begin{pmatrix}
\ddots& & \\
& \begin{pmatrix} \underset{(k \times k)}{M} \end{pmatrix} & \\
& & \ddots \end{pmatrix}.
\end{equation}
Assume for now that a given $k \times k$ principal submatrix, $M$, contains a nonsingular $r \times r$ principle submatrix $A$. We can apply a $k \times k$ unitary, $U$, to map the submatrix $M$ to the form in Eq.~\eqref{block_mat},
\begin{equation}
U M U^{\dagger}= \begin{pmatrix}
\underset{(r \times r)}{A}  & \underset{(k-r \times r)}{B^{\dagger}} \\
\underset{(r \times k-r)}{B}  & \underset{(k-r \times k-r)}{C} 
\end{pmatrix}.
\end{equation}
From Eq.~\eqref{Schur_rank}, since $\textrm{rank}(M) = \textrm{rank}(A) = r$, $\textrm{rank}(M/A) = 0$, and therefore $C = B A^{-1}B^{\dagger}$. This motivates our choice of $M$. If the measured elements make up $A$ and $B$ (and $B^{\dagger}$) then we can solve for $C$ and we have fully characterized $U M U^{\dagger}$, and therefore also $M$. An example application is considered in Appendix~\ref{app:constructions}. In general, an EP-POVM may measure multiple subspaces, $M_i$, and we can reconstruct $\rho$ only when the corresponding $A_i$, $B_i$, $C_i$ cover all elements of $\rho$. We label the set of all principle submatrices that are used to construct $\rho$ by $\bm{M} = \{M_i\}$. Since we can reconstruct a unique state within the set of $\mathcal{S}_r$ this is then a general description of a rank-$r$ complete EP-POVM.  The failure set, in which the measurement fails to reconstruct $\rho$, corresponds to the set of states that are singular on any of the $A_i$ subspaces. 

\subsection{Application to rank-$r$ strictly-complete} \label{ssec:EP_strict}
The framework defined above also allows us to determine if a given EP-POVM is strictly-complete. As an example, consider the rank-1 complete POVM in Eq.~\eqref{psi-complete}. Since $\rho/A=0$, by applying the inertia additivity formula to $\rho$ we obtain,
\begin{equation}
\textrm{In}(\rho) = \textrm{In}(A)+\textrm{In}(\rho/A)=\textrm{In}(A).
\end{equation}
This implies that $A$ is a positive semidefinite (PSD) matrix since $\rho$ is, by definition, a PSD matrix. For the POVM in Eq.~\eqref{psi-complete}, $A=\rho_{0,0}$, so this equation is a re-derivation of the trivial condition $\rho_{0,0}\geq0$. Let us assume that the POVM is not rank-$1$ strictly-complete. If so, there must exist a PSD matrix, $\sigma\geq0$, with $\textrm{rank}(\sigma)>1$, that has the same measurement vector and thus measured elements as $\rho$, but different unmeasured elements. We define this difference by $V\neq0$, and write
\begin{equation} \label{block_mat_sigma}
\sigma=  \left(
    \begin{array}{cccc}
{\rho_{0,0}} &  {\rho_{0,1}}& \cdots &{\rho_{0,d-1}}\\
\cline{2-4}\multicolumn{1}{c|}{\rho_{1,0}}&
      {} &{}& \multicolumn{1}{c|}{} \\
      \multicolumn{1}{c|}{\vdots}&
      {} &{\;\;\Large{\textit{C}}+\!\Large{\textit{V}}}& \multicolumn{1}{c|}{}\\
\multicolumn{1}{c|}{\rho_{d-1,0}}&
      {} &{}& \multicolumn{1}{c|}{}\\\cline{2-4}
    \end{array}
    \right)=\rho+\begin{pmatrix}
{0} &  {\bf 0}\\
{\bf 0} & V
\end{pmatrix}.
\end{equation}
Since $\sigma$ and $\rho$ have the same probabilities, for all $\mu$, $\textrm{Tr}(E_\mu\sigma)=\textrm{Tr}(E_\mu\rho)$. Summing over $\mu$ and using $\sum_\mu E_\mu=\mathds{1}$, we obtain that $\textrm{Tr}(\sigma)=\textrm{Tr}(\rho)$. This implies that $V$ must be a traceless Hermitian matrix, hence, $n_-(V) \geq 1$. Using the inertia additivity formula for $\sigma$ gives,
\begin{equation}
\textrm{In}(\sigma) = \textrm{In}(A)+\textrm{In}(\sigma/A).
\end{equation}
By definition, the Schur complement is
\begin{equation}
\sigma/A=C +V - B A^{-1} B^{\dagger}=\rho/A+V=V.
\end{equation}
The inertia additivity formula for  $\sigma$ thus reads,
\begin{equation}
\textrm{In}(\sigma) = \textrm{In}(A)+\textrm{In}(V).
\end{equation}
Since $A=\rho_{0,0}>0$, $n_-(\sigma) = n_-(V) \geq 1$ so $\sigma$ has at least one negative eigenvalue, in contradiction to the assumption that it is a PSD matrix. Therefore, $\sigma \not\geq 0$ and we conclude that the POVM in Eq.~\eqref{psi-complete} is rank-1 strictly-complete. 

A given POVM that is rank-$r$ complete is not necessarily rank-$r$ strictly-complete in the same way as the POVM in Eq.~\eqref{psi-complete}. For example, the bases in Eq.~\eqref{4gmb}, correspond to a rank-$1$ complete POVM, but not to a rank-$1$ strictly-complete POVM. For these bases, we can apply a similar analysis to show that there exists a quantum state $\sigma$  with $\textrm{rank}(\sigma)>1$ that matches the measured elements of $\rho$.

Given this structure, we derive the necessary and sufficient condition for a rank-$r$ complete EP-POVMs to be rank-$r$ strictly-complete. Using the notation introduced above, let us choose an arbitrary principal submatrix $M\in \bm{M}$ that was used to construct $\rho$. Such a matrix has the form of Eq.~\eqref{block_mat} where $C=BA^{-1}B^\dagger$. Let $\sigma$ be a higher-rank matrix that has the same measured elements as $\rho$, and let $\tilde{M}$ be the submatrix of $\sigma$ that spans the same subspace as $M$. Since $\sigma$ has the same measured elements as $\rho$, $\tilde{M}$ must have the form
\begin{equation}\label{Mtilde}
\tilde{M} = 
\begin{pmatrix}
 A & B^{\dagger} \\
 B & \tilde{C} 
 \end{pmatrix}\equiv\begin{pmatrix}
 A & B^{\dagger} \\
 B & C + V 
 \end{pmatrix}=M+\begin{pmatrix}
 {\bf 0} & {\bf 0} \\
 {\bf 0} & V 
 \end{pmatrix}.
 \end{equation}
Then, from Eq.~\eqref{Schur_iner}, $\textrm{In}(\tilde{M}) = \textrm{In}(A) + \textrm{In}(\tilde{M}/A) = \textrm{In}(A) + \textrm{In}(V)$, since $\tilde{M}/A = M/A + V = V$. A matrix is PSD if and only if all of its principal submatrices are PSD~\cite{Zhang2011}. Therefore, $\sigma \geq 0$ if and only if $\tilde{M}\geq 0$, and $\tilde{M} \geq 0$ if and only if $n_-(A) + n_-(V) = 0$. Since $\rho\geq0$, all of its principal submatrices are PSD, and in particular $A\geq0$. Therefore,  $\sigma \geq 0$ if and only if $n_-(V) = 0$. We can repeat this logic for all other submatrices $M \in \bm{M}$. Hence, we conclude that the measurement is rank-$r$ strictly-complete if and only if there exists at least one submatrix $M \in \bm{M}$ for which every $V$ that we may add (as in Eq.~\eqref{Mtilde}) has at least one negative eigenvalue. 

A sufficient condition for an EP-POVM to be rank-$r$ strictly-complete is given in the following proposition.
\begin{proposition} \label{prop1}
Assume that an EP-POVM is rank-$r$ complete. If its measurement outcomes determine the diagonal elements of the density matrix, then it is a rank-$r$ strictly-complete POVM.
\end{proposition}
{\em Proof.} Consider a Hermitian matrix $\sigma$ that has the same measurement probabilities as $\rho$, thus the same measured elements.  If we measure all diagonal elements of $\rho$ (and thus, of $\sigma$), then for any principal submatrix $\tilde{M}$ of $\sigma$, {\em cf.} Eq.~\eqref{Mtilde}, the corresponding $V$ is traceless because all the diagonal elements of $C$ are measured. Since $V$ is Hermitian and traceless it must have at least one negative eigenvalue, therefore, $\sigma$ is not PSD matrix and the POVM is rank-$r$ strictly-complete. $\square$ 

\noindent A useful corollary of this proposition is any EP-POVM that is rank-$r$ complete can be made rank-$r$ strictly-complete simply by adding POVM elements that determine the diagonal elements of the density matrix.

\section{Random bases} \label{sec:random_bases}
The final technique we consider for constructing strictly-complete POVMs is to measure a collection of random orthonormal bases. Measurement with random bases have been studied in the context of compressed sensing (see, e.g., in~\cite{Kueng2014,Acharya2016}). However, when taking into account the positivity of density matrices, we obtain strict-completeness with fewer measurements than required for compressed sensing~\cite{Kalev2015}. Therefore, strict-completeness is not equivalent to compressed sensing. While for quantum states, all compressed sensing measurements are strictly-complete~\cite{Kalev2015}, not all strictly-complete measurements satisfy the conditions required for compressed sensing estimators.

We perform the numerical experiments to determine rank-$r$ strictly-complete measurement for $r=1,2,3$.  To achieve this, we take the ideal case where the measurement outcomes are known exactly and the rank of the state is fixed. We consider two types of measurements on a variety of different dimensions: (i) a set of  Haar-random orthonormal bases on unary qudit systems with dimensions $d=11, 16, 21, 31, 41$, and $51$; and (ii) a set of local Haar-random orthonormal bases on a tensor product of $n$ qubits with $n=3,4,5$, and $6$, corresponding to $d=8, 16, 32$, and $64$, respectively. For each dimension, and for each rank, we generate $25d$ Haar-random states. For each state, we calculate the noiseless probability vector, $\bm{p}$,  with an increasing number of bases.  After each new basis measurement we use the constrained least-square (LS) program, Eq.~\eqref{general_norm_positive_CS}, where $\Vert\cdot\Vert$ is the $\ell_2$-norm, to produce an estimate of the state. We emphasize that the constrained LS finds the quantum state that is the most consistent with ${\bf p}$ without restrictions on the rank. The procedure is repeated until all estimates match the states used to generate the data (up to numerical error of $10^{-5}$ in infidelity). This indicates the random bases used correspond to a rank-$r$ strictly-complete POVM.

\begin{table}[ht]
\centering
\begin{tabular}{ cc|c|c|c|c|c|||c|c|c|c| }
& \multicolumn{9}{c}{\bf{Dimension}}\\ \cline{2-11}
 & \multicolumn{6}{|c|||}{Unary}  & \multicolumn{4}{c| }{Qubits}\\
 \multicolumn{1}{ c|| }{\bf{Rank}} &\bf{11} &\bf{16}&\bf{21} &\bf{31} &\bf{41}&\bf{51} &\bf{8} &\bf{16} &\bf{32}&\bf{64} \\
\hline\hline
\multicolumn{1}{ |c|| }{\bf{1}} &\multicolumn{6}{c|||}{6} & \multicolumn{4}{c|}{6}  \\\cline{1-11} 
\multicolumn{1}{ |c|| }{\bf{2}} & 7  & 8  & 8  & \multicolumn{3}{ c||| } {9}& \multicolumn{2}{ c| } {9} &  \multicolumn{2}{ c| } {10}  \\\cline{1-11}
 \multicolumn{1}{ |c|| }{\bf{3}} & 9  & 10  & 11  & 12 & 12  & 13 & 12 & \multicolumn{3}{ c| } {15}\\\cline{1-11}
\hline
\end{tabular}
\caption[Number of random orthonormal bases required for strict-completeness]{{\bf Number of random orthonormal bases corresponding to strict-completeness.}  Each cell lists the minimal number of measured bases for which the infidelity was below $10^{-5}$  for each of the tested states in the given dimensions and ranks. This indicates that a measurement of only few random bases is strictly-complete POVM.}\label{tbl:noiseless}
\end{table}  

We present our findings in Table~\ref{tbl:noiseless}.  For each dimension, we also tested fewer bases than listed in the table. These bases return infidelity below $10^{-5}$ for most states but not all. For example, in the unary system with $d=21$, using the measurement record from $5$ bases we can reconstruct all but one state with an infidelity below the threshold. The results indicate that measuring only few random bases, with weak dependence on the dimension, corresponds to a strictly-complete POVM for low-rank quantum states. Moreover, the difference between, say rank-1 and rank-2, amounts to measuring only a few more bases. This is important, as discussed below, in realistic scenarios when the state of the system is known to be close to pure. Finally, when considering local measurements on tensor products of qubits, more bases are required to account for strict-completeness when compared to unary system; see for example results for $d=16$.  We do not know if any of these bases suffer from a failure set but we see no evidence in our numerical simulations.

\section{Numerical studies of constructions with noise and errors} \label{sec:strict_comp_nums}
The techniques described in the previous sections allow for the construction of different rank-$r$ strictly-complete POVMs. However, we have yet to study how these measurements perform in the presence of noise and errors. In Sec.~\ref{ssec:rankr_strictcomp_robustness} we saw that rank-$r$ strictly-complete measurements are robust to all sources of noise and errors but we do not have an analytic form for the constant $\alpha$ in Eq.~\eqref{strict_comp_ineq} that describes the robustness. We can, however, use numerics to estimate this constant for a given POVM. 

To determine $\alpha$, we generate many pairs of quantum states, one rank-$r$, $\rho_r$, and one full-rank,  $\sigma$. To generate each state, we first select a random unitary $U$, from the Haar-measure and a $d$-dimensional vector, $\vec{\lambda}$, which has $r$ nonzero entries and $d-r$ zero entries. We renormalize $\vec{\lambda}$ such that $\sum_i \lambda_i = 1$. Then, the random rank-$r$ state is defined by, $\rho_r = U {\rm diag}[\vec{\lambda}] U^{\dagger}$, where the operation, ${\rm diag}[\cdot]$ puts the vector in the diagonal elements of the zero matrix. The full-rank state is generated by choosing all $d$ elements of $\vec{\lambda}$ to be nonzero, which is equivalent to generating a mixed state by the Hilbert-Schmidt measure. We then calculate the ratio of the HS-distance between the states to the distance between the measurement records, which is bounded by $1/\alpha$,
\begin{equation} \label{robust_ineq}
\frac{\| \rho_r - \sigma\|_2}{\| \mathcal{M}[ \rho _r - \sigma] \|_2} \leq \frac{1}{\alpha},
\end{equation}
where $\mathcal{M}$ represents the map of a rank-$r$ strictly-complete POVM. We test three different types of POVMs in various dimensions: a qudit measured with Haar-random bases for $d = 11,16,21,$ and 31, a collection of $n=3,4,5,$ and 6 qubits measured with a series of Haar-random bases on each qubit, and finally $n=3,4,5$ and 6 qubits measured with rank-$r$ generalization of the measurement proposed by Goyeneche~{\em et al.}~\cite{Goyeneche2015}, defined in Appendix~\ref{app:GMB}. A similar study was performed in Ref.~\cite{Carmeli2016}, for a different rank-1 strictly-complete measurement.

\begin{figure}[t]
\centering
\includegraphics[width=0.83\linewidth]{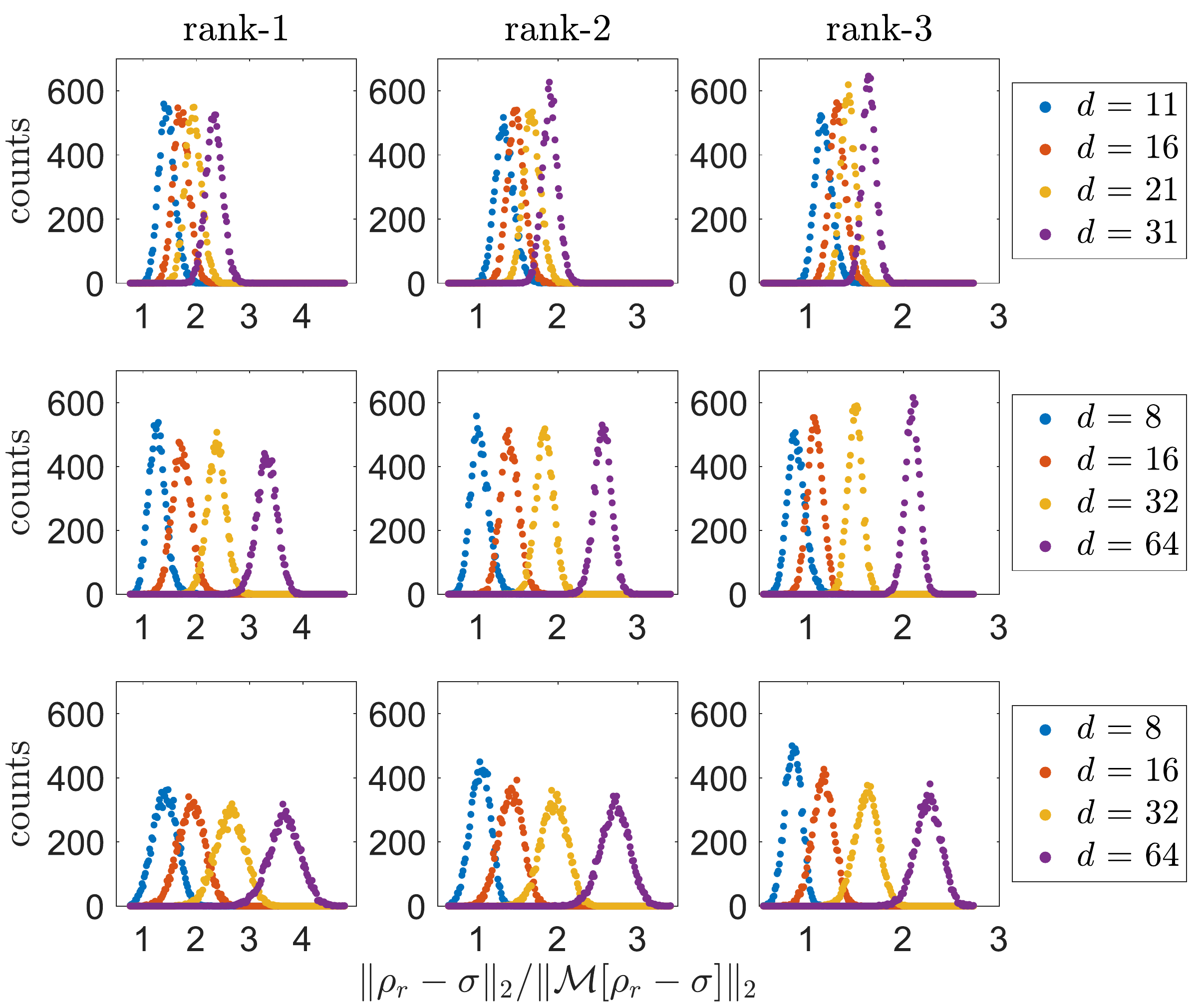}
\caption[Simulation of bounds for robustness inequality with random states]{{\bf Simulation of bounds for robustness inequality with random states.} We generated $10^4$ pairs of a rank-$r$ state and a full-rank state and calculate the ratio of the HS distance and the $\ell_2$-distance of the probabilities between the two. We repeat for three different types of measurements. {\bf Top row:} Haar-random bases for the unary system. {\bf Middle row:} Tensor products of Haar-random local bases on qubit subsystems. {\bf Bottom row:} The GMB construction given in Appendix~\ref{app:GMB}.}
\label{fig:bounds}
\end{figure}

By Definition~\ref{def:rankr_strictcomp}, we know $\alpha$ is not zero and therefore $1/\alpha$ is bounded. However, an arbitrarily large value of $1/\alpha$ makes the robustness bound in Corollary~\ref{cor:robustness} blow up. We see in Fig.~\ref{fig:bounds} that the values of $\| \rho_r - \sigma\|_2/\| \mathcal{M}[ \rho _r - \sigma] \|_2$ is concentrated in peaks. As the ratio goes to infinity the number of times we see that ratio in the numerics goes to zero. This means that it is very unlikely to get the largest values of the ratio. We can also see that the position of the peaks is very dependent on the dimension and rank. As dimension increases the peak shifts to to the right, i.e. larger ratios. As rank increases the peak shifts to the left, i.e. smaller ratios. If we let $1/\alpha_{\rm max}$ be the maximum value then from Fig.~\ref{fig:bounds}, we that the $1/\alpha_{\rm max}$ is not too large (the maximum value for all ranks, dimensions and measurements is $4.7784$). Therefore, the robustness bound will likely not blow up for the three different measurements considered. 

In order to determine the success of each measurement for QST, we perform a numerical study with realistic noise and errors. We simulate a realistic scenario where the state of the system is full-rank but high-purity and the experimental data contains statistical noise but no measurement errors. From Corollary~\ref{cor:robustness} we expect to obtain a robust estimation of the state by solving any convex estimator of the form of Eqs.~\eqref{general_positive_CS_noisy} and~\eqref{general_norm_positive_CS_noisy}. We calculate three estimates (using the MATLAB package CVX~\cite{cvx}) from the following programs: trace-minimization (given in Eq.~\eqref{Tr-min}), constrained least-squares (given in Eq.~\eqref{LS}), and maximum-likelihood (given in Eq.~\eqref{ML}). In the trace-minimization program the trace constraint is not included hence $\hat\rho=\hat{X}/\textrm{Tr}(\hat{X})$.

\begin{figure}[t]
\centering
\includegraphics[width=0.82\linewidth]{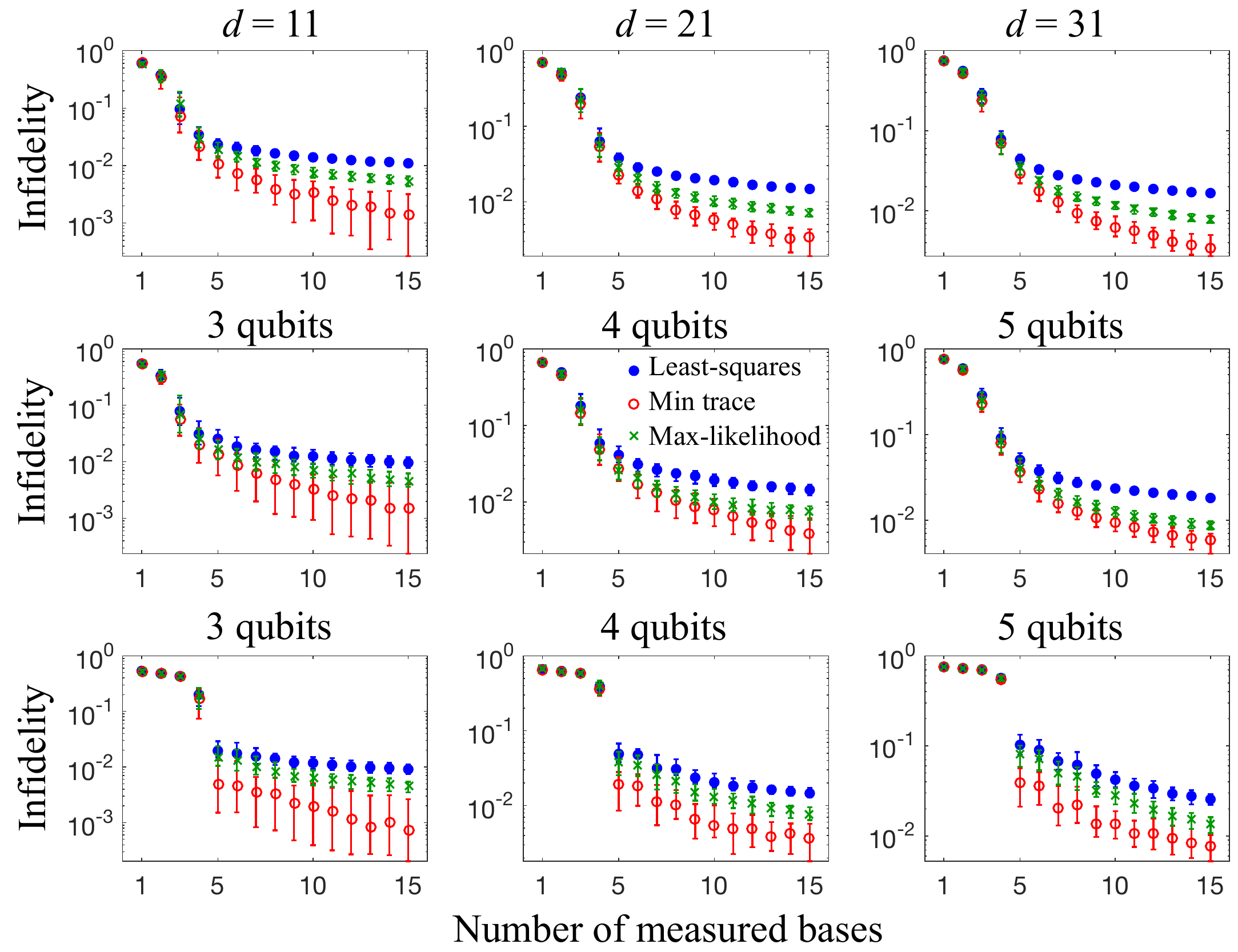}
\caption[Simulation of QST under realistic conditions]{{\bf Simulation of QST under realistic conditions.} We assume that the state of the system is a full-rank state close to a target pure state. We plot the median infidelity (on a log-scale) between the target pure state and its estimation as a function of measured bases for three different estimators Eqs.~\eqref{LS}-\eqref{ML}.  The error bars show the interquartile range (middle 50\%) of the infidelities found over 100 numerical experiments. {\bf Top row:} Haar-random measurement for a unary system. {\bf Middle row:} Tensor products of Haar-random local bases on qubit subsystems. {\bf Bottom row:} The GMB construction given in Appendix~\ref{app:GMB}. }
\label{fig:noisy}
\end{figure}

We apply the three measurements discussed above in three selected dimensions to a realistic system. For each measurement and dimension we generate 100 Haar-random pure-states (target states), $\{| \psi\rangle\}$, and create the actual prepared state, $\sigma =(1- q) | \psi \rangle \langle \psi | + q \tau$, where $q = 10^{-3}$, and $\tau$ is a random full-rank state generated from the Hilbert-Schmidt measure by the same procedure described above. The measurement vector, $\bm{f}$, is simulated by sampling $m = 300 d$ trials from the corresponding probability distribution. For each number of measured bases, we estimate the state with the three different convex optimization programs listed above. 

In Fig.~\ref{fig:noisy} we plot the average infidelity (over all tested states) between the target state, $ | \psi \rangle$, and its estimation, $\hat\rho$, $1-\overline{\langle\psi|\hat\rho| \psi \rangle}$. As ensured by Corollary~\ref{cor:robustness}, the three convex programs we used robustly estimate the state with a number of bases that correspond to rank-1 strictly-complete POVM, that is, six bases for the case of Haar-random basis measurements, and five bases based on the construction of Goyeneche et al. Ref.~\cite{Goyeneche2015}, reviewed in Appendix~\ref{app:GMB}. Furthermore, in accordance with our findings, if one includes the measurement outcomes of only a few more bases such that the overall POVM is rank-$2$ strictly-complete, or higher, we improve the estimation accordingly. The study does not provide evidence that the failure set impairs the estimation, despite the large magnitude of noise. The GMB construction is known to suffer from such a failure set but still produce a robust estimate. It is unknown whether the random bases suffer from such a failure set but both types of random bases produce robust estimates.

\section{Constructions for QDT} \label{sec:constructions_QDT}
As discussed in Sec.~\ref{ssec:br_QDT}, in QDT we typically have prior information about the POVM elements, for example, that they are rank-1 operators. We can apply the same techniques for constructing target POVMs for bounded-rank QST, to construct sets of probing states for bounded-rank QDT. We express each unknown POVM element in the rank-1 decomposition,
\begin{equation}
F_{\mu} = | \tilde{\phi}_{\mu} \rangle \langle \tilde{\phi}_{\mu} |,
\end{equation}
where $\ket{\tilde{\phi}_{\mu}} = \sum_k e_k^{(\mu)} \ket{k}$ is an unnormalized state vector. We use the letter $F$ for the unknown POVM element to differentiate it from the known POVM elements discussed in the previous sections for QST. In QDT, we measure the POVM elements by applying the unknown readout device to a set of known probing quantum state, $\{ \rho_{\nu} \}$. The conditional probability of getting outcome $\mu$ for the $\nu$th state is then,
\begin{equation}
p_{\mu,\nu} = \textrm{Tr}(E_{\mu} \rho_{\nu} ) = \langle \tilde{\phi}_{\mu} | \rho_{\nu} | \tilde{\phi}_{\mu} \rangle
\end{equation}
Therefore, constructing the set of probing states for bounded-rank QDT is very similar to constructing the POVM for bounded-rank QST. In fact, any POVM for QST can be translated to a set of probing states for QDT. For example, given $\{ E_{\mu} \}$, which is a POVM for QST, we can translate each element to to a probing state for QDT. Since each element is already positive, all that is required is normalization,
\begin{equation}
\rho_{\nu} = \frac{E_{\nu}}{\textrm{Tr}(E_{\nu})}.
\end{equation}
However, the translation does not guarantee that the informational completeness for $\{ \rho_{\nu} \}$ is the same as the informational completeness of $\{ E_{\nu} \}$. For example, if $\{ E_{\nu} \}$ is rank-1 strictly-complete for QST, the set $\{ \rho_{\nu} \}$ is not necessarily rank-1 strictly-complete for QDT.

Let us consider a concrete example, the POVM in Eq.~\eqref{psi-complete}. The translated probing states are then,
\begin{align}\label{QDT_psi-complete}
&\rho_0=\ket{0}\bra{0},\nonumber \\
&\rho_k=\frac{1}{d}(\mathds{1}+\ket{0}\bra{k}+\ket{k}\bra{0}),\; \;k=1,\ldots,d-1,\nonumber \\
&\widetilde{\rho}_k=\frac{1}{d}(\mathds{1}-\textrm{i} \ket{0}\bra{k}+\textrm{i} \ket{k}\bra{0}),\; \;k=1,\ldots,d-1,
\end{align}
where we omitted the translation of the final POVM element since it is not required in the proof of rank-1 completeness. Similar to the discussion in Sec.~\ref{sec:decomp_method}, we can reconstruct the amplitudes $\{ e_k^{(\mu)} \}$ from the probability of each outcome. When $e_0^{(\mu)}>0$, we find that  $e_0^{(\mu)}=\sqrt{p_0}$. The real and imaginary parts of $e_k^{(\mu)}$, $k=1,\ldots, d-1$, are related to the probabilities by $\textrm{Re}(e_k^{(\mu)})=\frac1{2e_0^{(\mu)}}(p_k-\textrm{Tr}(F_{\mu}))$ and $\textrm{Im}(e_k^{(\mu)})=\frac1{2e_0^{(\mu)}}(\tilde{p}_k-\textrm{Tr}(F_{\mu}))$. However, unlike with Eq.~\eqref{psi-complete} for QST, we cannot solve these equations for $\textrm{Re}(e_k^{(\mu)})$ and $\textrm{Im}(e_k^{(\mu)})$ since we do not know $\textrm{Tr}(F_{\mu})$. In QST, all POVMs measure the trace of the density matrix due to the constraint $\sum_{\mu} E_{\mu} = \mathds{1}$. However, from equation Eq.~\eqref{QDT_psi-complete}, $\sum_{\nu} \rho_{\nu} \neq \mathds{1}$. Therefore, in order to form a rank-1 complete set of probing states we need to complement Eq.~\eqref{QDT_psi-complete} with a set of probing states that measures the trace of the POVM element, $\textrm{Tr}(F_{\mu})$, for example the maximally mixed state $\rho = \frac{1}{d} \mathds{1}$. If it is not easy to create the maximally mixed state, the set $\rho_k = | k \rangle \langle k|$ for $k = 0,\dots d-1$ accomplishes the trace measurement as well.

The same translation can be applied to the other constructions provided in Sec.~\ref{sec:decomp_method} and Sec.~\ref{sec:EP}. We can also apply the same type of numerical analysis to QDT as was discussed in Sec.~\ref{sec:random_bases}. For QDT, instead of random bases, we generate random quantum states from some measure, e.g. Bures, Hilbert-Schmidt, etc., and use them to probe an bounded-rank POVM element with random trace. We produce the corresponding probability vector for various number of quantum states and apply the LS program to reconstruct the POVM element. When the reconstruction matches the original POVM element exactly then the set of probing states is likely rank-$r$ strictly-complete.

\section{Summary and conclusions}
We provided methods to construct rank-$r$ complete and rank-$r$ strictly-complete POVMs. Having multiple methods allows one to create the POVM that is best suited for a given experiment. We also provided a way of comparing rank-$r$ strictly-complete POVMs by numerically estimating the robustness constant, $\alpha$, in Eq.~\eqref{robust_ineq}. We generalized these results to QDT and showed how to translate a POVM for bounded-rank QST to a set of probing states for QDT.

In the previous chapter, we showed that rank-$r$ strictly-complete POVMs are compatible with convex optimization, and therefore offer an advantage over rank-$r$ complete POVMs. In this chapter, we saw that there is little difference in the number of POVM elements in rank-$r$ complete POVM vs the number in a rank-$r$ strictly-complete POVM. Therefore, there is no known advantage to rank-$r$ complete measurements, which enforces our conclusion that rank-$r$ strictly-complete measurements are superior for QST.

\chapter{Process tomography of unitary and near-unitary quantum maps} \label{ch:PT}
Quantum process tomography (QPT) is an even more demanding task than QST. In order to estimate an arbitrary quantum process, standard methods require $\mathcal{O}(d^4)$ measurements. This makes even small systems, e.g. three or more qubits, impractical for experimental application. However, in QPT, there is usually prior information about the applied quantum process, much like in QST where there is prior information about the quantum state. Most quantum information protocols require unitary maps, and therefore many experimental implementations try to engineer processes that are as close as possible to unitary.  Through previous diagnostic procedures, e.g., randomized benchmarking~\cite{Knill2008,Magesan2012}, there is usually have high confidence that the applied map is close to a target unitary. In this chapter we will demonstrate that such prior information can be used to drastically reduce the resources for QPT.

Previous workers have developed methods to diagnose devices that are designed to implement target unitary maps. Reich {\em  et al.}~\cite{Reich2013} showed that by choosing specially designed sets of probe states, one can efficiently estimate the fidelity between an applied quantum process and a target unitary map. Gutoski {\em  et al.}  \cite{Gutoski2014} showed that the measurement of $4d^2-2d-4$ Pauli-like Hermitian (i.e., two outcome) observables is sufficient to discriminate a unitary map from all other unitary maps, while identifying a unitary map from the set of all possible CPTP maps requires a measurement of $5d^2-3d-5$ such observables. These types of measurements, as well as the measurements we present in this chapter, are analogous to the rank-1 complete and rank-1 strictly-complete POVMs we discussed in the context of QST.  

In this chapter, we further study unitary QPT to establish the most efficient methods. We numerically show that some of these methods are equivalent to strict-completeness, and therefore robust to noise and errors. We additionally study the performance of these methods in the presence of noise and errors. We also find that while all estimators are robust, not all estimators behave the same with different sources of errors. We demonstrate that this difference can be used to diagnose sources of errors in the implementation of the unitary.

\section{Standard techniques for QPT}\label{sec:PT_review}
We begin by reviewing quantum processes and expanding on the basic definitions given in Sec.~\ref{ssec:process_intro}. An unknown quantum process, $\mathcal{E}[\cdot]$, which is a dynamical map on operator space, is represented by a process matrix $\chi$,
\begin{equation}\label{chiOp}
\chi=\sum_{\alpha,\beta=1}^{d^2}\chi_{\alpha,\beta} | \Upsilon_{\alpha} )( \Upsilon_{\beta}|,
\end{equation}
where $( \Upsilon_{\alpha} | \Upsilon_{\beta} ) =\delta_{\alpha, \beta}$ is an orthonormal basis. A completely positive (CP) quantum process is represented by a PSD process matrix. A trace preserving quantum process has process matrix that satisfies,
\begin{equation}
\sum_{\alpha, \beta} \chi_{\alpha, \beta} \Upsilon^{\dagger}_{\beta} \Upsilon_{\alpha} = \mathds{1}.
\end{equation}
A rank-1 process matrix corresponds to a unitary map, $\chi = | U )( U|$. 

The choice of basis for $\chi$ can have important consequences in estimation programs for QPT. One choice of $\{\Upsilon_{\alpha} \}$ is the ``standard'' basis $\{\Upsilon_{\alpha} =\Upsilon_{ij}=\ket{i}\bra{j}\}$ with the relabeling of $\alpha=1,\ldots,d^2$ is replaced by the pair $ij$ with $i,j=0,\ldots,d-1$. However, we can make many choices for $\{ \Upsilon_{\alpha} \}$, as will be the case in subsequent sections. Additionally, we can express $\chi$ in diagonal form,
\begin{equation}\label{chiOpDiag}
\chi=\sum_{\alpha=1}^{d^2}\lambda_{\alpha} | V_{\alpha} )( V_{\alpha}|,
\end{equation}
with eigenvalues $\lambda_{\alpha}$ and eigenvectors $| V_{\alpha} )$. 

In standard QPT, we prepare a set of $d^2$ input states, $\{ \rho^{\rm in}_{\nu} \}$, and evolve them with the unknown quantum process to a set of output states, $\mathcal{E}[\rho^{\rm in}_{\nu}] = \rho^{\rm out}_{\nu}$. The output states are then measured with a POVM, $\{ E_{\mu} \}$. The probability of observing an outcome $\mu$ for state $\rho^{\rm out}_{\nu}$,  is $p_{\mu,\nu} = \textrm{Tr}(\rho^{\rm out}_{\nu} E_{\mu})$, and expressed in terms of the process matrix using Eq.~\eqref{chiOp},
\begin{align} \label{measChi}
p_{\mu,\nu} &= \textrm{Tr} \left[ \sum_{\alpha, \beta =1}^{d^2}\chi_{\alpha,\beta}\Upsilon_{\alpha} \rho_{\nu}^{\rm in} \Upsilon_{\beta}^\dagger E_{\nu} \right], \nonumber \\
	&= \textrm{Tr}[\mathpzc{D}_{\mu,\nu} \chi].
\end{align}
Here $(\mathpzc{D}_{\mu,\nu})_{\alpha,\beta} \triangleq \textrm{Tr} [\rho_{\nu}^{\rm in}\, \Upsilon^{\dagger}_{\beta} E_{\mu} \Upsilon_{\alpha} ]$, are the elements of a $d^2 \times d^2$ matrix. The standard set of probing states were introduced in Ref.~\cite{Chuang1997}, as, 
\begin{align}\label{op basis nc}
&\ket{k},\; k=0,\ldots,d-1,\nonumber \\
&\frac{1}{\sqrt{2}}(\ket{k}+\ket{n}),\; k=0,\ldots,d-2,\;n=k+1,\ldots,d-1,\nonumber \\
&\frac{1}{\sqrt{2}}(\ket{k}+\ii\ket{n}),\; k=0,\ldots,d-2,\;n=k+1,\ldots,d-1,
\end{align}
and form a linearly independent set that spans the operator space. In standard QPT, we measure the output of such states with a full-IC POVM, such as the SIC or MUB, introduced in Sec.~\ref{ssec:noiseless_ST}. Therefore, standard QPT requires implementing at least $d^4$ POVM elements to reconstruct an arbitrary CP map.

\section{Numerical methods for QPT} \label{sec:PT_num}
In any application of QPT, there will necessarily exist sources of noise and errors that affect the measurement vector. Therefore, in order to characterize the quantum process in question, one must employ numerical estimators. The optimal solution for an estimator for QPT can be found using convex semidefinite programs (SDPs), given convex constraints $\chi \geq 0$ (CP constraint) and $\sum_{\alpha,\beta} \chi_{\alpha,\beta} \Upsilon_{\beta}^{\dagger} \Upsilon_{\alpha} = \mathds{1}$ (TP constraint). We consider three programs for QPT, two of which are based on the classical technique of compressed sensing. 

One of the original estimation techniques for QPT, was based on the classical maximum-likelihood principle~\cite{Fiurasek2001a,Sacchi2001}. However, in this chapter we consider the least-squares program, which can be seen as an approximation of maximum-likelihood. The LS program minimizes the (square of the) $\ell_2$-distance between the measurement vector and the expected probability vector subject to the CPTP constraints,
\begin{align}\label{PT_LS}
\underset{\chi}{\rm minimize:} \quad & \sum_{\mu,\nu}  \textrm{Tr} (\mathpzc{D}_{\mu,\nu} \chi)  - f_{\mu,\nu} |^2\nonumber \\
\textrm{subject to:} \quad &\sum_{\alpha,\beta} {\chi}_{\alpha,\beta} \Upsilon_{\beta}^{\dagger} \Upsilon_{\alpha} =\mathds{1} \nonumber \\
			&  \chi \geq 0. 
\end{align}
The estimated process matrix is $\hat{\chi}_{\textrm{LS}}$. The LS program does not include assumptions about the nature the of process matrix we are attempting to reconstruct, such as it being a unitary map. 

The second type of program we consider is Tr-norm program, originally proposed in the context of compressed sensing for quantum states~\cite{Gross2010}. We generalize the program here for QPT. In QPT, the trace of the process matrix is constrained by the one equation in the TP constraint. Therefore, to minimize the trace we must drop that part of the TP constraint. In order to maintain the maximal number of constraint equations, we take the basis as the set of traceless Hermitian matrices, $\{ H_{\alpha} \}$, thereby ensuring that there is only one equation related to the trace of the process matrix, which is dropped. We thus define the Tr-norm program for QPT as follows:
\begin{align} \label{PT_Tr}
\underset{\chi}{\rm minimize:} \quad & |\textrm{Tr} ( {\chi} ) \nonumber \\
\textrm{subject to:} \quad
 & \sum_{\mu,\nu} |\textrm{Tr} (\mathpzc{D}_{\mu,\nu} {\chi} ) - f_{\mu,\nu} |^2 \leq \varepsilon \nonumber \\
& \sum_{\alpha,\beta \neq 1} {\chi}_{\alpha,\beta} H_{\beta}^{\dagger} H_{\alpha} =0 \nonumber \\
& \chi \geq 0,
\end{align}
where now $\chi$ and $\mathpzc{D}_{\mu,\nu}$  are represented in a basis with $H_1 = \frac{1}{\sqrt{d}} \mathds{1}$ and the elements $H_{\alpha \neq 1}$ are orthogonal traceless Hermitian matrices. The sum in the second constraint include all the terms except $\alpha=\beta=1$. The first constraint equation now requires that the probabilities from our optimization variable should match our measurement frequencies up to some threshold $\varepsilon$. The threshold is chosen based on a physical model of the statistical noise sources for the measurements. The estimated process matrix, $\hat{\chi}_{\textrm{Tr}}$, must be renormalized such that $\textrm{Tr}[\hat{\chi}]=d$.

The final program we consider is the $\ell_1$-norm program, originally proposed for QPT in Ref.~\cite{Kosut2008}, and also inspired by compressed sensing techniques. The $\ell_1$-norm program is advantageous when the process matrix is sparse in a known basis. This is common in many applications. For example, if the goal is to implement a quantum logic gate, one is attempting to build a target unitary map $U_{\rm t}$. We therefore expect that if the error in implementation is small, when expressed in an orthogonal basis on operator space, $\{ V_{\alpha} \}$, that includes the target process as a member, $V_1 = U_{\rm t}$, the process matrix describing the applied map will be close to a sparse matrix. This in turn implies that the $\ell_1$-norm optimization algorithm can efficiently estimate the applied process. We define the $\ell_1$-norm estimator as follows:
\begin{align} \label{PT_L1}
\underset{\chi}{\rm minimize:} \quad & \| {\chi} \|_{1} \nonumber \\
\textrm{subject to:} \quad
&\sum_{\mu,\nu} | \textrm{Tr} (\mathpzc{D}_{\mu,\nu} {\chi} ) - f_{\mu,\nu} |^2 \leq \varepsilon, \nonumber \\
& \sum_{\alpha,\beta} {\chi}_{\alpha,\beta} V_{\beta}^{\dagger} V_{\alpha} =\mathds{1}, \nonumber \\
& \chi \geq 0.
\end{align}
We take the basis $\{V_{\alpha} \}=\{U_{\rm t}, U_{\rm t}H_2, \ldots, U_{\rm t}H_{d^2-1}\}$, where $\{H_{\alpha} \}$ are the basis of traceless Hermitian observables.   We express $\mathpzc{D}_{\mu,\nu}$ also in the $\{ V_{\alpha} \}$ basis such that $\mathpzc{D}_{\mu,\nu, \alpha,\beta } = \textrm{Tr} [\rho_{\nu}^{\rm in} V^{\dagger}_{\alpha} E_{\mu}V_{\beta} ]$.  Again, we constrain the probabilities to match the measurement vector, with some threshold $\varepsilon$ based on the noise and errors present.

We can regard the representation of the applied process matrix in this basis as a transformation into the ``interaction picture'' with respect to the target map; any deviation of  the applied process matrix from the projection onto $| U_{\rm t} )$ indicates an error. Therefore the ${\ell_1}$-norm program directly estimates the error matrix studied in detail in Ref.~\cite{Korotkov2013}. This feature holds also if the target map is not a unitary map. In this situation, we represent the applied map in the eigenbasis of the target map.

To determine the success of the QPT we use the process fidelity~\cite{Gilchrist2005}. The process fidelity between two arbitrary process matrices $\chi_1$ and $\chi_2$, is defined as,
\begin{equation}\label{PT_fidelity}
F(\chi_1, \chi_2) =\frac1{d^2}\left({\rm Tr}\sqrt{\sqrt{\chi_1} \chi_2 \sqrt{\chi_2 }}\right)^2.
\end{equation}
The best measure of the success of QPT is the fidelity between the actual quantum process to the estimated process. However, in a real application of QPT, we do not know the actual process. Therefore, in the numerical simulations below, we also compare the estimated process to the target process. In this chapter the target process is always a unitary such that, $\chi_{\rm t} = | U_{\rm t})( U_{\rm t}|$. Therefore, the process fidelity between the target and an estimated process matrix $\hat{\chi}$ is,
\begin{equation} \label{PT_U_fidelity}
F(\hat{\chi}, U_{\rm t}) = \frac1{d^2}  ( U_{\rm t} | \hat{\chi} | U_{\rm t}).
\end{equation}

\section{Reconstruction of unitary processes}\label{sec:unitary}
We begin by studying unitary QPT, or rank-1 QPT, in analogy to the previous chapters on rank-1 QST. In this section, we assume the ideal setting for QPT, when we know the density matrices that describe the input states and the probabilities of each outcome exactly. Since a quantum process is determined by both states and POVMs, we simplify the discussion by assuming each output state is measured with an informationally complete (IC) POVM. Here, we define IC as any POVM that is either full-IC, rank-$r$ complete, or rank-$r$ strictly-complete. Therefore, the question of whether the unitary map can be reconstructed in the ideal situation is only dependent on the input states that probe the process.

\subsection{Minimal sets of input states}
If reconstruction of a set of output states uniquely identifies an arbitrary unitary map within the set of all unitary maps, we call the set ``unitarily informationally complete'' (UIC). This is analogous to rank-1 complete POVMs for QST, which uniquely identify a pure state within the set of all pure states. A similar problem was studied by Reich {\em  et al.} \cite{Reich2013}. They developed an algebraic framework to identify sets of input states from which one can discriminate any two unitary maps given the corresponding output states. In particular, a set of input states, $\{\rho^{\rm in}_{\nu} \}$, provides sufficient information to discriminate any two unitary maps if and only if the identity operator is the only operator that commutes with all $\rho^{\rm in}_{\nu}$'s in this set. If the reconstruction of the input states discriminate any two unitary maps then they also uniquely identify any unitary map within the set of all unitary maps. Therefore, these sets of states are UIC.

An example of such a set on a $d$-level system consist of the two states,
\begin{equation}
\mathcal{S}=\left\{\rho_0^{\rm in}=\sum_{n=0}^{d-1}\lambda_n \ket{n}\bra{n},\,\, \rho_1^{\rm in}=\ket{+}\bra{+}\right\},
\end{equation}
where the eigenvalues of $\rho_0^{\rm in}$ are nondegenerate,  $\{\ket{n}\}$ is an orthonormal basis for the Hilbert space, and $\ket{+} = \frac{1}{\sqrt{d}}\sum_{n=0}^{d-1} \ket{n}$. Reich {\em  et al.} \cite{Reich2013} considered $\mathcal{S}$ in order to set numerical bounds on the average fidelity between a specific unitary map and a random CPTP map. In fact, $\mathcal{S}$ is the minimal UIC set of states for QPT of a unitary map on a $d$-dimensional Hilbert space.  

To see that $\mathcal{S}$ is a UIC set, we write the unitary map as a transformation from the orthonormal basis $\{\ket{n}\}$ to its image basis $\{\ket{u_n}\}$,
\begin{equation}\label{unitary}
U=\sum_{n=0}^{d-1}\ket{u_n}\bra{n}.
\end{equation}
In its essence, the task in QPT of a unitary map is to fully characterize the basis $\{\ket{u_n}\}$, along with the relative phases of the summands $\{\ket{u_n}\bra{n}\}$. By probing the map with $\rho_0^{\rm in}$, we obtain the output state $\rho_0^{\rm out}=U\rho_0^{\rm in}U^\dagger=\sum_{n=0}^{d-1}\lambda_n \ket{u_n}\bra{u_n}$, which we measure with a IC POVM to obtain a full reconstruction. We then diagonalize $\rho_0^{\rm out}$ and learn $\{\ket{u_n}\bra{u_n}\}$. Without loss of generality, we take the global phase of $\ket{u_0}$ to be zero. Next, we probe the map with $\rho_1^{\rm in}$, and fully characterize the output state  $\rho_1^{\rm out}=\frac{1}{d}\sum_{n,m=0}^{d-1} \ket{u_n}\bra{u_m}$ with a full-IC POVM (this state requires a full-IC POVM since it is full rank). The $\{\ket{u_n}\}$ are calculated according to the relation, $\ket{u_n}\bra{u_n}\rho_1^{\rm out}\ket{u_0}=\frac{1}{d}\ket{u_n}$. This procedure identifies a unique orthonormal basis $\{\ket{u_n}\}$ if and only if the map is a unitary map and requires a total of $d^2 + 2d$ POVM elements.

While $\mathcal{S}$ is the minimal UIC set, in practice we may not have reliable methods to produce a desired mixed state, $\rho^{\rm in}_0$.  We thus turn our attention to minimal UIC sets that are composed only of pure states (arbitrary pure states can be reliably produced using the tools of quantum control~\cite{Smith2012}). Such UIC sets are composed of $d$ pure states that form a nonorthogonal vector basis for the $d$-dimensional Hilbert space. For example, the set
\begin{align}\label{n+}
\ket{\psi_n}&=\ket{n},\; n=0,\ldots,d-2,\nonumber \\
\ket{\psi_{d{-}1}}&=\ket{+}=\frac{1}{\sqrt{d}}\sum_{n=0}^{d-1}\ket{n},
\end{align}
is a minimal UIC set of pure states. A similar set (with $d+1$ elements) was considered in Ref.~\cite{Reich2013}. Here, we focus on a different set of $d$ pure states that is UIC,
\begin{align}\label{0+n}
\ket{\psi_0}&=\ket{0}, \nonumber \\
\ket{\psi_n}&=\frac1{\sqrt2}(\ket{0}+\ket{n}),\; n=1,\ldots,d-1.
\end{align}
This is a subset of the standard states used in QPT from Eq.~(\ref{op basis nc}). The only operator that commutes with all of the projectors $\{\ket{\psi_n}\bra{\psi_n}\}$, $n=0,\ldots,d-1$, is the identity.  

With Eq.~\eqref{unitary}, we can show that the set of probing states in Eq.~\eqref{0+n} is UIC. Starting with the first state, $\ket{\psi_0}$ the output state is then $U\ket{\psi_0}=\ket{u_0}$, which we characterize with an IC POVM. From the eigendecomposition we obtain the state $\ket{u_0}$ (up to a global phase that we can set to zero). Next, we act the unitary map on $\ket{\psi_1}$ and perform an IC POVM on the output state $U\ket{\psi_1}\bra{\psi_1}U^\dagger$. From the relation $U\ket{\psi_1}\bra{\psi_1}U^\dagger\ket{u_0}=\frac1{2}(\ket{u_0}+\ket{u_1})$ we obtain the state $\ket{u_1}$, including its phase relative to $\ket{u_0}$. We repeat this procedure for every state $\ket{\psi_n}$ with $n{=}1,{\ldots},d{-}1$, thereby obtaining all the information about the basis $\{\ket{u_n}\}$, including the relative phases in the sum of Eq.~\eqref{unitary}, and completing the tomography procedure for a unitary map. Since the unitary operator is uniquely identified by a series of linear equations then the set of states and POVM elements are rank-1 complete for QPT. If we choose the minimal rank-1 strictly-complete POVM in Eq.~\eqref{psi-complete} , which has $2d$ elements, to apply to each of the $d$ output states then this procedure requires a total of $2d^2$ POVM elements. This approach is substantially more efficient than standard QPT. 

We can further reduce the resources required for unitary process tomography by including the trace constraint. In this case the trace constraint assures that the states $\ket{u_n}$ are orthonormal. In the procedure considered above, we did not take this fact into account. By leveraging this constraint, we can reduce the number of required measurement outcomes on each output state. The first step is, as before, to use $\ket{\psi_0}$ as a probe state, and perform an informationally complete measurement, which has $2d$ outcomes, on the output  state,  $\ket{u_0}$. This procedure fails for states with $c_{00}=0$, a set of measure zero. Next, we probe the unitary map with $\ket{\psi_1}$ of Eq.~(\ref{0+n}), and perform an IC PVOM on the output state,  $\frac1{\sqrt2}(\ket{u_0}+\ket{u_1})$. However, since $\ket{u_1}$ is orthogonal to $\ket{u_0}$, it is sufficient to make a POVM that yields only the first $d-1$ probability amplitudes $c_{1n}=\braket{n | u_1}$, $n=0,\ldots,d-2$ and then use the orthogonality condition $\braket{u_0 | u_1}=\braket{u_1 |u_0}=0$ to calculate the $d$th amplitude, $c_{1,d{-}1}$. A measurement with $2d-2$ outcomes can be, for example, the measurement of Eq.~\eqref{psi-complete}, but with  $n=0,\ldots,d-2$. Therefore, to measure the state $\ket{u_k}$, $k=0,\ldots,d-1$ we perform a measurement with $2d-2k$ outcomes, and use $2k$ orthogonality relations. This leads to a total requirement of $d^2+d$ POVM elements. 

\subsection{Reconstruction for unitary QPT}
UIC sets provide an efficient way to characterize a unitary process, since they require only $d$ states instead of the standard $d^2$. However, in analogy to rank-1 QST, there may be other sets of input states that uniquely identifies the the unitary within the set of {\em all} CPTP maps. The corresponding output states must then be measured with either a rank-1 strictly-complete or full-IC POVM. These states would then be analogous to rank-1 strictly-complete POVMs in QST, and therefore are advantageous for QPT since they would be compatible with convex optimization. 

Riech {\em et al.} proved that the set of states in Eq.~\eqref{n+} plus the additional state $\ket{\psi_{d-1}}=\ket{d-1}$, uniquely identify a unitary within the set of all CPTP maps. In this section we give numerical evidence that the set of states in Eq.~\eqref{n+} and~\eqref{0+n}, when measured with a rank-1 strictly-complete POVM is in fact rank-1 strictly-complete for QPT. We generate a set of 100 Haar-random unitary maps for a 5-dimensional Hilbert space. We then evolve the three different sets of states: (dotted red) the standard set of states for QPT, given in Eq.~\eqref{op basis nc}, (dashed green) the UIC set of states given in Eq.~\eqref{n+} supplemented with $d^2-d$ other linearly independent states, and (solid blue) the UIC set given in Eq.~\eqref{0+n} supplemented with $d^2-d$ other linearly independent states. The output states are then measured with the POVM in Eq.~\eqref{psi-complete}, which we proved is rank-1 strictly-complete in Sec.~\ref{sec:EP}. We assume the ideal situation for QPT, when we have direct access to the probabilities of each outcome and use the LS program to reconstruct the process matrix. We then compare the estimated process matrix to the process matrix of the Haar-random unitary that was used to generate the probabilities.

\begin{figure}[ht]
\centering
\includegraphics[width=0.9\linewidth]{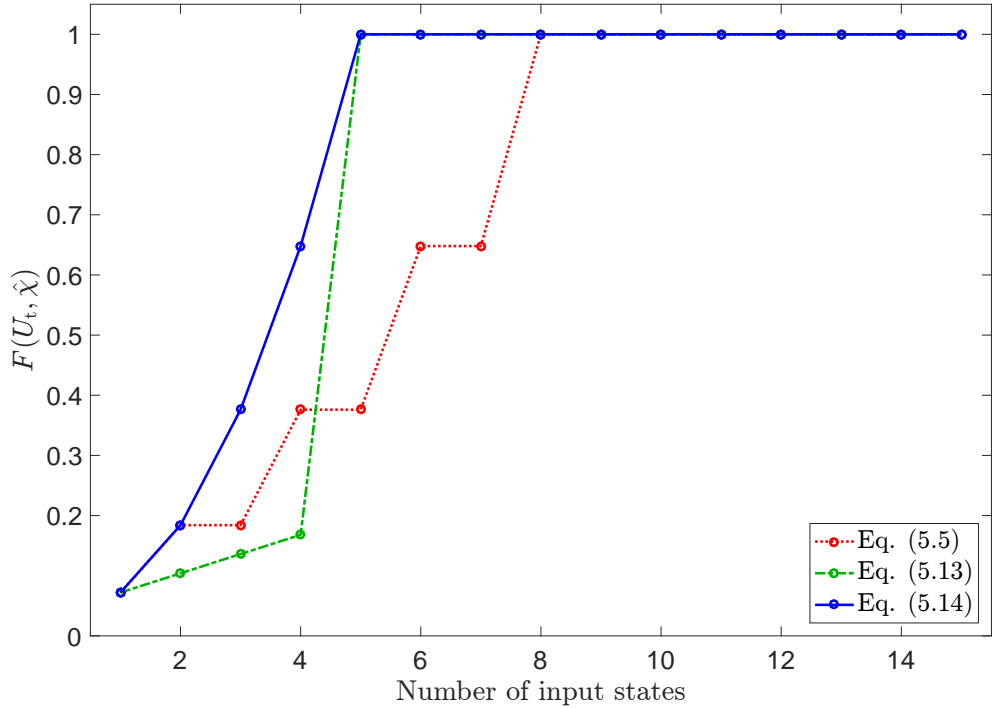}
\caption[Comparison of different UIC sets of states]{ {\bf Comparison of different UIC sets of states} Fidelity between  a unitary map on a $d=5$ dimensional Hilbert space and the LS estimate of the process matrix from (red) the standard order of quantum states given in Eq.~\eqref{op basis nc}, (green) the UIC set of states given in Eq.~\eqref{n+}, and (blue) the UIC set given in Eq.~\eqref{0+n}. Each point is an average of 100 Haar-random target unitary maps and has zero standard deviation.}
\label{fig:unitary_diff_sets}
\end{figure}

The results of the numerical study are plotted in Fig.~\ref{fig:unitary_diff_sets}. We can see that each set of probing states reaches unit fidelity well before $d^2 = 25$ states. This indicates that at this point the set of probing states uniquely identifies the unitary within the set of all CPTP maps. It is important to note that while each point is an average over 100 Haar-random target unitary maps, there is zero variance in the fidelity, i.e. the resulting fidelity is independent of the applied unitary. For the states in Eq.~\eqref{n+} and Eq.~\eqref{0+n}, we see unit fidelity is reached for $d = 5$ probing states. This corresponds to the UIC set, which was proven rank-1 complete in the previous section. The standard set of states, shown by the red dotted line, have a unique, plateau feature. These indicate that we gain information only from particular states in that set, while others do not provide additional information. The positions of the plateaus occur for the same input state for each of the sampled unitary map, and they are independent of their details. To see this point more clearly, take for example the two input states with $k=0$ and $n=1$ of Eq.~\eqref{op basis nc}, $\frac1{\sqrt2}(|0\rangle+|1\rangle)$ and $\frac1{\sqrt2}(|0\rangle+{\rm i}|1\rangle)$. Probing a unitary map with these two states giving us the same information, namely the image $|u_1\rangle$. Since probing the unitary map with either states gives the same information, for efficient reconstruction it is sufficient to probe the map only with one of the states.

\section{Near-unitary process tomography} \label{sec:QPT_werrors}
While the previous section established the notion of UIC sets of states to reconstruct unitary processes, in any physical implementation the process is never exactly unitary due to errors in the apparatus. We call such errors, process errors and they cause the applied map to differ from our target unitary map. Process errors are similar in nature to preparation errors in QST, considered in Sec.~\ref{sec:estimation_wnoise}. Therefore, since rank-1 strictly-complete POVMs are robust to preparation errors, we expect the UIC set considered in Eq.~\eqref{0+n} to also be robust to process errors in QPT. However, the robustness property does not guarantee that each estimation program behaves the same, and in fact the behavior of the programs is very dependent on the type of process error present. We asses these differences in this section.

We consider two types of process errors in the implementation of a quantum map: ``coherent'' errors and ``incoherent'' errors.  A coherent error is one where the applied map is also unitary, but ``rotated'' from the target. All other errors are defined to be incoherent errors, for example, statistical mixtures of different unitary maps arising from inhomogeneous control or decoherence. We define the target unitary map ${\cal E}_{\rm t} = | U_{\rm t} )( U_{\rm t} |$, with corresponding process matrix $\chi_{\rm t}$. The actual applied map in the experiment has errors. We denote it, ${\cal E}_{\rm a}$, with corresponding process matrix $\chi_{\rm a}$.  We assume good experimental control so that the implementation errors are low, hence, $\chi_{\rm a}$ is close to $\chi_{\rm t}$. 

For both types of errors, we numerically model the applied process, $\mathcal{E}_{\rm a}$, by a composing the target process and an error process,
\begin{equation} \label{coh_err}
{\cal E}_{\rm a} = {\cal E}_{\rm err}\circ{\cal E}_{\rm t}.
\end{equation}
For coherent errors,
\begin{equation}
{\cal E}_{\rm err}[\cdot]=U_{\rm err}[\cdot]U_{\rm err}^\dagger,
\end{equation}
where the unitary error map is generated by a random, trace one, Hermitian matrix, selected by the Hilbert-Schmidt measure $U_{\rm err}=e^{\ii\eta H}$, with $\eta\geq0$. Such that the applied map is
\begin{equation} \label{coh_err}
{\cal E}_{\rm a}[\cdot]=U_{\rm err}U_{\rm t} [\cdot] U_{\rm t}^{\dagger} U_{\rm err}^\dagger.
\end{equation}
We numerically generate an incoherent error as
\begin{equation} 
{\cal E}_{\rm err}[\cdot]=(1-\xi)[\cdot]+\xi\sum_{n=1}^{d^2}A_n[\cdot]A_n^\dagger,
\end{equation}
which is not the only type of incoherent error possible but serves our numerical study. The applied map is then given by,
\begin{equation}\label{inc_err}
{\cal E}_{\rm a}[\cdot]=(1-\xi)U_{\rm t}[\cdot]U_{\rm t}^\dagger+\xi\sum_{n=1}^{d^2}A_nU_{\rm t}[\cdot]U_{\rm t}^\dagger A_n^\dagger.
\end{equation}
The set $\{A_n U_{\rm t}\}$ are Kraus operators associated with a CP map and $\xi \in [0,1]$. The $\{A_n\}$'s are generated by choosing a Haar-random unitary matrix $U$ of dimension $d^3$, and a random pure state of dimension $d^2$ from the Hilbert-Schmidt measure, $\ket{\nu}$, such that $ A_n= \bra{n}U\ket{\nu}$ where the set $\{\ket{n}\}$ is a computational basis~\cite{Bruzda2009}.  

\begin{figure}[t]
\centering
\includegraphics[width=0.9\linewidth]{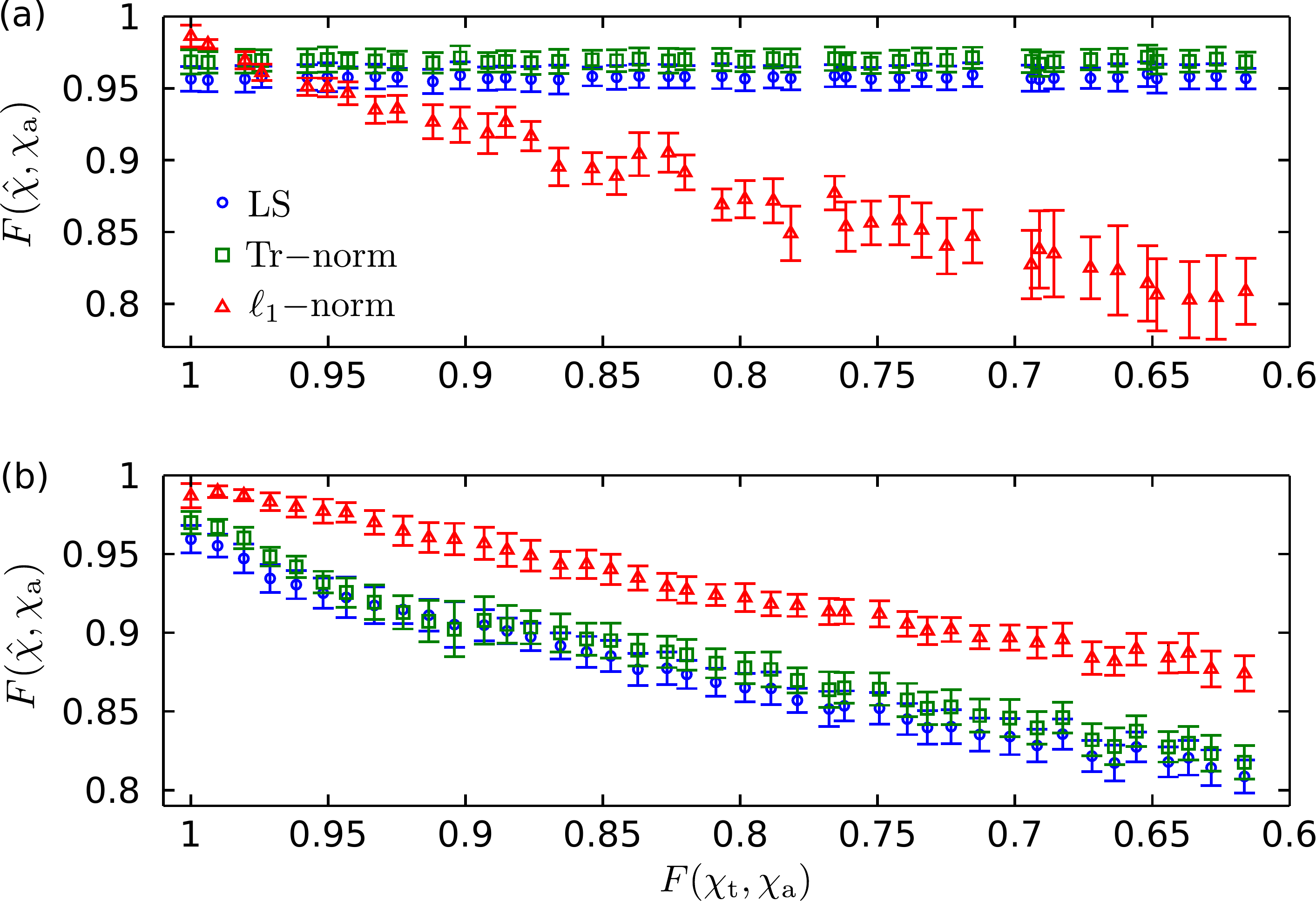} 
\caption[Comparison of reconstruction for UIC sets of states]{{\bf Comparison of reconstruction for UIC sets of states.} The fidelity between the estimate and the applied map as a function of the fidelity between the target map and the applied map for the case of (a) coherent errors, given by Eq.~\eqref{coh_err} with $\eta\in[0,3]$, and (b) and incoherent errors given by Eq.~\eqref{inc_err} with $\xi \in [0,0.6]$. The error bars represent the standard deviation. The estimates are obtained with data from only the five states of Eq.~(\ref{0+n}). Each data point in the plots is obtained by an average over 50 random target unitary maps each with a random error map.} 
 \label{fig:fidFid}
\end{figure}

We first numerically test the sensitivity of the all three programs to the type of preparation error and magnitude. We choose $d=5$ and prepare the five input quantum states defined by Eq.~\eqref{0+n}. We evolve each state with 50 randomly chosen applied processes, once with coherent errors and once with incoherent errors. We measure each output state with the MUB~\cite{Wootters1989}, introduced in Sec.~\ref{ssec:noiseless_ST}, and include noise from finite sampling. In Fig.~\ref{fig:fidFid}, we plot the fidelity between the applied process matrix, $\chi_{\rm a}$, and the estimated matrices, $\hat{\chi}$, determined by each of the three estimators, as a function of the fidelity between the applied process, $\chi_{\rm a}$, and the target, $\chi_{\rm t}$. The latter fidelity, $F(\chi_{\rm t},\chi_{\rm a})$,  is a measure of the magnitude of the error in the applied process. As expected by the robustness bound, all of the estimators return reconstructions that have high fidelity with the applied map when the applied map is close to the target unitary map ${\cal E}_{\rm a}[\cdot]\approx U_{\rm t}[\cdot]U_{\rm t}^\dagger$. In particular in our simulations $F(\hat{\chi},\chi_{\rm a})\gtrsim 0.95$ when  $F(\chi_{\rm t},\chi_{\rm a})\gtrsim 0.97$.  

However, as the magnitude of the preparation error increases, i.e. $F(\chi_{\rm t},\chi_{\rm p})$ decreases, the performance of the three estimators depends strongly on the nature of the errors.  The ${\ell_1}$-norm program is more sensitive to coherent errors than the Tr-norm and LS program, as seen in Fig.~\ref{fig:fidFid}a.   Using the data from five input states, the fidelity between the ${\ell_1}$-norm estimate and the applied map begins to fall below $\sim\!90\%$  in these simulations for $F(\chi_{\rm t},\chi_{\rm a})\lesssim 0.9$  while the Tr-norm and LS estimators maintain their high fidelity. This trend is reversed for incoherent errors, as seen in Fig.~\ref{fig:fidFid}b. The ${\ell_1}$-norm program is more robust to incoherent errors of the form of Eq.~\eqref{inc_err} than either the Tr-norm or LS programs because the process matrix is no longer close to a low-rank matrix, but it is still relatively close to a sparse matrix in the preferred basis. As the incoherent error magnitude increases, the  ${\ell_1}$-norm program returns an estimate with (on average) higher fidelity with the applied map than either the Tr-norm or the LS estimates. We thus conclude that when the applied map is sufficiently far from the target unitary map the performance of the three estimators varies in a manner that depends on the type of the error.

\begin{figure}[t]
\centering
\includegraphics[width=0.91\textwidth]{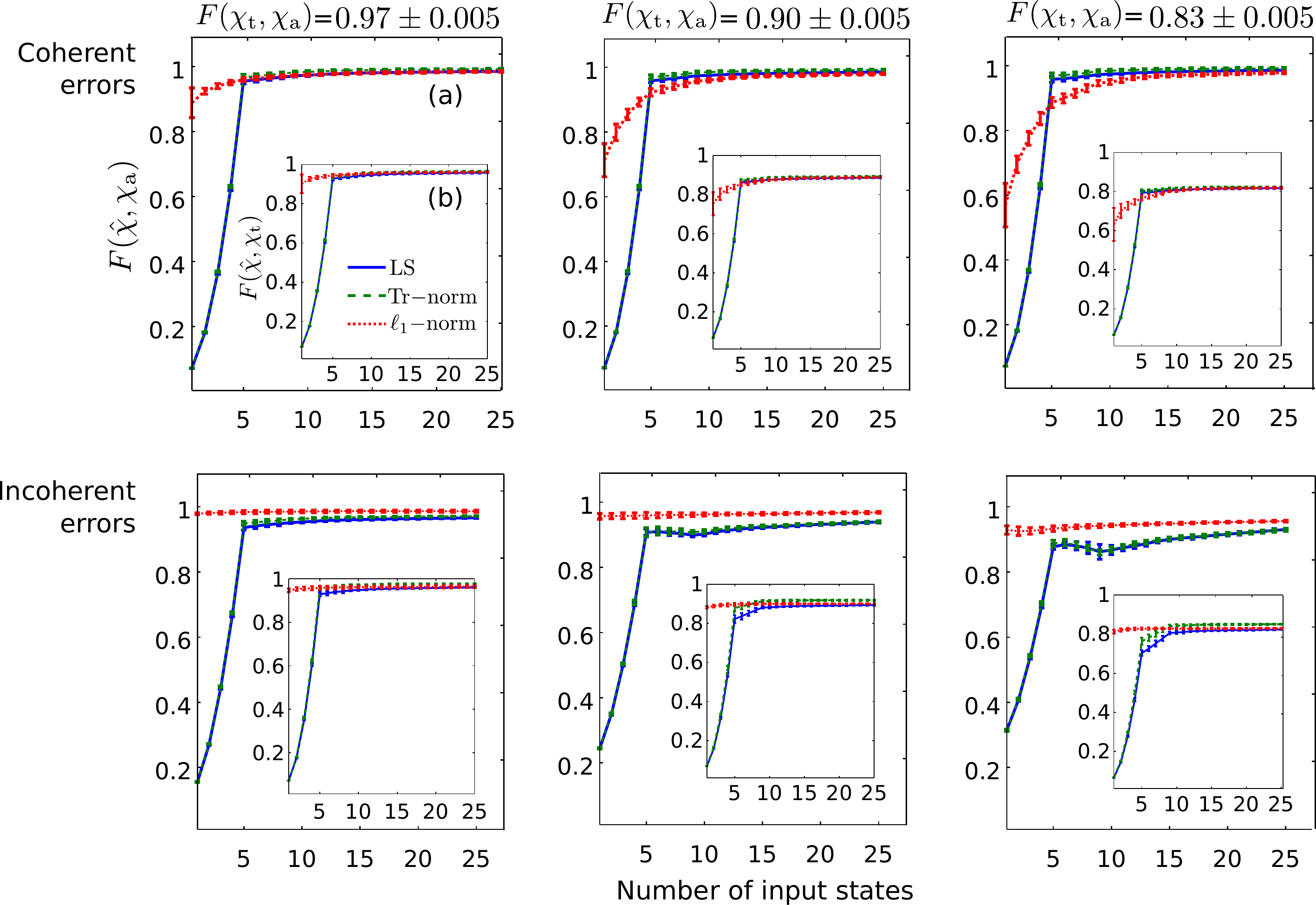} 
\caption[Comparison of estimators for QPT on a UIC set of states]{{\bf Comparison of estimators for QPT on a UIC set of states.} The fidelity between the estimate and the applied map, $F(\hat{\chi}, \chi_{\rm a})$ (a), and the estimate and the target map, $F(\hat{\chi}, \chi_{\rm t})$, inset (b), as a function of the number of input states, averaged over 50 applied processes, using different estimators, and under different error models for the applied map. Error bars represent the standard deviation. Each column corresponds to a different magnitude of implementation error, represented by the fidelity between the applied and target map, $F(\chi_{\rm t},\chi_{\rm a})$. Top row: Coherent errors as in Fig.~\ref{fig:fidFid}a. and bottom row: Incoherent errors  as in Fig.~\ref{fig:fidFid}b.} 
 \label{fig:fidStates}
\end{figure}

We can use the different behavior of the estimators as an indicator of the type of error that occurred in the applied process. This is seen when comparing the fidelity between the estimate and the applied map (Fig.~\ref{fig:fidStates}a)  and between the estimate and the target map (Fig.~\ref{fig:fidStates}b) as a function of the number of input states. We plot the fidelity averaged over 50 Haar-random applied maps. The plots on the top and bottom rows correspond to different levels of coherent and incoherent errors. Again, the robustness property is verified since the fidelity after $d=5$ input states is proportional to the error magnitude. While the Tr-norm and the LS estimators require the $d$ states in Eq.~\eqref{0+n} to reliably characterize the applied map, with proper formulation, the  ${\ell_1}$-norm program returns a reliable estimate with information obtained from a single input state. 

The estimator based on the ${\ell_1}$-norm is somewhat unstable when the reconstruction is based on data taken from very few input states. To overcome this instability,  one can use the same data obtained from the first $d$ input states, in different orders, to estimate the process, and then average over the resulting processes. This reduces the sensitivity to the specific choice of state of the first input state. In Fig.~\ref{fig:fidStates} we have used such averaging for estimating the process based of $1,2,\ldots, d=5$ input states.  Each estimated process is an average of the 5 reconstructed process matrices, each based on the data associated with an informationally complete measurement record on the 5 states  $\ket{\psi_n}$, $n=0,1,\ldots,4$ of Eq.~(\ref{0+n}), taken in cyclic permutations. 

If the error in the applied map is not small, we can infer the dominant source of the imperfection by examining the behavior of the different estimators.  As seen in Fig.~\ref{fig:fidStates}a, with $F(\chi_{\rm t},\chi_{\rm a})= 0.83\pm 0.005$,  when employing  the ${\ell_1}$-norm program,  a large coherent error results in a curvature in $F(\hat{\chi},\chi_{\rm t})$ as a function of the number of  input states.  Additionally, for the same data, using the Tr-norm and LS programs, we see that  $F(\hat{\chi},\chi_{\rm t})$ exhibits a sharp cusp after $d$ UIC probe states. In contrast, when the errors are dominantly incoherent, we see that when employing the ${\ell_1}$-norm program, $F(\hat\chi,\chi_{\rm t})$ is more or less a constant function of the number of input states.  In addition, there is a more gradual increase of $F(\hat{\chi},\chi_{\rm t})$ for the Tr-norm and  LS estimators around $d$ states; the cusp behavior is smoothed. These variations are signatures of the nature of the error in implementing the target unitary map.
 
In the regime $0.90\lesssim F(\chi_{\rm t},\chi_{\rm a})\lesssim 0.97$  it is difficult to distinguish, with high confidence, the nature of errors based solely on the behavior of $F(\hat\chi,\chi_{\rm t})$  as a function of input state,  and additional methods will be required to diagnose process matrix.  Nonetheless, a low fidelity of $F(\hat{\chi},\chi_{\rm t})\lesssim 0.95$ after $d$ input states challenges the validity of our assumptions and indicates the presence of noise.

A similar procedure could be adapted for strictly-complete POVMs in QST, or a set of strictly-complete probing states for QDT. In these cases the Tr-norm and $\ell_1$-norm programs can be used to diagnose the type of preparation error or measurement error present. In Sec.~\ref{sec:numerical_methods}, we introduced the Tr-norm program for QST, and the form for QDT is similar. The $\ell_1$-norm program can easily be translated to both QST and QDT as well.

\section{Summary and conclusions}
We have studied the problem of QPT under the assumption that the applied process is a unitary or close to a unitary map. We found that by probing a unitary map on a $d$-level system with $d$ specially chosen pure input states (which we called UIC set of states), one can discriminate it from any other arbitrary unitary map given the corresponding output states. In the ideal case of no errors, we can use a UIC set of states and a rank-1 strictly-complete POVM to characterize an unknown unitary with $2d^2$ POVM elements. We then numerically demonstrated that this combination is rank-1 strictly-complete for QPT. 

We used the methods of efficient unitary map reconstruction to analyze a more realistic scenario where the applied map is close to a target unitary map and the collected data includes statistical errors. Under this assumption, we studied the performance of three convex-optimization programs, the LS, Tr-norm and $\ell_1$-norm.  For each of these programs we estimated the applied process from the same simulated measurement vector obtained by probing the map with pure input states, the first $d$ of which form a UIC set. We considered two types of errors that may occur on the target map, coherent errors, for which the applied map is a unitary map but slightly ``rotated'' from the target map, and incoherent errors in which the applied map is full rank but with high purity. In our simulation, shown in Sec.~\ref{sec:QPT_werrors}, we used the states of Eq.~\eqref{0+n} to probe a randomly generated (applied) map with the desired properties verifying the robustness property applies for QPT. Our analysis suggests that when the prior assumptions are valid the three estimators yield high-fidelity estimates with the applied map using only the input UIC set of states.  We found that the sensitivity of these methods for various types of errors yields important information about the validity of the prior assumptions and about the nature of the errors that occurred in the applied map. In particular, probing the map with a UIC set of $d$ pure states and obtaining low fidelity between the estimates and the target map indicates that the errors are actually not small and the applied map is not close to the target unitary map. Furthermore, the performance of the different estimators under coherent and incoherent noise, enables the identification of the dominant error type. One can then take this this information into account to further improve the implementation of the desire map.

\chapter{Experimental comparison of methods in quantum tomography} \label{ch:experiment}
We have introduced many different methods for QT that offer theoretical advantages in efficiency and robustness. However, QT is designed to be a diagnostic tool for experiments in quantum information, so the performance of these methods must be verified in this context. In this chapter, we compare different methods for QST and QPT in an experimental platform, conducted in collaboration with the group of Prof. Poul Jessen at the University of Arizona. The quantum system we study is the hyperfine spin of $^{133}$Cs atoms in their electronic ground state, which corresponds to a 16-dimensional Hilbert space for encoding quantum states, processes, and POVMs. In this system, there are many ways to accomplish QT. For this chapter we compare different choices to illustrate the tradeoff between efficiency and robustness in QT protocols. Experiments were performed at the University of Arizona by Hector Sosa-Martinez and Nathan Lysne~\cite{SosaMartinez2016}.

\section{Physical system}
The physical system is an ensemble of approximately $10^6$ laser-cooled cesium atoms in ground manifold, $6S_{1/2}$. Each atom is (almost) identically prepared and addressed such that the quantum state, $\rho_N$, that describes the entire system is well approximated by a tensor product of the internal state of each atom, $\rho_N = \rho^{\otimes N}$. The internal state is confined to the hyperfine ground manifold with Hilbert space described by a tensor product of the nuclear and electron spin states, $\mathcal{H} = \mathcal{H}_I \otimes \mathcal{H}_S$. Cesium has a nuclear spin of $I= 7/2$ and, since it is an alkali metal, it has a single valence electron, so $S= 1/2$. The ground manifold is then described by a $(d=16)$-dimensional Hilbert space, which provides a large testbed for QT protocols. The Hilbert space can also be expressed as the direct sum of two hyperfine subspaces, $\mathcal{H} = \mathcal{H}_+ \oplus \mathcal{H}_-$, where $\mathcal{H}_{\pm}$ are the Hilbert spaces corresponding to the total hyperfine angular momentum quantum numbers, $F^{(+)} = 4$ and $F^{(-)} = 3$ spins. Each spin has $2F^{(\pm)} + 1$ degenerate magnetic sublevels. 

\subsection{Quantum control of the cesium system} 
In order to prepare quantum states, create evolutions, and readout information, we need control over the hyperfine manifold. In the experiment, the system is controlled with time-dependent magnetic fields applied approximately uniformly to the ensemble. The Hamiltonian that describes the dynamics is,
\begin{equation}
H = A {\bf I} \cdot {\bf S} - \bm{\mu} \cdot {\bf B}(t),
\end{equation}
where the first term is due to the hyperfine interaction (since $L = 0$ in the ground manifold) and $A$ is the hyperfine coupling. The second term describes the interaction with the time-dependent external magnetic fields, ${\bf B}(t)$, where $\bm{\mu}$ is the atomic magnetic moment. We can express the hyperfine interaction in terms of the total angular momentum, $F^{(\pm)}$, since ${\bf I} \cdot {\bf S} = \frac{1}{2} \left( {\bf F}^2 - {\bf I}^2 - {\bf S}^2 \right)$, and therefore $A {\bf I} \cdot {\bf S} = \frac{\Delta E_{HF}}{2}(\Pi^{(+)} - \Pi^{(-)})$, where $\Delta E_{HF} \triangleq A F^{(+)}$ is the hyperfine splitting, and $\Pi^{(\pm)}$ is the projection onto the plus or minus spin subspace. 

The interaction between the atoms and the magnetic fields is defined by the atomic magnetic moment, which is the sum of the spin and nuclear magnetic moments, $\bm{\mu} = \bm{\mu}_S + \bm{\mu}_I$. The contribution from the nuclear moment is extremely small, $\mu_B \gg \mu_N$, i.e. the Bohr magneton is much larger than the nuclear magneton. Therefore, we approximate the total magnetic moment as $-\bm{\mu} \cdot {\bf B}(t) \approx \mu_B g_s {\bf S} \cdot {\bf B}(t)$. When $\mu_B |{\bf B}(t) | \ll A$, i.e., the magnitude of the external field is much weaker than the hyperfine coupling, we can apply the Land\'{e} projection theorem to express the interaction in terms of the total angular momentum,
\begin{align}
\mu_B g_s {\bf S} \cdot {\bf B}(t) &\approx g_s \mu_B \sum_{i = \pm} \frac{{\bf S} \cdot {\bf F}^{(i)}}{F^{(i)} (F^{(i)} + 1)} {\bf F}^{(i)} \cdot {\bf B}(t), \nonumber \\
	& = g_+ \mu_B {\bf F^{(+)}} \cdot {\bf B}(t) + g_- \mu_B {\bf F^{(-)} } \cdot {\bf B}(t)
\end{align}
where $g_{\pm} \triangleq \pm \frac{g_S}{2 F^{(+)}} = \pm \frac{1}{F^{(+)}}$.

\begin{figure}[t]
\centering
\includegraphics[width=\linewidth]{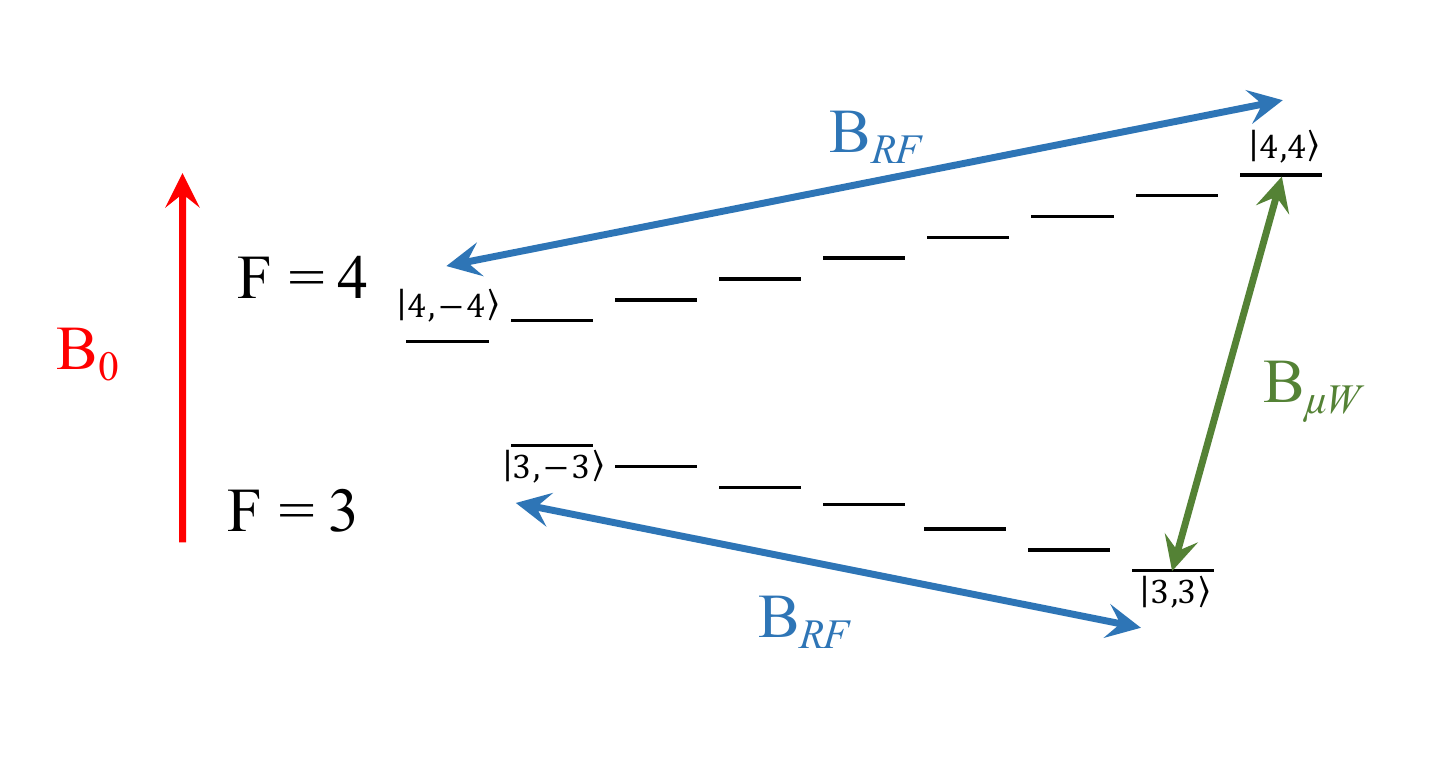}
\caption[Level-diagram of the hyperfine ground state of cesium with control fields]{{\bf Level-diagram of the hyperfine ground state of cesium with control fields} }
\label{fig:cesium}
\end{figure}

The magnetic field is applied as three separate fields,
\begin{equation}
{\bf B}(t) = B_0 {\bf e}_z + {\bf B}_{RF}(t) + {\bf B}_{\mu W}(t),
\end{equation}
where the first term is the bias field, which breaks the degeneracy of each hyperfine sublevel, shown in Fig.~\ref{fig:cesium}. The other two fields oscillates at radio (RF) and micro-wave ($\mu W$) frequencies respectively. Since the bias field is time-independent, we group it with the hyperfine interaction to form the ``drift'' Hamiltonian,
\begin{equation}
H_0 = \frac{\Delta E_{HF}}{2} \left(\Pi^{(+)} - \Pi^{(-)} \right) + \Omega_0 \left( F^{(+)}_z - F^{(-)}_z \right)
\end{equation}
where $\Omega_0 \triangleq \frac{\mu_B B_0}{F^{(+)}}$. 

The $RF$ field is a combination of two fields in the ${\bf e}_x$ and ${\bf e}_y$ direction,
\begin{equation}
{\bf B}_{RF}(t) = B_x {\bf e}_x \cos\left[ \omega_{RF} t + \phi_x(t) \right] + B_y {\bf e}_y \cos\left[ \omega_{RF} t + \phi_y(t) \right].
\end{equation}
The RF fields cause Larmor precession on the $F^{(+)}= 4$ and $F^{(-)}= 3$ spins, shown in Fig.~\ref{fig:cesium}, with corresponding Hamiltonian,
\begin{align}
H_{RF}(t) &= \Omega_x \left(F^{(+)}_x - F^{(-)}_x \right) \cos\left[ \omega_{RF} t + \phi_x(t) \right]\nonumber \\
	& + \Omega_y \left(F^{(+)}_y - F^{(-)}_y \right) \cos\left[ \omega_{RF} t + \phi_y(t) \right],
\end{align}
where $\Omega_{x,y} \triangleq \frac{\mu_B B_{x,y}}{F^{(+)}}$. 

The $\mu W$ field is tuned to resonance with the $\ket{4,4}$ and $\ket{3,3}$ transition, as shown in Fig.~\ref{fig:cesium}. This causes Rabi oscillations between the two magnetic sublevels, with resulting Hamiltonian,
\begin{equation}
H_{\mu W}(t) = \Omega_{\mu W} \sigma_x \cos\left[ \omega_{\mu W} t + \phi_x(t) \right],
\end{equation}
where $\Omega_{\mu W}$ is the Rabi frequency of the $\mu W$ interaction and $\sigma_x = | 4,4 \rangle \langle 3,3| + | 3,3 \rangle \langle 4,4|$.

In the experiment, the system is controlled by varying the the phases $\phi_x(t)$, $\phi_y(t)$, and $\phi_{\mu W}(t)$, which are referred to as the ``control parameters.'' To eliminate the time dependence in the Hamiltonian, outside of the control parameters, we move to the rotating frame defined by the unitary,
\begin{equation}
U = \textrm{exp}\left[ -i \omega_{RF} t \left( F^{(+)}_z - F^{(-)}_z \right) \right] \textrm{exp}\left[ -it\frac{\omega_{\mu W} - 7 \omega_{RF}}{2} \left( \Pi^{(+)} - \Pi^{(-)} \right) \right].
\end{equation}
We apply $U$ to each term in the total Hamiltonian such that $H' = U^{\dagger} H U$. By the rotating wave approximation, all terms proportional to $\cos(2 \omega_{RF}t)$ and $\cos(2 \omega_{\mu W} t)$ are approximately zero. We include the term, $H_{\rm rot.} = -iU^{\dagger} \frac{d U}{dt}$, which is due to the rotating frame, with the drift Hamiltonian, since it is also time independent. Therefore,
\begin{align}
H_0' &\approx U^{\dagger} H_0 U - iU^{\dagger} \frac{d U}{dt}, \nonumber \\
&= \frac{\Delta_{\mu W}}{2} \left( \Pi^{(+)} - \Pi^{(-)} \right) + \Delta_{RF} \left( F^{(+)}_z - F^{(-)}_z \right), 
\end{align}
where $\Delta_{\mu W} \triangleq \omega_{\mu W} - \Delta E_{HF} - 7 \omega_{RF}$ and $\Delta_{RF} \triangleq \omega_{RF} - \Omega_0$.  The $RF$ and $\mu W$ control Hamiltonians are
\begin{align}
H_{RF}'(t) &\approx \frac{\Omega_x}{2} \cos \left[ \phi_x(t) \right] \left( F^{(+)}_x - F^{(-)}_x \right) -\frac{\Omega_x}{2} \sin \left[ \phi_x(t) \right]\left( F^{(+)}_y + F^{(-)}_y \right) \nonumber \\
			&+ \frac{\Omega_y}{2} \cos\left[ \phi_y(t) \right]\left( F^{(+)}_y - F^{(-)}_y \right) +  \frac{\Omega_y}{2} \sin\left[ \phi_y(t) \right]\left( F^{(+)}_x + F^{(-)}_x \right), \nonumber \\
H_{\mu W}'(t) &\approx  \frac{\Omega_{\mu W}}{2} \left( \cos \left[\phi_{\mu W}(t)\right] \sigma_x + \sin \left[ \phi_{\mu W}(t)\right]\sigma_y \right).
\end{align}
where $\sigma_y$ is the Pauli-y operator across the $\ket{4,4}$ and $\ket{3,3}$ subspace. Higher order terms in the rotating wave approximation were derived in Ref.~\cite{Riofrio2011}.

Full controllability with the $RF$ and $\mu W$ magnetic fields was proven in Ref.~\cite{Merkel2008}. Therefore, the magnetic fields can create any unitary map in $\textrm{SU}(16)$. To produce a given unitary, we numerically search for a set of the phases, $\phi_x(t), \phi_y(t),$ and $\phi_{\mu W}(t)$, that optimize an objective function (further details are given in Appendix~\ref{app:control} and Ref.~\cite{Anderson2013}). The procedure is also made robust to inhomogeneities in the control fields to minimize errors~\cite{Anderson2013,Smith2012}. Therefore, we can prepare any pure state or apply any unitary evolution to the 16-dimensional Hilbert space.

\subsection{Stern-Gerlach measurement} \label{sec:SG_measurement}
\begin{figure}[t]
\centering
\includegraphics[width=\linewidth]{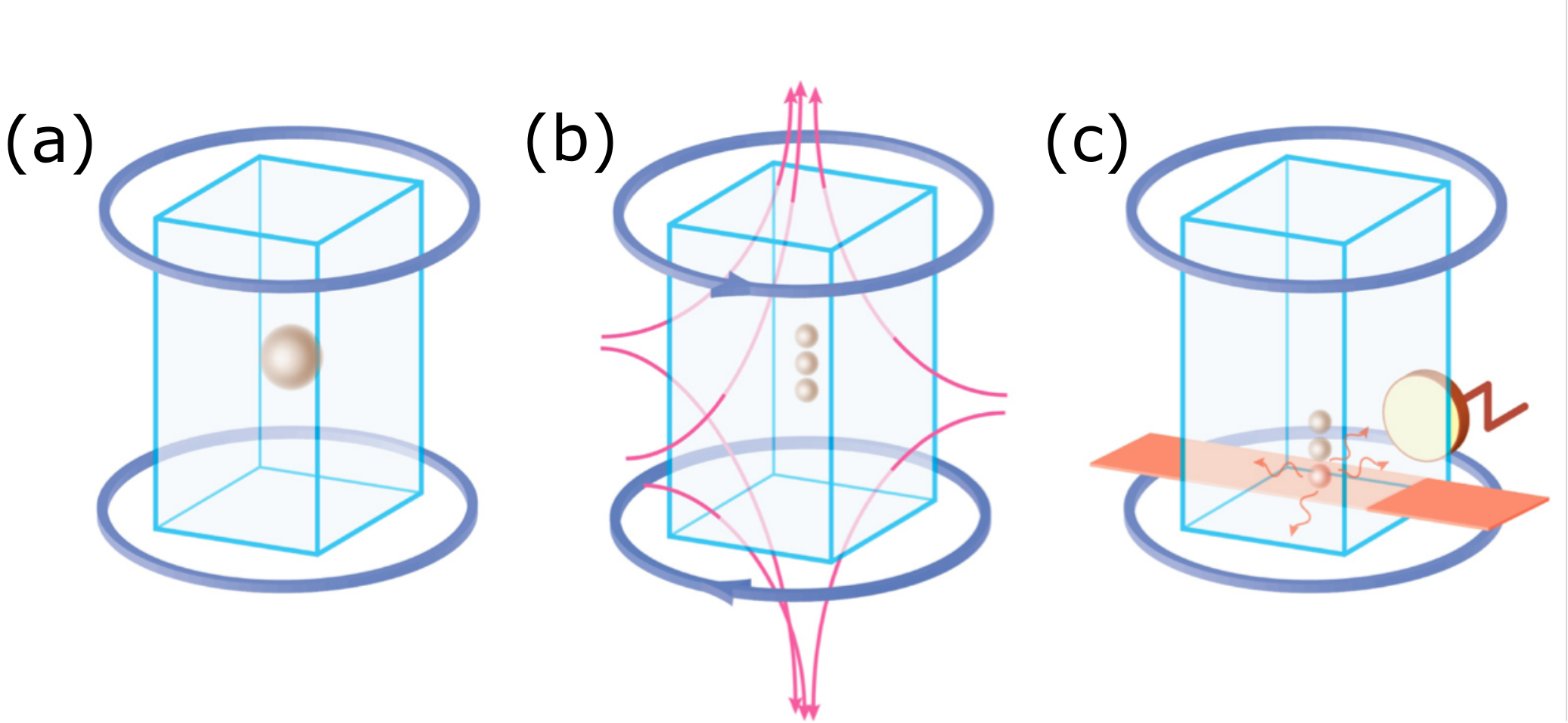}
\caption[Stern-Gerlach analyzer]{{\bf Stern-Gerlach analyzer} (Figure provided by Hector Sosa-Martinez) (a) The atoms are initially prepared in a single cloud. (b) A gradient magnetic field is applied to spacial separate the atoms based on the internal spin thus entangling their position with the spin state. (c) The atoms are dropped and fall through a sheet of laser light and the fluorescence signal is detected.}
\label{fig:sg_steps}
\end{figure}

\begin{figure}[t]
\centering
\includegraphics[width=\linewidth]{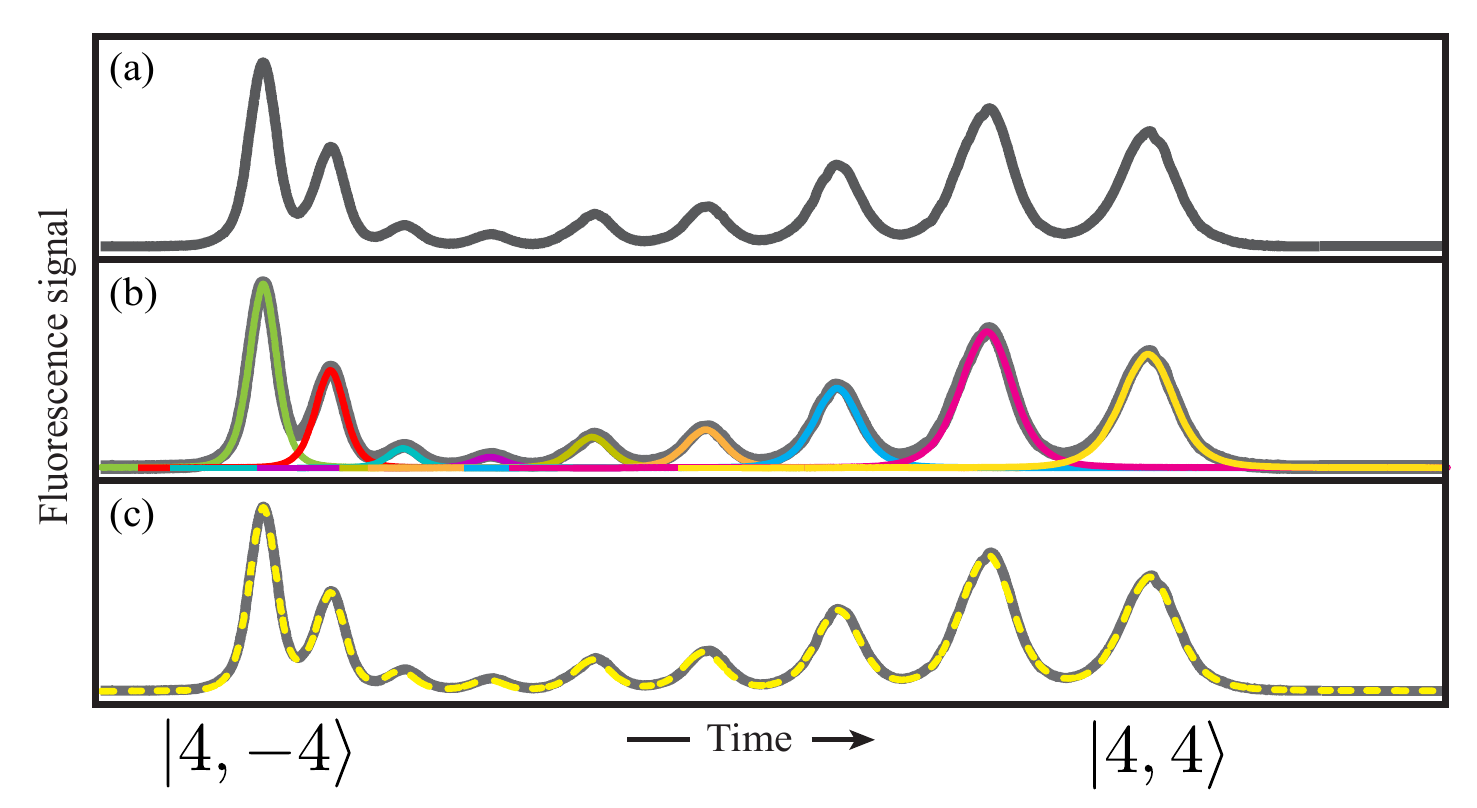}
\caption[Sample Stern-Gerlach signal for the $F= 4$ subspace]{{\bf Sample Stern-Gerlach signal for the $F= 4$ subspace} (Figure provided by Hector Sosa-Martinez) (a) The raw signal from from the detector. (b) The nine fit functions for each of the magnetic sublevels overlapping the signal shown in various colors. (c) The resulting fit to the date shown by the dashed yellow line.}
\label{fig:sg_signal}
\end{figure}

The quantum state is measured with a Stern-Gerlach analyzer that creates a signal proportional to the population of each magnetic sublevel. This is accomplished by applying a gradient magnetic field in the $z$-direction, parallel to the bias field, that separates the atoms spatially, shown in Fig.~\ref{fig:sg_steps}b. The separation is proportional to the spin projection, $m_F$, in the $z$-direction. The atoms are then dropped, and fall through a sheet of laser light that is resonant with a transition between the hyperfine-ground manifold and an excited state, shown if Fig.~\ref{fig:sg_steps}c. This causes fluorescence that is detected in discrete time bins. Each bin corresponds to a time-of-flight measurement of the atoms trajectory. Since the signal is proportional to $m_F$, the procedure is repeated for each submanifold, $F^{(\pm)}$. An example signal is shown in Fig.~\ref{fig:sg_signal} for $F^{(+)}$. 

The gradient magnetic field acts to entangle the internal spin projection with the position of the atoms. This interaction can be described by a unitary map, $U_{\bf B}$ on tensor product space of the atoms internal spin (labelled with subscript $s$) and position (labelled with subscript $p$),
\begin{equation}
\rho_{sp} = U_{\bf B} \left( \rho_s \otimes \rho_p\right) U_{\bf B}^{\dagger}.
\end{equation}
The time-of-flight signal is then related to the position of each atom by classical dynamics. Therefore, the POVM elements that describe the measurement are projectors onto the position. Since the signal is discrete, the POVM elements correspond to a projection onto a range of positions, $\delta x$,
 \begin{equation} \label{x_POVM}
 E_x = \int_x^{x + \delta x} dx | x \rangle \langle x|.
 \end{equation}
In principle, we could accomplish QST with the raw signal from the time-of-flight measurement and the POVM elements in Eq.~\eqref{x_POVM}. However, the number of time bins that correspond to the different measurement outcomes is potentially very large, and therefore implementing the estimation required for QST may be expensive.

Instead of directly using the discrete time-of-flight signal and POVM in Eq.~\eqref{x_POVM}, we perform ``two-step'' QST by first extracting most of the information from the signal, then using this information for state estimation. The signal for each $F^{(\pm)}$ submanifold contains $2F^{(\pm)} + 1$ peaks that correspond to the $2F^{(\pm)} + 1$ magnetic sublevels, which are pulled apart into separate clouds by the gradient magnetic field. The distribution is proportional to the $2F^{(\pm)} + 1$ real numbers that specify the population in each magnetic sublevel. In order to extract these real numbers, we fit the signal to $2F^{(\pm)} + 1$ template functions. A numerical description of the template functions is found by applying the Stern-Gerlach analyzer to the ensemble without the gradient magnetic field. The result is a single peak that describes the distribution of the falling atoms. The template function is then fit to each of the 16 peaks when the gradient field is applied, with fitting parameters proportional to the height, width, and center of each fit function, as shown in Fig.~\ref{fig:sg_signal}b. There are additional parameters that scale the template function proportional to the trajectory of the different magnetic sublevels. 

The result of the fitting is used to estimate the fraction of spins in each magnetic sublevel, which is then related to the populations, $\{ \rho_{F,m_F} \}$. Therefore, we have extracted the measurement vector that corresponds to the $F_z$-basis measurement,
\begin{equation}
\mathbbm{B}_z = \{ \ket{4,-4}, \dots, \ket{4,4}, \ket{3,-3}, \dots, \ket{3,3} \}.
\end{equation}
where the states are written as $| F, m_F \rangle$ basis. While we are not directly, making a projective measurement in this basis, the measurement vector returned by the fitting program is in good approximation the populations we seek. 

We can use the control Hamiltonians to also determine the measurement vector of an arbitrary orthonormal basis measurement. In the following discussion, we label $\ket{\mu} = \ket{F, m_F} $, such that $\mu = 0,\dots,15$, referred to as the ``standard basis,'' for convenience. An arbitrary orthonormal basis measurement, $ \mathbbm{B}_{\psi} = \{ | \psi_{\mu} \rangle \}$, has probability of each outcome
\begin{equation}
p_{\mu} = \langle \psi_{\mu} | \rho | \psi_{\mu} \rangle = \langle \mu | W^{\dagger} \rho W | \mu \rangle
\end{equation}
where $| \psi_{\mu} \rangle = W |\mu \rangle$. Therefore, to approximate the measurement vector of the basis $\mathbbm{B}_{\psi}$, we apply the unitary map, $W^{\dagger} = \sum_{\mu} | \mu \rangle \langle \psi_{\mu} |$, before the Stern-Gerlach analyzer. Then, the measurement vector from the fitting procedure is an approximation of the probability of each outcome from the basis $\mathbbm{B}_{\psi}$. We can search for a control parameters that implements the unitary $W^{\dagger}$ with a unitary control objective, which is described in Appendix~\ref{app:control}.

We can also approximate the measurement vector of a basis on $s$-dimensional subspaces of the total Hilbert space, $\mathbbm{B}_{\psi} = \{ | \psi_1 \rangle, \dots, | \psi_{s} \rangle \}$ with a partial isometry. A partial isometry is a mapping of $s < d$ orthonormal states to another set of $s$ orthonormal states. Partial isometries require less total time to implement then a full unitary map, which is advantageous when there exist sources of errors or noise that compound with time. In order to implement $\mathbbm{B}_{\psi}$, we need the partial isometry mapping, $ \{ | \psi_{\mu} \rangle\}  \rightarrow \{ | \mu \rangle \}$ for $s$ orthonormal pairs of states. Partial isometry control objectives are described in detail in Appendix~\ref{app:control}.

We can additionally use partial isometries to determine the measurement vector of a POVM with $N \leq 16$ rank-1 elements, nonorthogonal elements $E_{\nu} = | \tilde{\phi}_{\nu} \rangle \langle \tilde{\phi}_{\nu} |$, on an $s$-dimensional subspace of the total Hilbert space by the Neumark extension~\cite{Preskill1998}. The Neumark extension is accomplished by determining a basis measurement on the 16-dimensional Hilbert space whose projection onto the $s$-dimensional subspace is the desired POVM elements. That is, a basis, $\mathbbm{B}_{\phi} = \{ \phi_1,\dots, \phi_{16}\}$, accomplishes the Neumark extension if $\Pi_s | \phi_{\nu} \rangle \langle \phi_{\nu} | \Pi_s =  | \tilde{\phi}_{\nu} \rangle \langle \tilde{\phi}_{\nu} |$, where $\Pi_s$ is the projection onto the $s$-dimensional subspace.

The choice of $\mathbbm{B}_{\phi}$ is not unique, in that there are many bases that have the same projection onto the $s$-dimensional subspace. This freedom can be exploited to make more accurate POVMs by using a partial isometry control. To see this, we organize the set of vectors $\{ |\phi_{\nu} \rangle \}$ into an $s \times N$ matrix,
\begin{equation}
V = \begin{pmatrix}
\uparrow & & \uparrow \\
|\tilde{\phi}_{1} \rangle& \cdots &| \tilde{\phi}_N \rangle \\
\downarrow & & \downarrow
\end{pmatrix} = 
\begin{pmatrix}
\leftarrow & \bra{\psi_1} & \rightarrow \\
	& \vdots & 	\\
\leftarrow & \bra{\psi_N} & \rightarrow
\end{pmatrix}
\end{equation}
where $\bra{\psi_{\alpha} } = \sum_{\nu} V_{\alpha,\nu} \bra{\nu}$ are the rows of $V$ and $V_{\alpha,\nu} = \langle \alpha | \tilde{\phi}_{\nu} \rangle$ are the elements of $V$. The rows of $V$ are orthonormal,
\begin{equation}
\langle \psi_{\alpha} | \psi_{\beta} \rangle = \sum_{\mu,\nu} V_{\alpha,\mu} V_{\beta,\nu}^* \delta_{\mu, \nu} = \sum_{\nu}\langle \alpha | \tilde{\phi}_{\nu} \rangle \langle  \tilde{\phi}_{\nu} | \beta \rangle = \delta_{\alpha, \beta},
\end{equation}
due to the POVM condition, $\sum_{\nu} | \tilde{\phi}_{\nu} \rangle \langle  \tilde{\phi}_{\nu} |  = \mathds{1}$. We then implement the partial isometry mapping, $\{ \ket{\alpha} \} \rightarrow \{ \ket{\psi_{\alpha}} \}$ for $s$ states, where $\{\ket{\alpha} \}$ is the standard basis elements that span the $s$-dimensional subspace. Therefore, the partial isometry maps, $\rho^{\rm in} = \sum_{\alpha,\beta} \rho_{\alpha,\beta} | \alpha \rangle \langle \beta |$ to $\rho^{\rm out} = \sum_{\alpha,\beta } \rho_{\alpha,\beta} |\psi_{\alpha} \rangle \langle \psi_{\beta} |$, where $\rho_{\alpha,\beta} = \langle \alpha | \rho^{\rm in} | \beta \rangle$. Then, after applying the Stern-Gerlach analyzer, the probability of getting $\mu$th outcome is,
\begin{align}
p_{\mu} &= \langle \mu | \rho^{\rm out} | \mu \rangle, \nonumber \\
	&= \sum_{\alpha, \beta} \underbrace{\langle \mu | \psi_{\alpha} \rangle \langle \alpha |}_{V^*_{\alpha,\mu} \bra{\alpha}} \rho^{\rm in} \underbrace{| \beta \rangle \langle \psi_{\beta} | \mu \rangle}_{V_{\beta,\mu} \ket{\beta}}, 
\end{align}
by the definition of $V$, $\sum_{\alpha} V^*_{\alpha,\mu} \bra{\alpha} = \bra{\tilde{\phi}_{\mu}}$ and $V_{\beta,\mu} \ket{\beta} = \ket{\tilde{\phi}_{\mu}}$. Therefore,
\begin{equation}
p_{\mu}  = \langle \tilde{\phi}_{\mu} | \rho^{\rm in} | \tilde{\phi}_{\mu} \rangle,
\end{equation}
such that the probability of each outcome of the standard basis measurement is equal to the probability of each POVM outcome. Therefore, if we apply the partial isometry mapping before the Stern-Gerlach analyzer, the measurement vector  approximates the $N$-element POVM, $\{ E_{\nu} \}$.

\subsection{Sources of noise and errors} \label{ssec:errors_expt}
There are several sources of noise and errors that can reduce the accuracy of QT in the cesium spin system. One major source of error comes from inexact implementation of the control fields that perform the unitary maps and partial isometries. This error is caused by inhomogeneities and variation from the ideal control fields. The inhomogeneities are a result of nonuniform magnetic fields across the ensemble of atoms in the extended cloud. Therefore, each atom does not interact with the magnetic field equally and the final state of the system will not be in exact tensor product of $N$ identical states. The effect of inhomogeneities is reduced by the robustness procedures used in the control~\cite{Anderson2013, Mischuck2010}. The other source of control error comes from variation in the phases that are actually applied to the atomic ensemble from the ideal behavior. As discussed in Appendix~\ref{app:control}, the control phases ($\phi_x(t)$, $\phi_y(t)$, and $\phi_{\mu W}(t)$) are designed to be a piecewise-constant functions of time. However, this is not exactly true in the experiment because of the finite response time of the controllers. Deviation from the piecewise-constant function causes coherent control errors.

The magnitude of all control errors was previously characterized by a randomized benchmarking inspired technique in Refs.~\cite{Anderson2015, Smith2012}. This procedure determines the average fidelity, $\mathcal{F}$, associated with all unitary mappings and is independent of other sources of errors, such as errors in the Stern-Gerlach analyzer. It was found that each unitary on the 16-dimensional space has average fidelity, $\mathcal{F} = 0.975$~\cite{Anderson2015}. A similar method was used to determine the error associated with partial isometry mappings of any dimension; for example, in Ref.~\cite{Smith2012} the state preparation fidelity was measured as $\mathcal{F} = 0.995$.

There will also necessarily exists statistical noise from finite sampling, since there are a finite number of atoms that are measured. While there are approximately $10^6$ atoms prepared, only a fraction of fluorescing atoms produce photons that are detected. However, the effect of the finite sampling is still much smaller than the other control errors. In previous diagnostics, it was seen that the final measurement vector barely fluctuates between repetitions of the Stern-Gerlach analyzer with the same controls, so finite sampling effects are negligible.

There are other sources of of noise and errors present in the experiment. One possible source of errors is decoherence due to interactions with stray light and background magnetic fields. There may be other errors associated with implementing the Stern-Gerlach analyzer that are less well understood. For example, the gradient magnetic field may not be exactly aligned with the bias field meaning the POVM that we think describes the measurement is slightly rotated. Also, there is shot noise due to measuring the fluorescence of the photons in the signal and other technical noise associated with the photon detection. However, we believe all these other sources to be small compared to the control errors.

\section{Implemented POVMs} \label{sec:exp_POVMs}
Several different POVMs were implemented in the experiment. All measurement vectors were produced via the two-step procedure described in Sec.~\ref{sec:SG_measurement} and extracted by a least-squares-fit  to the time-of-flight signal.  If we precede the Stern-Gerlach analyzer by the appropriate unitary map, we can extract the measurement vector in an arbitrary basis on the full 16-dimensional Hilbert space, or on any subspace.  Alternatively, via the Neumark extension, we can implement any POVM with up to 16 rank-1 POVM elements with states that live is an $s<d$-dimensional Hilbert space. The study was carried out for Hilbert spaces in $d=4$ and $d = 16$ dimensions.

The measurement vector of the following full-IC POVMs were approximated via the two-step method described in Sec.~\ref{sec:SG_measurement}:
\begin{itemize}
\item {\bf Symmetric informationally complete POVM:} (SIC POVM), originally proposed in Ref.~\cite{Renes2004a} and described in further detail in Sec.~\ref{ssec:noiseless_ST}. The SIC POVM has $d^2$ outcomes and has a known construction for all dimensions $d \leq 67$~\cite{Scott2010}. In the experiment, the SIC is applied only for $d= 4$, and 16 POVM elements are implemented with the Neumark extension. 

\item {\bf Mutually unbiased bases:} (MUB) originally proposed in Ref.~\cite{Wootters1989} and described in further detail in Sec.~\ref{ssec:noiseless_ST}. The MUB consists of $d+1$ orthonormal bases and has an analytic expression for  dimensions that are primes or powers of primes~\cite{Wootters1989}. Therefore, the MUB can be implemented in both the $d = 4$ and 16 systems. 

\item {\bf Gell-Mann bases:} (GMB) an extension of the five bases proposed in Ref.~\cite{Goyeneche2015} and discussed Sec.~\ref{sec:decomp_method}. The GMB consists of $2d-1$ orthonormal bases for dimensions that are powers of two with algorithm provided in Appendix~\ref{app:GMB}. (Constructions for other dimensions exist but are more complicated.) The GMB is applied for both $d = 4$ and $d= 16$ in the experiment.
 \end{itemize}
 
The measurement vector of the following rank-1 strictly-complete POVMs was estimated by the two-step procedure in Sec.~\ref{sec:SG_measurement}:
\begin{itemize}
\item {\bf PSI-complete:} (PSI) originally proposed Ref.~\cite{Flammia2005} and proven to be rank-1 complete therein. The PSI-complete POVM has $3d-2$ rank-1 POVM elements in any dimension. A construction of the the POVM and a proof that it is rank-1 strictly-complete is provided in Appendix~\ref{app:constructions}. We implement the PSI-complete POVM for $d=4$ via the Neumark extension. 

\item {\bf Five Gell-Mann bases:} (5GMB) originally proposed in Ref.~\cite{Goyeneche2015} as a rank-$1$ complete measurement. The 5GMB are the first five orthonormal bases of the GMB. In Appendix~\ref{app:GMB}, we use the matrix completion method from Sec.~\ref{sec:EP} to prove that the 5GMB are also rank-$1$ strictly-complete. Since the GMB are applied for both $d = 4$ and 16 we can also apply the 5GMB for both systems. 

\item {\bf Five polynomial bases:} (5PB) originally proposed in Ref.~\cite{Carmeli2016} and proven to be rank-1 strictly-complete therein. The 5PB are the 4PB, discussed below, plus the basis, $\mathbbm{B}_z$. The remaining four bases are constructed from a set of orthogonal polynomials. The construction applies for any dimension, so we implement them for both $d=4$ and 16. We provide an explicit construction of the five bases in Appendix~\ref{app:constructions}. The measurement vector for the first bases is taken from the first bases of the 5GMB, since it is the same basis measurement.

\item {\bf Five Mutually unbiased bases:} (5MUB) Numerical simulations, similar to the ones performed for random bases in Sec.~\ref{sec:random_bases}, indicate that the first five bases of the MUB correspond to a rank-1 strictly-complete POVM. We only apply the 5MUB for $d= 16$ since the 5MUB in $d=4$ is full-IC. The measurement vector for the 5MUB is the same measurement vector as the first five bases of the MUB.
\end{itemize}

Finally, the measurement vector for the following rank-1 complete POVMs were estimated by the two-step procedure described in Sec.~\ref{sec:SG_measurement}:
\begin{itemize}
\item {\bf Four GMB:} (4GMB) originally proposed in Ref.~\cite{Goyeneche2015} and given in Eq.~\eqref{4gmb}. The 4GMB are four of the orthonormal bases that make up the GMB. Since the GMBs can be implemented for $d = 4$ and 16 the first four can be implemented in both dimensions as well. The measurement vector for the 4GMB is the measurement vector from four bases of 5GMB.

\item {\bf Four polynomial bases:} (4PB) originally proposed in Ref.~\cite{Carmeli2015}. The 4PB consists of four orthonormal bases that are constructed based on a set of orthogonal polynomials for any dimension. We generate the 4PB based on the Hermite polynomial and apply them for both $d = 4$ and 16. We provide an explicit definition of the four bases in Appendix~\ref{app:constructions}. The measurement vector for the 4PB is the measurement vector from the last four bases of the 5PB.
\end{itemize}

\section{Quantum state tomography results} \label{sec:ST_exp}
In the experiments performed in the Jessen lab, each measurement was applied to a fixed set of 20 different Haar-random pure states prepared by the state-to-state mapping described in Appendix~\ref{app:control}. To determine how each measurement and estimation procedure performs, we would like to compare the actual prepared state, $\rho_{\textrm{a}}$, to the estimated state, $\hat{\rho}$. However, as with any QT experiment, we do not know the actual prepared state, so instead, we compare the estimated state to the target state, $\rho_{\textrm{t}}$. From Ref.~\cite{Smith2012}, we know that the prepared state is close to the target state (the average fidelity of state preparation is $\mathcal{F} = 0.995$), and therefore this comparison is a reasonable measure of QT. We use the infidelity between two quantum states, which is the standard measure of QST, to quantify the comparison,
\begin{equation} \label{fidelity}
1-F(\rho_{\rm t}, \hat{\rho}) = 1-\left| \langle \psi_{\rm t} | \hat{\rho} | \psi_{\rm t} \rangle \right|
\end{equation}
where $\rho_{\rm t} = | \psi_{\rm t} \rangle \langle \psi_{\rm t} |$, since in the experiment all target states are pure. (We use the script letter, $\mathcal{F}$, to denote the average fidelity but the capital $F$ to denote fidelity between two particular quantum states.)

\begin{figure}[]
\centering
\includegraphics[width = \linewidth]{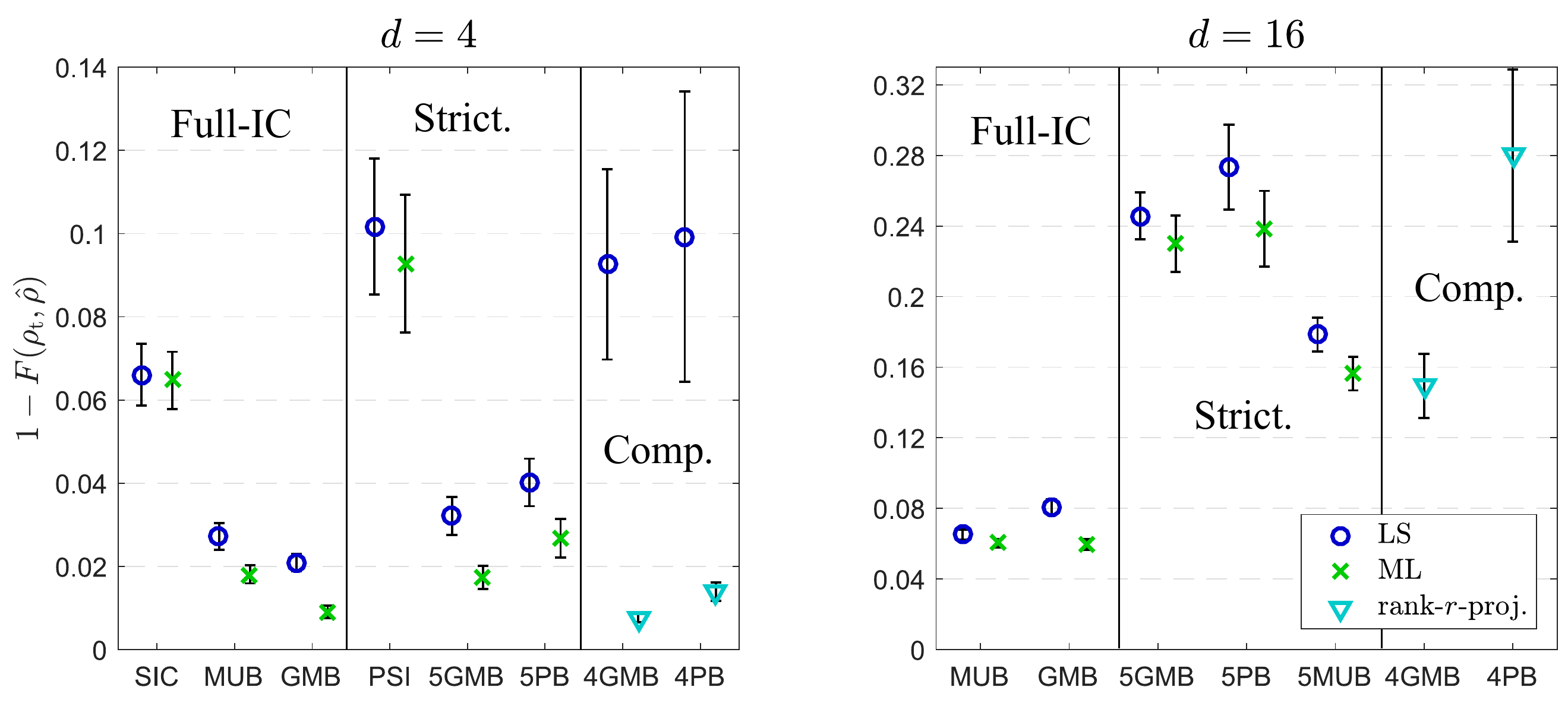}
\caption[Mean infidelity of estimate and target state for each POVM]{{\bf Mean infidelity of estimate and target state for each POVM} Estimates are created with LS (blue circles), ML (green exes) , and the rank-$r$-projection algorithm in Sec.~\ref{sec:noiseless_rankr_comp} (cyan triangles). The markers indicate the average over the 20 Haar-random pure states and the error bars correspond to the standard error in the mean. The POVM used to create the measurement vector is given on the x-axis. The top figure corresponds to $d = 4$, while the bottom corresponds to $d=16$. The type of POVM is labelled on the graph. For $d = 16$, with the 4GMB and 4PB the LS estimator is omitted since it produces very large infidelity. }
\label{fig:IC}
\end{figure}

The value of $\hat{\rho}$ will depend on the type of estimator chosen. In Sec.~\ref{sec:numerical_methods}, we presented many estimators for QST such as, linear-inversion (LI), least-squares (LS), maximum-likelihood (ML), and trace-norm minimization (Tr-norm). In this section we forgo applying the LI estimator since it does not produce a physical state so Eq.~\eqref{fidelity} does not apply. We determine the estimate from LS and ML with the CVX package~\cite{cvx} in MATLAB for each of the full-IC and rank-1 strictly-complete discussed in Sec.~\ref{sec:exp_POVMs}. For the rank-1 complete POVMs, we use the LS program and rank-$r$-projection algorithm described in Sec.~\ref{sec:noiseless_rankr_comp}. We discuss the Tr-norm estimates in further detail later.  The results are plotted in Fig.~\ref{fig:IC}

All POVMs and estimators produce low infidelity estimates of the quantum state. The ML estimate produces a consistently lower infidelity than the LS. This would be expected if the experiment was limited by finite sampling, since LS differs from ML when there are a finite number of copies~\cite{James2001}. However, the measurements in this experiment are not limited by finite sampling, so we do not believe the difference is caused by this effect. Instead, the difference is likely due to the positivity condition that changes the shape of the likelihood function. This effect will be studied in future work that investigates how each estimator behaves in the presence of the positivity constraint. While the estimate from ML has lower infidelities, it does require greater computational effort to produce the ML estimate since the log-likelihood function is less smooth, so more difficult to optimize over. We find that when computational effort is not a limitation, ML is the best estimator, though the gain is modest. 

Interestingly, for $d = 4$, Fig.~\ref{fig:IC} shows that the estimates from the rank-$r$-projection algorithm with the rank-1 complete POVMs yield the lowest infidelities. However, it is not fair to compare different POVMs when the data is fed into different estimators. To make a fair comparison, we apply the rank-$r$-projection algorithm to the data from the rank-1 strictly-complete POVMs. In this case, we find the infidelities of the rank-1 strictly-complete POVMs to be $1-F_{\rm 5GMB} = 0.0068 (0.0010)$ and $1-F_{\rm 5PB} = 0.0115(0.0019)$, which are comparable to the values found for the 4GMB, $1-F_{\rm 4GMB} = 0.0931(0.0010)$ (standard error in the mean are in parentheses). Therefore, the low infidelities produced by the rank-1 complete POVMs should be attributed to the rank-$r$ projection algorithm that is required for this type of POVM. However, this algorithm is not always desirable. Beyond the issue of convergence, discussed in Sec.~\ref{sec:noiseless_rankr_comp}, the algorithm is, by construction, biased to produce only pure states. Therefore, the estimate lacks information about preparation errors that may cause the actual state to be mixed. Moreover, since the estimate is always pure it may be much closer to the target state, which is also pure, than the actual state. In this case, comparing the different POVMs by the infidelity between the target state and the estimate would not be an accurate measure of success. For these reasons, we do not consider rank-1 complete POVMs for the remainder of the discussion.


We focus on LS (blue circles) in order to compare the different POVMs, since ML and LS have similar trends. For both dimensions, the results mostly match what we expect based on previous discussions of informational completeness. Since we have prior information that the state is near, but certainly not exactly pure, we predict the full-IC POVMs to perform the best, since they can characterize an arbitrary full-rank state. However, since the state is near-pure, we expect the strictly-complete POVM to perform almost as well, since these POVMs are robust to preparation errors. For $d=4$, we see that the full-IC POVMs, GMB and MUB, indeed produce the lowest infidelity estimates, followed by the  strictly-complete POVMs, 5GMB and the 5PB, which matches our predictions. The results are similar for $d=16$, the MUB and GMB produce the lowest infidelity estimates, followed by the 5MUB, 5GMB and then the 5PB. However, for $d=4$, the SIC curiously performs worse than the 5GMB and the 5PB despite the fact that it is full-IC. The PSI also performs much worse than expected. Therefore, there is some other difference between the POVMs, besides their informational completeness properties.  

\begin{figure}[]
\centering
\includegraphics[width = \linewidth]{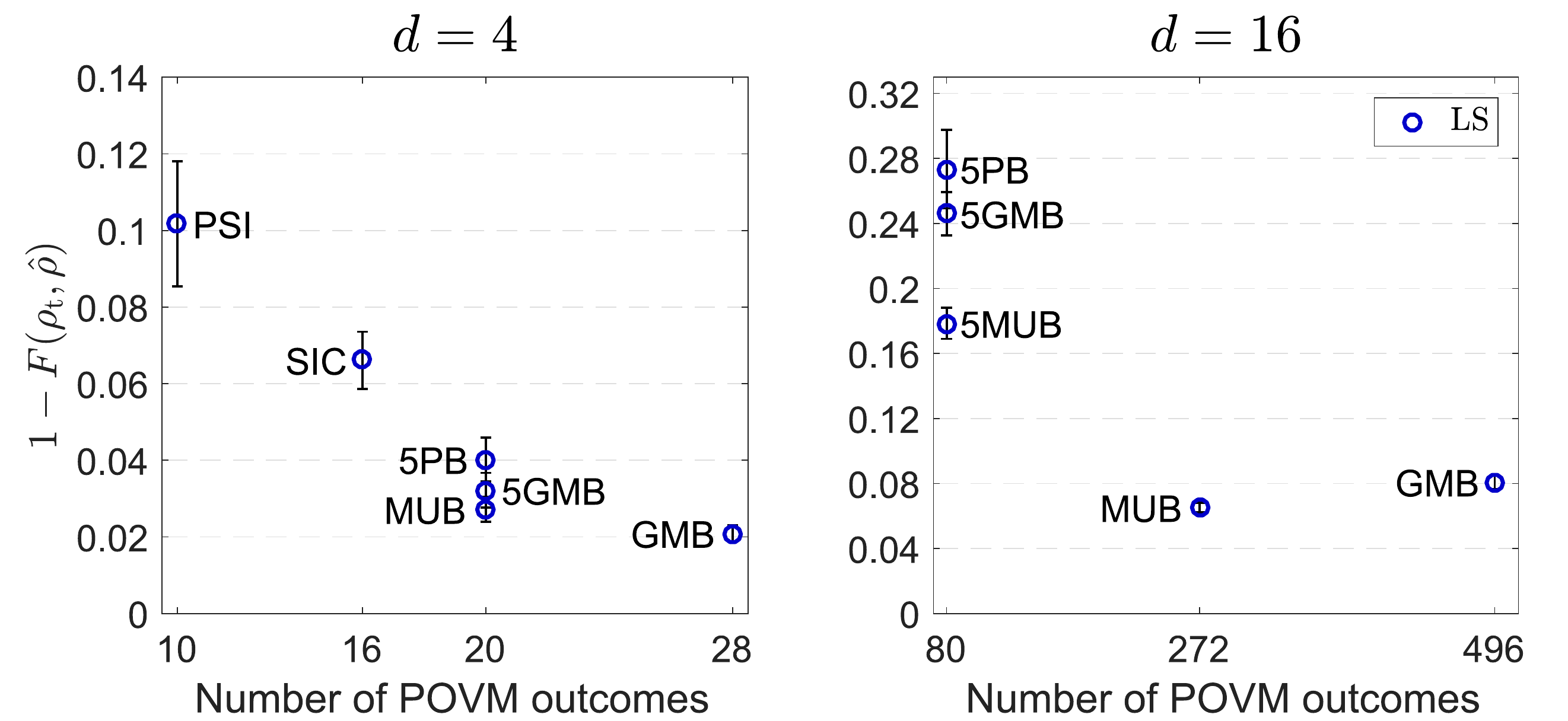}
\caption[Mean infidelity of LS estimate versus the number of POVM elements]{{\bf Mean infidelity of LS estimate versus the number of POVM elements} Each point represents the infidelity between the LS estimate and the target state averaged over the 20 different target states. The error bars show the standard error in the mean. The POVM used is labelled next to the point.}
\label{fig:infid_v_ele}
\end{figure}


In order to shed light on the differences between the POVMs, we re-plot the infidelity for the LS estimate from Fig.~\ref{fig:IC} as a function of the number of POVM outcomes in Fig.~\ref{fig:infid_v_ele}. We omit the rank-1 complete POVMs for the reasons discussed above. The results show a strong correlation between the infidelity of the LS estimate and the number of POVM elements. POVMs that contain more elements, like the GMB and MUB, produce the lowest infidelity estimates, while the ones with the less elements, like SIC and PSI, produce the highest infidelity estimates. This correlation arises because POVMs with more elements contain repeated information, which has the effect of reducing the noise and error level. Since the state is near-pure it is almost entirely described by the $2d-2$ free parameters that describe a pure state. The relation between these free parameters and the measured outcomes is nonlinear and very complex. However, POVMs with more than $2d-2$ elements provide repeated information that is averaged to reduce the noise and error level. It is important to keep in mind that in this experiment the dominant source of noise and errors is the control errors, which are essentially random but fixed for a given control field. Therefore, POVMs that contain more than $2d-2$ elements produced with multiple control fields should perform the best, since the effect of the control errors will be reduced by the redundant information. 

With this understanding, we can explain the results in Fig.~\ref{fig:infid_v_ele}. The PSI POVM contains the fewest elements ($2d$) and is implemented with a single control field. Therefore, it contains no averaging of the control errors, explaining the high infidelity. The SIC, which has redundancy in the $d^2$ elements, performs poorly since it is implemented with a single control field, and therefore has no averaging of the errors. The other POVMs contain at least $5d$ elements that are implemented with at least five control fields, so have some averaging of the errors. For $d=16$, the GMB perform worse than the MUB, which is contrary to this understanding. The reason could be related to the fact that GMB requires many more bases and the state preparation may have drifted slightly for each measurement.

The results in Fig.~\ref{fig:infid_v_ele} demonstrate that, in practice, not all rank-1 strictly-complete POVMs are equal. Some rank-1 strictly-complete POVMs produce lower fidelity estimates in the presence of noise and errors, such as the 5GMB or 5PB than others, such as the PSI. However, the lower fidelity comes at the price of more measurements. This demonstrates an important concept in the practical implementation of QST, which is the tradeoff between efficiency and robustness. For example, while the GMB is robust, i.e., produces a low infidelity estimate, it is not very efficient, i.e., it requires the most POVM elements. As shown in Fig.~\ref{fig:infid_v_ele}, after a certain point the gains in robustness are modest and must be weighed with the loss of efficiency. This effect becomes more pronounced as the dimension increases, as seen with $d=4$ versus the $d=16$ plots in Fig.~\ref{fig:infid_v_ele}. To accomplish effective QST, it is desirable to look for POVMs that have a satisfactory tradeoff between efficiency and robustness, such as the MUB for $d=16$.

\begin{figure}[]
\centering
\includegraphics[width = \linewidth]{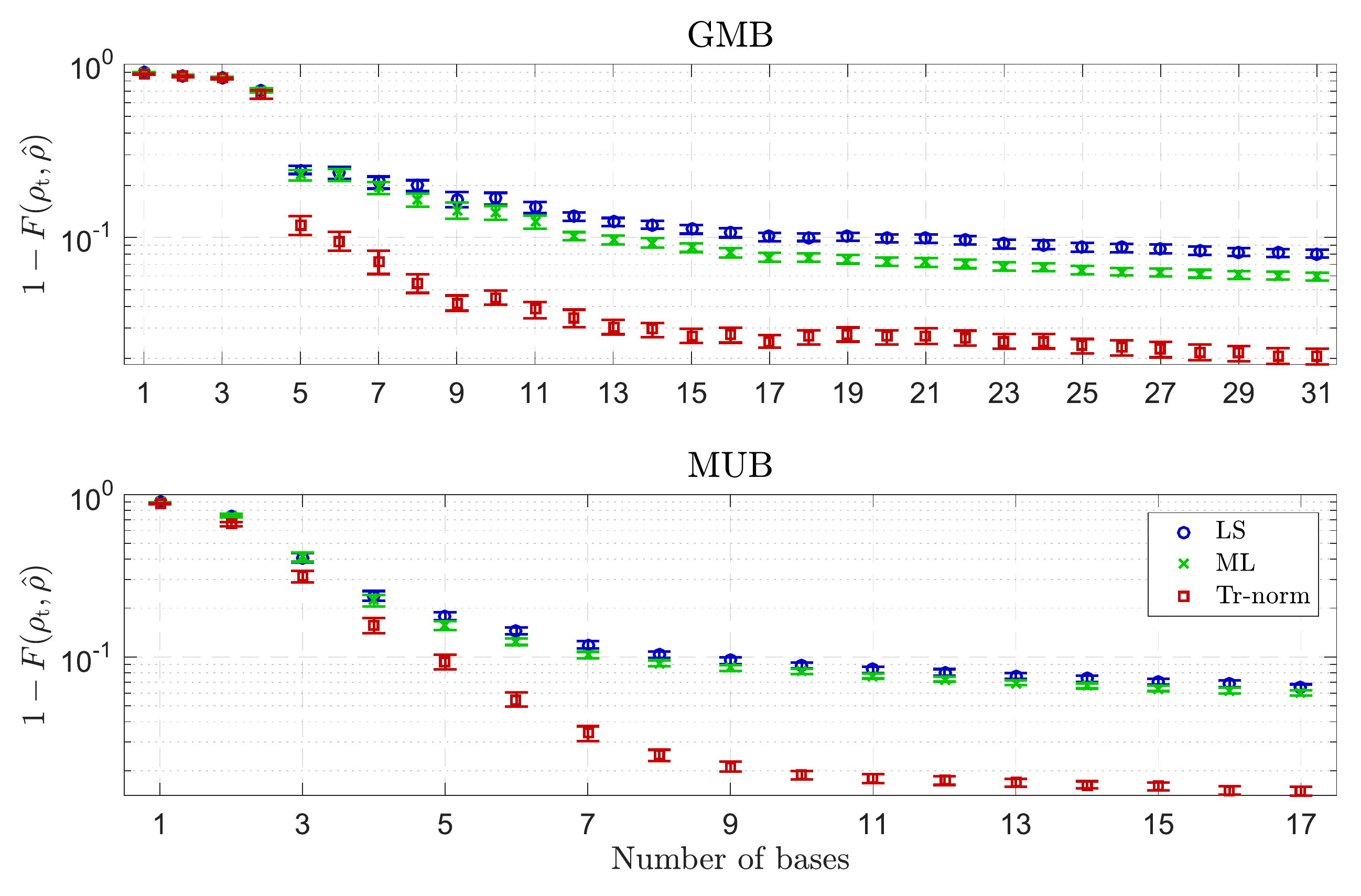}
\caption[Infidelity of estimate and target state as a function of number of bases]{{\bf Fidelity of estimate and target state as a function of number of bases} Estimates are created with (blue circles) LS, (green exes) ML, and (red squares) Tr-norm. The error bars correspond to the standard error of the mean. The POVM used to create the measurement vector is {\bf (top)} the GMB and {\bf (bottom)} the MUB}
\label{fig:fid_v_b}
\end{figure}

We can also see the tradeoff of efficiency and robustness in POVMs that consist of multiple basis measurements, such as the GMB and MUB. For this case, we study the infidelity of the estimate as a function of the number of orthonormal bases measured. After each basis that makes up the GMB and MUB, we compile the measurement vector for all previous bases and apply the two programs, LS and ML. For this comparison, we also estimate the state with the Tr-norm program, since this situation is similar to the quantum compressed-sensing for which Tr-norm was proposed. The value of $\varepsilon$ in the Tr-norm program was created by numerically modeling the types of errors expected in the control fields used to produce the unitary maps~\cite{SosaMartinez2016}. We calculate the infidelity between the estimate and the target state. The results are plotted in Fig.~\ref{fig:fid_v_b}.

As expected both POVMs and all estimators produce a low infidelity estimate with five bases, since for both, five bases form a rank-1 strictly-complete measurement and all three programs satisfy the form given in Corollary~\ref{cor:robustness}. This is only possible due to the positivity constraint on quantum states, and thus the existence of rank-1 strictly-complete measurements. By measuring more bases, we refine the estimate and the infidelity slowly decreases. This matches the comparison shown in Fig.~\ref{fig:infid_v_ele} that demonstrated the tradeoff between robustness and efficiency for different POVMs. After a certain number of measurements, the gain in robustness,  i.e. decrease in infidelity, is modest. 

We see that the Tr-norm estimator has slightly lower infidelity than LS and ML. This is due to the ``bias'' in the estimator towards pure states as was true with the rank-$r$-projection algorithm. While the Tr-norm is not as biased as the rank-$r$-projection algorithm (it does not constrain the state to be pure) it has a similar effect when we compare the estimate to the target state. Since the estimate is biased towards pure states it may be closer to the target state, which is also pure, than the actual state, which is likely full rank. Therefore, we conclude that this type of estimator is not desirable for QST, when we are looking to diagnose preparation errors.

Another factor that may impact the performance of the rank-1 strictly-complete and rank-1 complete POVMs is the failure set discussed in Chapters~\ref{ch:new_IC} and~\ref{ch:constructions}. This is the subset of measure zero within the set of quantum states where the probabilities from the POVM cannot uniquely identify the quantum state. The PSI, 5GMB and 4GMB suffer from such failure sets. Since the set has zero volume, we do not expect to randomly select states that are within such sets, and in fact none of the 20 Haar-random states are within any of the failure sets. However, Finkelstein~\cite{Finkelstein2004} showed that in the presence of noise and errors, the failure set in fact has a finite measure. Therefore, the failure set will have a non-negligible impact on estimation in QST. The failure set for each POVM is not the same, as shown in Chapter~\ref{ch:constructions} and Appendix~\ref{app:constructions}. Some POVMs have more complicated failure sets, which translate to smaller finite volume sets in the presence of noise and errors.  For example, we show in Appendix~\ref{app:constructions} that the 5GMB have a complicated failure set, while the PSI has a very simple set. Therefore, we expect that the PSI would suffer from such a failure set more than the 5GMB.  However, there is no clear indication of this effect in the experimental results. While the PSI, does perform worse than all other POVMs, this could also be explained by the fact that the PSI has the least number of POVM elements and is implemented with a single control field. Moreover, while the 5GMB suffers from the failure set, it still produces a lower infidelity estimate than the 5PB, which has the same number of elements but no failure set. Therefore, while the failure set may be effecting the estimation, we do not see any clear indication, and therefore do not believe it to be a practical limitation of such POVMs.

\section{Comparison of POVMs with Hilbert-Schmidt distance} \label{sec:HS_analysis}
We saw in the previous section that the number of POVM elements correlates with the success of the POVM. In this section, we formalize this relation by studying the structure of POVMs. The results are presented in terms of Hilbert-Schmidt (HS) distance squared, 
\begin{equation} \label{Delta2}
\Delta^2(\hat{\rho}, \rho_{\rm t} ) = \|  \hat{\rho}  - \rho_{\textrm{t}}\|_2^2,
\end{equation}
which we define in terms of the target state, $\rho_{\rm t}$, since we do not know the actual prepared state. For comparison, the HS-distance squared ranges in values from $\Delta^2(\hat{\rho}, \rho_{\rm t} ) = 0$, when the states are identical, to $\Delta^2(\hat{\rho}, \rho_{\rm t} ) = 2$, which occurs with two orthonormal pure states. The HS-distance offers the advantage that it is more straightforward to study analytically. We have used the HS-distance frequently in previous chapters. For example, in Chapter~\ref{ch:background} and~\ref{ch:new_IC}, we derived the robustness bound for full-IC and rank-$r$ strictly-complete POVMs based on HS-distance. Moreover, Scott~\cite{Scott2006} showed that the estimate returned from certain POVMs, referred to as ``tight,'' minimize the expected HS-distance over all realizations of the experiment when the measurement is only limited by finite sampling. These POVMs are, therefore, optimal for this particular situation. Two common examples of tight POVMs were implemented in the experiment, the SIC and the MUB. We reassess the experimental results with respect to the HS-distance and compare the results to theoretical predictions.

\subsection{Comparison of full-IC POVMs} \label{ssec:HS_full-IC}
We start with the full-IC POVMs and only consider the linear-inversion estimate, $\hat{\rho} = \hat{R}$. While this estimate is not necessarily a quantum state, and thus not appropriate for many applications, it is a useful mathematical tool for comparing POVMs. This was the approach taken by Scott~\cite{Scott2006}, who showed that tight POVMs, such as the SIC and MUB, produce the lowest, average HS-distance squared when there is a fixed number of copies and the experiment is only limited by the resulting finite sampling noise. The experimental results with the HS-distance for the full-IC POVMs and the linear-inversion estimate are given in Table~\ref{tbl:full_IC}.
\begin{table}[h]
\centering
\def\arraystretch{2}
\begin{tabular}{lll}
\cline{2-3}
\multicolumn{1}{l|}{}      & \multicolumn{1}{c|}{$d=4$}  & \multicolumn{1}{c|}{$d=16$}  \\ \hline
\multicolumn{1}{|l|}{SIC}  & \multicolumn{1}{c|}{0.0466 (0.0048)} & \multicolumn{1}{c|}{-}         \\ \hline
\multicolumn{1}{|l|}{MUB} & \multicolumn{1}{c|}{0.0111 (0.0011)} & \multicolumn{1}{c|}{0.0586 (0.0035)}  \\ \hline
\multicolumn{1}{|l|}{GMB} & \multicolumn{1}{c|}{0.0067 (0.0008)} & \multicolumn{1}{c|}{0.0710 (0.0020)}  \\ \hline
\end{tabular}
\caption[Experimental value of HS-distance squared for full-IC POVMs]{{\bf Experimental value of HS-distance squared for full-IC POVMs} Each cell gives the HS-distances squared between the linear-inversion estimate and the target state averaged over all 20 Haar-random pure states, with standard error of the mean given in parentheses.}
\label{tbl:full_IC}
\end{table}  
From the table, we see that for $d = 4$ the GMB produce the minimum average HS-distance squared, followed by the MUB, and then the SIC POVM. This matches the infidelity results in the previous section. For $d=16$, the MUB produces the lower value of the HS-distance than the GMB, which is the same as when we compared infidelity.  Therefore, the HS-distance results match the same trends we saw with infidelity.

While the HS-distances follows a similar trend as infidelity, they do not match the result by Scott~\cite{Scott2006} for two reasons. First, we know that the experiment is not limited by finite sampling since repeating the Stern-Gerlach analyzer with the same control parameters produces a nearly identical measurement vector. Instead, the measurements are actually limited by errors in the control fields. Second, the SIC POVM is implemented with a single run of the Stern-Gerlach analyzer while the MUB and GMB are done with many runs. Therefore, if the experiment was limited by finite sampling, it would be as if the SIC used $m$ copies, the MUB used $(d+1)m$ copies, and the GMB used $(2d-1)m$ copies. While we know the finite sampling noise is negligible, the difference in the number of applications of the Stern-Gerlach analyzer still has an impact on the accuracy of the POVM. 

To gain better insight into the experimental results, we construct a general framework for predicting the HS-distance in the presence of arbitrary noise and errors. This framework is an extension of the work by Scott~\cite{Scott2006} and will  allow us to compare arbitrary full-IC POVMs in the presence of any type of noise or error. In any experiment, random noise may determine the exact value of the HS-distance squared. So instead of studying the HS-distance squared, we focus on the expected HS-distance. This is defined by,
\begin{equation} \label{exp_Delta2}
\bar{\Delta}^2(\hat{R}, \rho_{\rm t}) = \mathbb{E} \left[ \| \hat{R}  -  \rho_{\textrm{t}} \|_2^2 \right],
\end{equation}
where the expectation value is over all realizations of the experiment. We wish to relate the value of $\bar{\Delta}^2(\hat{R}, \rho_{\rm t})$ to the POVM in order to compare different POVMs. The linear-inversion estimate provides a method to make such a relation, which motivates the choice of linear-inversion in this section. It can be expressed in terms of the reconstruction operators, $\{ Q_{\mu} \}$, which can be derived from $\Xi^+$, discussed in Sec.~\ref{sec:numerical_methods}, 
\begin{equation} \label{LI_expand}
\hat{R} = \sum_{\mu} f_{\mu} Q_{\mu}.
\end{equation}
For the SIC and MUB the reconstruction operators, $\{ Q_{\mu} \}$, have a ``painless'' form and are only proportional to the POVM elements~\cite{Scott2006}. We can also express the target state in terms of these reconstruction operators, $\rho_{\rm t} = \sum_{\mu} p_{\mu} Q_{\mu}$, where $\{ p_{\mu} \}$ are the probability of each outcome given the target state.  If we substitute Eq.~\eqref{LI_expand} into Eq.~\eqref{exp_Delta2} then,
\begin{align} \label{Delta_exp}
\bar{\Delta}^2(\hat{R}, \rho_{\rm t}) &= \mathbb{E} \left[ \textrm{Tr} \left( \sum_{\mu, \nu} \left( f_{\mu} Q_{\mu} - p_{\mu} Q_{\mu} \right)^{\dagger} \left( f_{\nu} Q_{\nu} - p_{\nu} Q_{\nu} \right) \right) \right], \nonumber \\
		&= \sum_{\mu, \nu} \mathbb{E} \left[ (f_{\mu} - p_{\mu} )( f_{\nu} - p_{\nu} ) \right] \textrm{Tr}[ Q_{\mu} Q_{\nu} ].
\end{align}
We define $G_{\mu,\nu} = \textrm{Tr}[Q_{\mu} Q_{\nu}]$ as elements of the ``Gramian matrix,'' $G$. Similarly, we define $Y_{\mu, \nu}(\rho_{\rm t}) = \mathbb{E} \left[ (f_{\mu} - p_{\mu} )( f_{\nu} - p_{\nu} ) \right]$ as elements of another matrix, called the ``noise/error matrix,'' $Y(\rho_{\rm t})$. We call this the noise/error matrix because the noise and errors perturb the measurement vector from the probabilities. The noise/error matrix is a function of $\rho_{\rm t}$, since the noise and errors may be state dependent, as is the case with finite sampling noise. This leads to the following compact equation,
\begin{equation} \label{Delta2}
\bar{\Delta}^2(\rho_{\rm t}) = \textrm{Tr} \left[ Y(\rho_{\rm t} )G \right],
\end{equation}
where we have dropped the dependence on $\hat{R}$ since it is contained within $Y(\rho_{\rm t})$. We have thus cleanly related $\bar{\Delta}^2(\rho_{\rm t})$ to two matrices, one that is only dependent on the POVM, $G$, and one that is only dependent on the noise/errors present, $Y$. In general, POVMs that have small values of $G$ will produce smaller $\bar{\Delta}^2(\hat{R}, \rho_{\rm t})$. This matches with Table~\ref{tbl:full_IC}, since the $G$ matrix from the GMB has the smallest elements, followed by the MUB, and then the SIC. However, this does not explain why, for $d =16$, the MUB produces a lower value of the HS-distance squared than the GMB. The reason is likely related to the fact that $\bar{\Delta}^2(\hat{R},\rho_{\rm t})$ is also dependent on the type of noise and errors present, which is contained in $Y(\rho_{\rm t})$. Therefore, in order truly compare POVMs, we also need to know the form of the noise and errors. 

In the cesium spin system, the dominate source of error is in the imperfections in the implementation of unitary maps that produce each basis measurement, which defines each POVM. These errors are only dependent on the control field, and therefore independent of the state that is measured, so $Y(\rho_{\rm t}) = Y$ for all $\rho_{\rm t}$. Moreover, for a given control field, the errors are constant, i.e. systematic errors. This was verified experimentally by repeating the measurement with the same control field and determining that the measurement vector is constant between repetitions. Then, $Y$ is a constant for all realizations of the experiment with the given control field. The value of $Y_{\mu,\nu}$ is proportional to errors in the unitary map, which not a straightforward relation. Therefore, without further study the control errors, we cannot exactly determine the form of $Y$. 

While we do not know the exact form of $Y$ for the cesium spin system, the University of Arizona group performed an additional experiment that gives insight into the magnitude of the errors present. This experiment was based on repeating the SIC POVM but each repetition was done with a different control fields. Due to the nature of numerical control optimization, there exist infinitely other control fields that implement the same unitary. Different control fields may have different errors associated with them. Therefore, if we repeat the same POVM but implement it with different control fields, we effectively randomize over some control errors. There may be some errors that cannot be randomized over, such as decoherence. These errors then remain systematic errors. Based on the behavior of $\Delta^2(\hat{R},\rho_{\rm t})$, we can determine the magnitude of the random errors compared to the systematic errors.

To accomplish this, we build two theoretical models of the behavior of $\bar{\Delta}^2(\hat{R},\rho_{\rm t})$ and compare them to the experimental results. Since we assume that $Y$ is independent of the measured state, we denote $\bar{\Delta}^2 = \bar{\Delta}^2(\hat{R},\rho_{\rm t})$. The two models we consider are two different forms of the noise/error matrix. The first is that the control errors are totally random. Then, the value of $f_{\mu} - p_{\mu}= e_{\mu}$ is a random variable with zero mean, and $Y_{\mu,\nu} = \mathbb{E}[e_{\mu} e_{\nu}]$ is the covariance matrix. In this case, we label $Y=C$, for covariance. With random errors, repeating the SIC POVM $n$ times will decrease $\bar{\Delta}^2$ by a factor of $1/n$ since covariance matrices add, and we average over the $n$ repetitions. Therefore, $\bar{\Delta}^2$ is a function of the number of repetitions,
\begin{equation} \label{Delta2_rand}
\bar{\Delta}^2(n) = \frac{1}{n} \textrm{Tr} \left[ C G \right] = \frac{x_{\textrm{rand}}}{n},
\end{equation}
where $x_{\textrm{rand}} \triangleq \textrm{Tr} \left[ C G \right]$. 

The second model we consider is when there exists both random and systematic errors in the control fields. In this case, $f_{\mu} - p_{\mu} = e_{\mu} + k_{\mu}$, where $e_{\mu}$ represents the random errors and $k_{\mu}$ represents the systematic errors, and we have assumed that both error sources are uncorrelated. The systematic errors, $k_{\mu}$, are constant for all realizations of the experiment, such that $\mathbb{E}[f_{\mu} - p_{\mu}] = k_{\mu}$. The elements of $Y_{\mu,\nu}$ are then,
\begin{equation}
Y_{\mu,\nu} = \mathbb{E}[ (f_{\mu} - p_{\mu})(f_{\nu} - p_{\nu}) ] = \mathbb{E}[(e_{\mu} + k_{\mu})(e_{\nu} + k_{\nu})] = \mathbb{E}[e_{\mu}e_{\nu}] +  k_{\mu} k_{\nu},
\end{equation}
where the middle terms are zero, since $\mathbb{E}[e_{\mu}] = 0$ and we previously assumed the errors are uncorrelated. The first term is the covariance of the random errors, so we again label the elements as $C_{\mu,\nu}$. The second term is constant for all realization due to the definition of the systematic errors and we label the elements as $K_{\mu,\nu} = k_{\mu} k_{\nu}$. We now plug this expression into Eq.~\eqref{Delta2},
\begin{equation} \label{Delta2_both}
\bar{\Delta}^2(n) =  \frac{1}{n} \textrm{Tr} \left[ C G \right] + \textrm{Tr} \left[ K G \right] =  \frac{x_{\textrm{rand}}}{n} +  x_{\textrm{sys}}.
\end{equation}
where $x_{\textrm{sys}} \triangleq \textrm{Tr} \left[ K G \right]$.

\begin{figure}[t]
\centering
\includegraphics[width=0.95\linewidth]{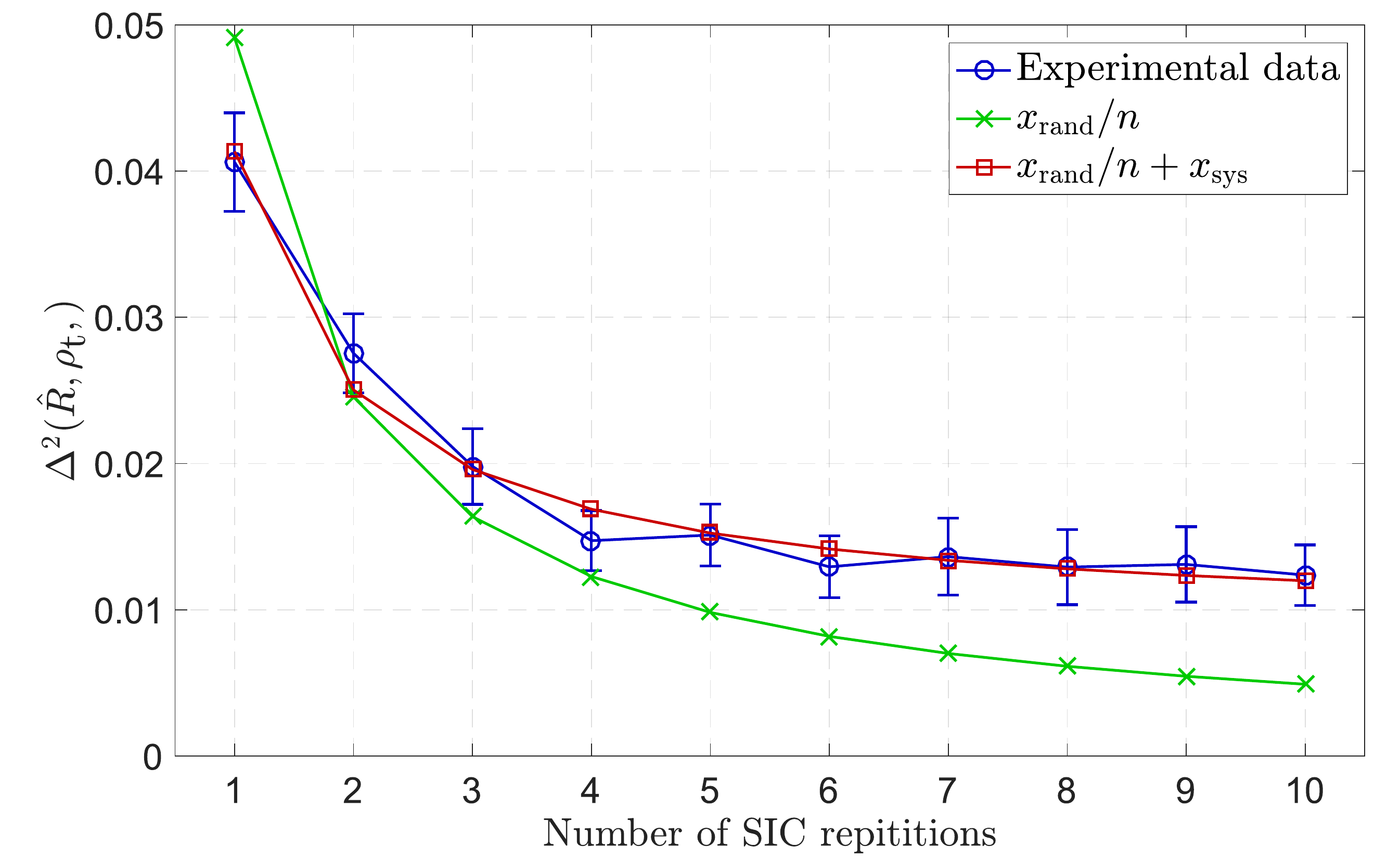}
\caption[Repetitions of SIC POVM with different control fields]{{\bf Repetitions of SIC POVM with different control fields} Experiment data and theoretical predictions for $\Delta^2$ between the target state and estimate from linear inversion as a function of the number of repetitions of the SIC POVM. (Blue) experimental data averaged over 10 Haar-random pure states, with point corresponding to the mean of all $\Delta^2$ values and error bars corresponding to standard error in the mean. (Green) Theoretical model for the behavior when there only exists random errors, fit to the experimental data. (Red) Theoretical model for the behavior when there exists both random and systematic errors, fit to the experimental data based. }
\label{fig:SIC_rep}
\end{figure}

In the experiment, the SIC POVM was implemented with 10 different control fields and each was applied to 10 Haar-random pure states. Since $\bar{\Delta}^2(n)$ is the same for any target state, we can approximate the value of $\bar{\Delta}^2(n)$ by averaging over the 10 experimentally measured values of $\Delta^2(\hat{R},\rho_{\rm t})$ for each repetition. We denote the average as $\Delta^2(n)$ for $n$ repetitions of the SIC POVM. In Fig.~\ref{fig:SIC_rep}, we plot $\Delta^2(n)$ as a function of the number of repetitions of the SIC POVM (blue). Fig.~\ref{fig:SIC_rep} also contains two fits to the Eq.~\eqref{Delta2_rand} (green) and Eq.~\eqref{Delta2_both} (red), created with MATLAB's fit function. For the fit to Eq.~\eqref{Delta2_rand}, we find,
\begin{equation}
x_{\textrm{rand}} =  \textrm{Tr} \left[ C G \right] = 0.0491 \, (0.03772, 0.0605),
\end{equation}
with $r^2 = 0.5283$ (parentheses contain 95\% confidence interval). For the fit to Eq.~\eqref{Delta2_both}, we find,
\begin{align} \label{error_mag}
x_{\textrm{rand}} = \textrm{Tr}[C G] = 0.0326 \, (0.0290, 0.0363), \nonumber \\
x_{\textrm{sys}}  = \textrm{Tr}[K G] = 0.0087  \, (0.0073, 0.0101),
\end{align}
with $r^2 = 0.9816$. From the figure we see that averaging over the control errors does decrease the HS-distance; however, the experimental data more closely matches the model that contains both random and systematic errors. In this model, the random term dominates but the systematic term has a significant contribution. Therefore, we can conclude that a majority of the control errors are due to effectively random sources associated with each control fields. This means, in principle, that the estimation from any of the POVMs can be improved by repeating the POVM with a different control fields and averaging the results. 

An important contribution to the systematic error term in the analysis above is due to preparation errors, and therefore not really a systematic error {\em per se}. The fidelity of state preparation was measured to be $\mathcal{F} = 0.995$~\cite{Smith2012}, which corresponds to $\| \rho_{\textrm{t}} - \rho_{\textrm{a}} \|_2 = 0.01 - (1- \textrm{Tr}[\rho_{\textrm{p}}^2]) \leq 0.01$. We believe $\rho_{\textrm{a}}$ is highly pure ($\textrm{Tr}[\rho_{\textrm{a}}^2] \approx 1$), but even with a small amount of impurity, $(1- \textrm{Tr}[\rho_{\textrm{p}}^2])$ the preparation error has a non-negligible contribution to the value of $x_{\rm sys}$. Therefore, the estimate found may contain information about the preparation errors that can be used to develop better state preparation procedures.

\subsection{Comparison of rank-1 strictly-complete POVMs}
In Chapter~\ref{ch:new_IC}, we proved that the estimate produced from the measurement vector of rank-1 strictly-complete POVMs are robust to noise and errors. The robustness bound was given in terms of the HS-distance between the target state and an estimate returned by a convex program in the form given in Corollary~\ref{cor:robustness}. We calculate the HS-distance squared with the experimental results to see if they are consistent with the robustness bound.  We only compare the HS-distance with the estimate from the LS program for simplicity. The experimental results are given in Table~\ref{tbl:strict}. We see from the table that all POVMs in both dimensions produce a low average HS-distance squared. For $d=4$, the 5GMB produce the smallest value of the HS-distance squared, followed by the 5PB, and then the PSI. These trends match the infidelity results shown in Fig.~\ref{fig:IC}. For $d=16$, the 5MUB produce the smallest value, followed by the 5PB, and then the 5GMB. This is counter to the infidelity results, which showed the 5GMB perform better than the 5PB. However, in both cases the values for the 5GMB and 5PB are very similar, and the standard error in the mean overlap. Therefore, the experimental results are consistent with the robustness bound, and the infidelity and HS-distance results roughly agree.

\begin{table}[t]
\centering
\def\arraystretch{2}
\begin{tabular}{lll}
\cline{2-3}
\multicolumn{1}{l|}{}      & \multicolumn{1}{c|}{$d=4$}  & \multicolumn{1}{c|}{$d=16$}  \\ \hline
\multicolumn{1}{|l|}{PSI} & \multicolumn{1}{c|}{0.0710 (0.0160)} & \multicolumn{1}{c|}{-}  \\ \hline
\multicolumn{1}{|l|}{5GMB}  & \multicolumn{1}{c|}{0.0092 (0.0013)} & \multicolumn{1}{c|}{0.2119 (0.0180)}         \\ \hline
\multicolumn{1}{|l|}{5PB} & \multicolumn{1}{c|}{0.0124 (0.0017)} & \multicolumn{1}{c|}{0.1991 (0.0256)}  \\ \hline
\multicolumn{1}{|l|}{5MUB} & \multicolumn{1}{c|}{-} & \multicolumn{1}{c|}{0.0905 (0.0065)}  \\ \hline
\end{tabular}
\caption[Experimental values of HS-distance squared for rank-1 strictly-complete POVMs]{{\bf Experimental values of HS-distance squared for rank-1 strictly-complete POVMs} The result is an average over all twenty random pure states with standard error of the mean given in parentheses.}
\label{tbl:strict}
\end{table}  

\begin{figure}[t]
\centering
\includegraphics[width=\linewidth]{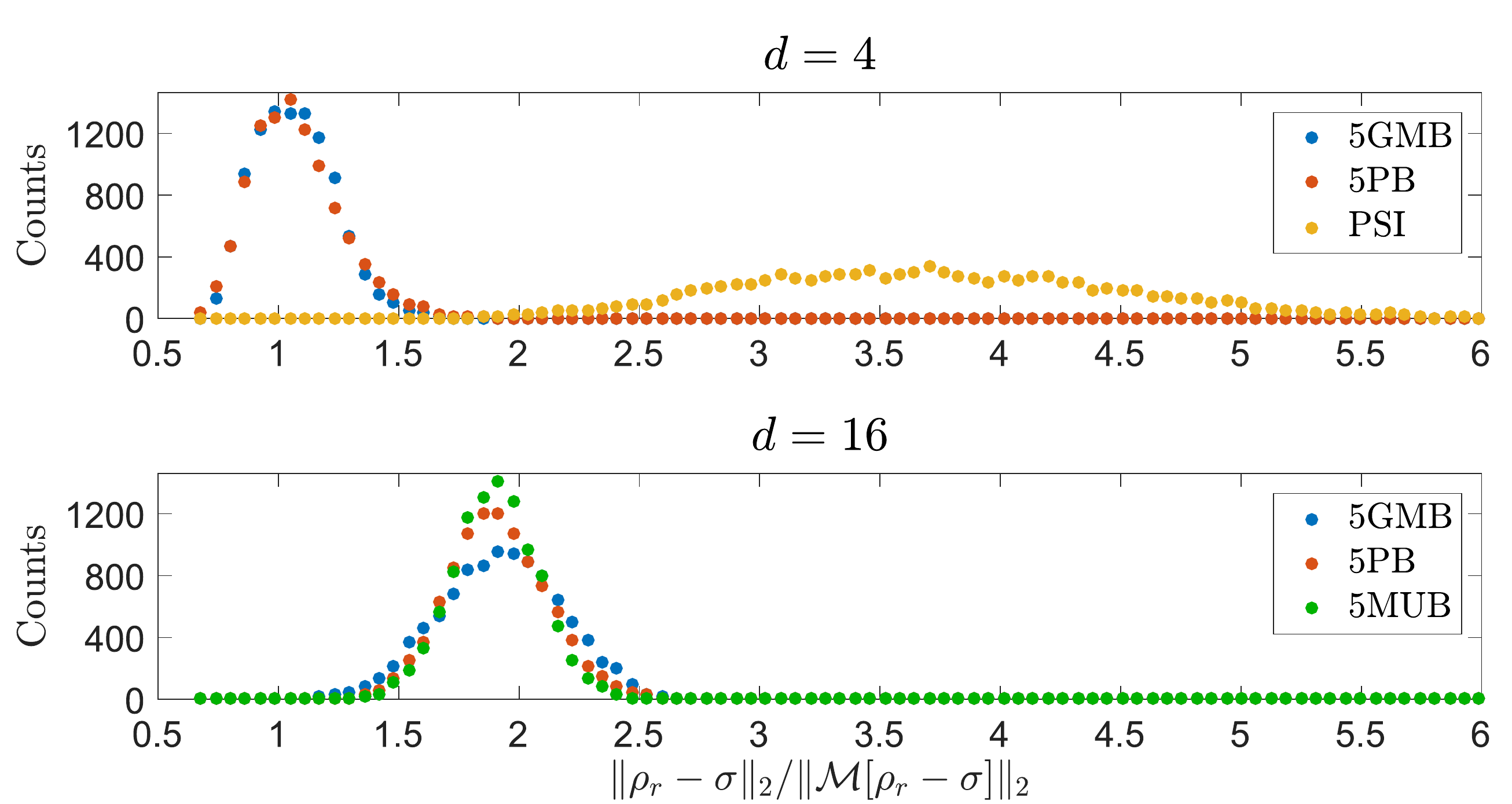}
\caption[Numerical simulation of robustness parameters for rank-1 strictly-complete POVMs]{{\bf Numerical simulation of robustness parameters rank-1 strictly-complete POVMs} We generate $10^4$ pairs of with one Haar random pure states and one random mixed state with the HS-measure. We then calculate the ratio between the HS-distance and the $\ell_2$ norm of the measurement record, given in Eq.~\eqref{alpha_ratio}. We repeat for the 5GMB, 5PB, PSI-complete ($d=4$ only), and 5MUB ($d=16$ only) for {\bf (top)} $d=4$ and {\bf (bottom)} $d = 16$. The results are binned and the number of occurrences of each bin is plotted.}
\label{fig:strict}
\end{figure}

As with the full-IC POVMs, we would like a method to compare different POVMs based on their structure. The approach taken in the previous section, based on theoretically predicting $\bar{\Delta}^2$, cannot be extended to rank-1 strictly-complete POVMs since the linear-inversion estimate is not unique. Instead, we compare different rank-1 strictly-complete POVMs by the robustness constants, $\alpha$, used in the derivation of Corollary~\ref{cor:robustness},
\begin{equation} \label{alpha_ratio}
\frac{\| \rho_1 - \rho_2 \|}{\|\mathcal{M}[ \rho_1 - \rho_2] \| }  \leq \frac{1}{\alpha}.
\end{equation}
where the constant $\alpha$ contributes to the robustness bound,
\begin{equation} \label{robustness_bound}
\| \rho_{\rm a} - \hat{\rho} \|  \leq C_1 \varepsilon + 2C_2 \upsilon
\end{equation}
where $C_1 = \frac{2}{\alpha}$ and $C_2 = \frac{\beta}{\alpha}$. There is no known analytic form for this constant, but in Sec.~\ref{sec:strict_comp_nums}, we presented a numerical method for estimation. To accomplish this, we generate $10^4$ pairs of density matrices, one that is rank-1 and one that is full-rank, chosen by the method described in Sec.~\ref{sec:strict_comp_nums}. We then calculate the ratio of the HS-distance between the two density matrices, to the $\ell_2$-distance between the probability vectors of a strictly-complete POVM, that is the LHS term in Eq.~\eqref{alpha_ratio}. We then bin the number of times each ratio is determined numerically. The results are shown in Fig.~\ref{fig:strict} for both $d = 4$ and~16 and follow the same trend outlined in Sec.~\ref{sec:strict_comp_nums}, where the distributions are centered around a peak.  It should be noted that since each pair of states is randomly generated, the numerical test is very unlikely to sample a state from the failure set, which has zero volume. Therefore, this numerical test is independent of this failure set and the results offer a way to compare POVMs separate from this effect.

For $d = 4$, the 5GMB and 5PB produce approximately the same distribution. The distribution for the PSI is, however, significantly shifted to larger ratios and wider. This means that the robustness constant, $1/\alpha$, is much larger. The bound in Eq.~\eqref{robustness_bound} is then much larger for the PSI POVM. Therefore, we expect that the HS-distance between the estimated state and the actual state for PSI is much larger than the same measure for the other POVMs. This matches with the experimental data shown in Table~\ref{tbl:strict}. Numerically, we find that the range for the ratio for the 5GMB and 5PB is reasonable, $0.6117- 2.3769$, such that the robustness constants for both are not too large and the bound in Eq.~\eqref{robustness_bound} is small. For $d =16$, the three distributions are roughly centered on the same value and have a similar range of the ratio, $0.9277 - 2.7747$. However, the width of each distribution is different, where the 5MUB is the narrowest, followed by the 5PB, and the 5GMB. We expect that a narrower distribution will have smaller values for $1/\alpha$. Therefore, narrow distributions correspond to measurements that produce a smaller bound in Eq.~\eqref{robustness_bound}, and thus better estimation. This matches the experimental values of $\Delta^2$ in Table~\ref{tbl:strict}, which show the 5MUB produce the smallest value, followed by the 5PB, and then the 5GMB. The difference between the 5PB and 5GMB is small, which is reflected in the similar distribution in Fig.~\ref{fig:strict}. 

The results of the numerical test match both the experimental results for HS-distance and the intuition established in Sec.~\ref{sec:ST_exp} about the tradeoff of efficiency and robustness. The reason that the PSI performs so badly is likely related to the fact that it has much fewer elements. However, since we do not have an analytic expression for the robustness constants it is not currently possible to formalize this relation. The numerical test also provides further evidence that the failure set is not what is limits the PSI or effects the 5GMB since the results are independent of the failure set. Therefore, we conclude that the failure set is not a practical limitation for strictly-complete POVMs.

\section{Process tomography} \label{sec:PT_exp}
Sosa-Martinez {\em et al.} also experimentally tested the efficient methods for QPT that were outlined in Chapter~\ref{ch:PT}. In the experiment, the target quantum process was a unitary map. However, due to control errors, or other sources outlined in Sec.~\ref{ssec:errors_expt}, the applied process is not exactly unitary. From previous tests, such as the randomized benchmarking inspired protocol~\cite{Anderson2015}, we know that the magnitude of these errors is small. Therefore, we have strong evidence that the applied process is near-unitary and the methods for QPT with UIC sets of states should produce a robust estimate.

\begin{figure}[t]
\centering
\includegraphics[width=\linewidth]{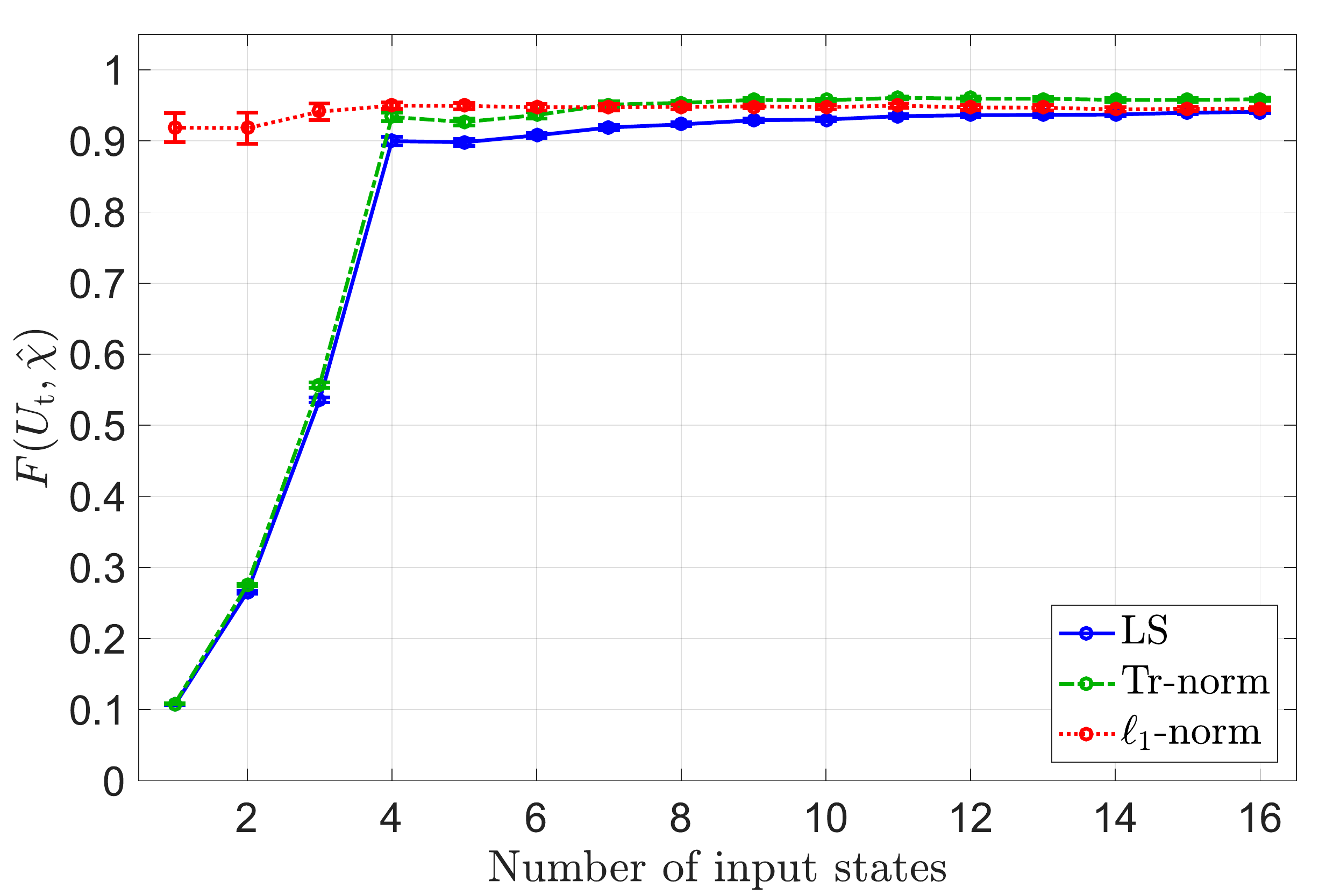}
\caption[Experimental results for QPT of a $4$-dimensional Hilbert space with efficient order of probing states]{{\bf Experimental results for QPT of a $4$-dimensional Hilbert space with efficient order of probing states} The quantum process was estimated after each ordered input state with (solid blue) the LS program given in Eq.~\eqref{PT_LS}, (dashed green) Tr-norm minimization given in Eq.~\eqref{PT_Tr}, and (dotted red) $\ell_1$-norm minimization given in Eq.~\eqref{PT_L1}. }
\label{fig:PT_d4}
\end{figure}

QPT was implemented for both the $d = 4$ subspace and the full $d=16$ Hilbert space. For $d=4$, Sosa-Martinez {\em et al.} generated 10 Haar-random unitary maps as the target processes. The actual near-unitary processes are probed with the $d$ UIC set of states given in Eq.~\eqref{0+n}, supplemented with a set of $d^2-d$ linearly independent states from Eq.~\eqref{op basis nc}. The output was then measured with the MUB. The measurement vector was analyzed with the three estimation programs, LS, Tr-norm and $\ell_1$-norm, outlined in Sec.~\ref{sec:PT_num}.  The estimated processes were then compared to the target processes to determine the process fidelity, given in Eq.~\eqref{PT_U_fidelity}, after each input state is measured. The results are plotted in Fig.~\ref{fig:PT_d4}.

In Fig.~\ref{fig:PT_d4}, we see that the estimation programs follow similar trends to what was discussed in Sec.~\ref{sec:QPT_werrors}. The LS and Tr-norm produce high fidelity estimates after $d = 4$ input states. This verifies that the applied process is in fact near-unitary. The $\ell_1$-norm program produces a high fidelity estimate for all input states since it has more prior information about the applied process. As outlined in Sec.~\ref{sec:QPT_werrors}, the fact that the $\ell_1$-norm estimate has near constant fidelity for all input states, and the Tr-norm program produces estimates with slightly higher fidelity, indicates that the applied process has incoherent errors. This is consistent with the types of errors seen in QST and discussed in Sec.~\ref{ssec:HS_full-IC}. Therefore, we conclude that (1) the UIC set of input states accomplishes efficient QPT in an experimental setting and (2) that the dominant error in the each unitary map implemented in the experiment is likely incoherent due to averaging the ensemble over random local Hamiltonians, e.g., random bias magnetic fields.

\begin{figure}[t]
\centering
\includegraphics[width=\linewidth]{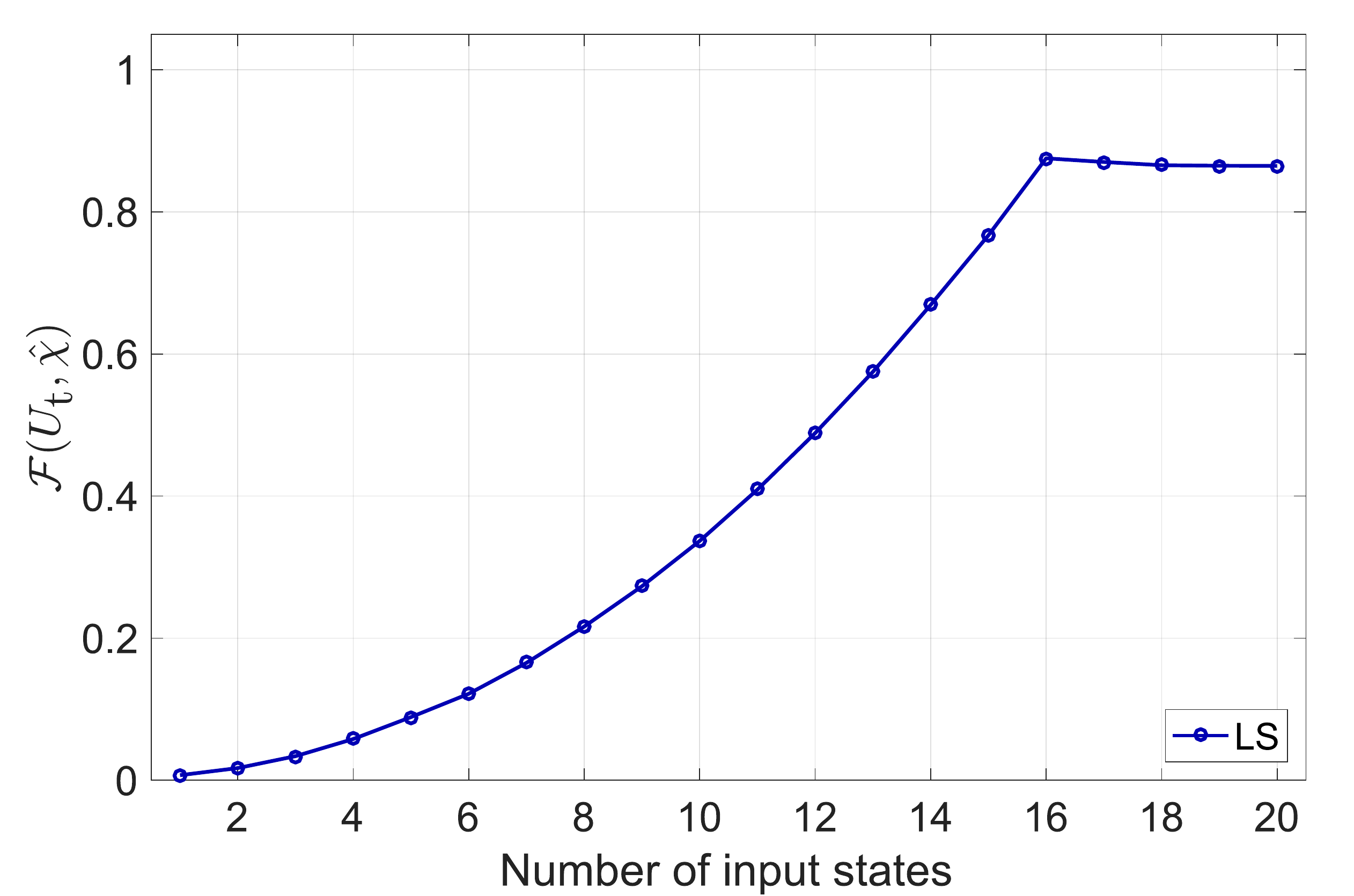}
\caption[Experimental results for QPT of a $16$-dimensional Hilbert space with efficient order of probing states]{{\bf Experimental results for QPT of a $16$-dimensional Hilbert space with efficient order of probing states} The quantum process was estimated after each ordered input state with the LS program given in Eq.~\eqref{PT_LS}, and implemented with a gradient-projection algorithm. }
\label{fig:PT_d16}
\end{figure}

Sosa-Martinez {\em et al.} also implemented QPT for the full $d=16$ Hilbert space. For $d=16$, it is not practically feasible to evolve $d^2 = 256$ input states, which are required for standard QPT. For one, it would take a very long time to perform an experiment with 256 states and effects, such as drift in the experimental settings, may contaminate the results. Also, the classical computation required to produce an estimate for such a large system is not possible with current convex optimization algorithms. Therefore, the efficient set of UIC states is mandatory for QPT of such a large system. In the experiment, the 16 states from Eq.~\eqref{0+n}, along with four extra states for comparison, were used to probe a single, near-unitary process. After measuring each state, we calculate an estimate with only the LS program. Instead of using the CVX package for MATLAB, as was done for $d= 4$, we applied a gradient-projection algorithm. This algorithm is different form the rank-$r$-projection algorithm discussed above. The gradient-projection algorithm used for $d=16$ QPT only projects to the set of CPTP quantum process, i.e. PSD matrices with proper TP constraint. Therefore, it does not have the same type of bias issues associated with the rank-$r$-projection algorithm for rank-1 complete POVMs and is an implementation of the standard LS program. We then compared the estimate from this algorithm to the target unitary with the process fidelity given in Eq~\eqref{PT_U_fidelity}. The results are plotted in Fig.~\ref{fig:PT_d16}.

For $d = 16$, we are still able to reconstruct a high-fidelity estimate of the quantum process with the UIC set of states given in Eq.~\eqref{0+n}. This is the largest Hilbert space that QPT has been implemented and is only made possible by using the UIC set. Given the large amount of data it is difficult to implement the Tr-norm and $\ell_1$-norm estimation programs in order to determine the type of errors present. However, the estimation provided by the LS program is still useful for diagnosing errors in the map by other methods, such as the ones discussed in Refs.~\cite{Korotkov2013,Kofman2009}.

\section{Summary and Conclusions}
The experiments performed by Sosa-Martinez {\em et al.} demonstrated many different methods for measurement and estimation in QT. We found that full-IC POVMs produce the lowest infidelity estimation of the quantum state with ML. However, rank-1 strictly-complete POVMs also produce low infidelity estimates even for larger systems. This demonstrates the tradeoff between efficiency and robustness in QST. While the full-IC POVMs produce the lowest infidelity estimate, they require many POVM elements. Conversely, we saw that while rank-1 strictly-complete POVMs theoretically offer both efficiency and robustness, some constructions, such as PSI, are not accurate enough for practical use. Therefore, it is important to choose POVMs for real implementations of QST that offer sufficient robustness and are still efficient.

The experiment also demonstrates that estimators that are required for rank-1 complete POVMs are biased. In order to reliably produce an estimate for these POVMs, we must use the rank-$r$-projection algorithm, which projects to pure states. This algorithm produces estimates that have much lower infidelity with the target state than expected. Therefore, these estimates cannot be trusted for QST. This is another drawback of rank-1 complete POVMs.

We also discussed methods of comparing the structure of POVMs for QST based on the HS-distance. For full-IC POVMs, we presented a mathematical framework that can predict how each POVM will perform when there exists knowledge about the noise and errors that effect the experiment. Currently, in the cesium spin experiment, we do not know the exact form the noise and errors, and therefore cannot apply this result. We were able to determine the magnitude of random control errors and systematic errors by studying the experimental results of the repeated SIC POVM. This test showed that random control errors dominate but systematic errors do have a non-negligible contribution. For the rank-1 strictly-complete and complete POVMs, we applied a numerical study in order to estimate the robustness parameters to understand how each POVM performs. We saw that this method matched with the experimental results for the rank-1 strictly-complete POVMs.


The QPT results by Sosa-Martinez {\em et al.} show that UIC sets of states produce efficient and robust estimates. Moreover, different estimations strategies for QPT were used to determine that incoherent errors likely dominate the processes. For the $d=16$ system, the UIC set serves as an example of the power of efficient QT techniques. QPT in this system would not be possible with standard techniques; however, with the UIC set we are able to produce high fidelity estimates of a near-unitary process.

\chapter{Conclusions and outlook} \label{ch:conclusions}
In this dissertation, we introduced new methods for quantum tomography (QT) that are more efficient to implement and robust to noise and errors. We showed that these methods are made possible by applying prior information about the quantum system that is consistent with the goals of most quantum information processing experiments. Specifically, for quantum state tomography (QST) the prior information is that the quantum state is close to pure and for quantum process tomography (QPT) it is the process is close to unitary. Pure states and unitary processes are required for most quantum information processing protocols, and therefore most experiments work to engineer states and processes near this regime. We showed that the new methods for QST and QPT produce robust estimates even if the states are not exactly pure and the processes are not exactly unitary. Therefore, these results offer a way to accomplish QT in larger dimensional Hilbert spaces than were previously possible with standard techniques. 

We began the dissertation by outlining the mathematical framework for standard QT in Chapter~\ref{ch:background}. Standard QT is defined by the notion of full informational completeness (full-IC). We reviewed how this notion applies to QST, QPT, and QDT in the ideal case where we have direct access to the probabilities. We showed that, in this case, QT is a linear algebra problem where the probabilities are linearly related to the free parameters that describe an arbitrary state, process, or POVM. However, in any real application of QT, there necessarily exist noise and errors, and therefore we do not have direct access to the probabilities. To study this case, we formalized the effect of noise and errors in QT. We also presented previously proposed numerical algorithms for estimating the quantum states, processes, and detectors in this situation. We showed that the standard methods are robust to such noise and errors. However, standard QT requires resources that scale polynomially with the dimension of the Hilbert space, and therefore are limited to small systems.

In order to accomplish QT more efficiently, we devised methods to incorporate prior information about the the quantum system into the measurements and estimation. We began by focusing on QST in Chapter~\ref{ch:new_IC}. We showed that there exists POVMs that fully characterize pure states with less elements than needed for standard QST in the ideal setting when we have direct access to the probabilities. We defined two types of these POVMs: rank-1 complete and rank-1 strictly complete. Rank-1 complete POVMs uniquely identify pure states from within the set of all pure states while rank-1 strictly-complete POVMs uniquely identify pure states from within the set of all quantum states. The notion of rank-1 strictly-complete POVMs is only made possible by the positivity constraint on quantum states, i.e., all density matrices are constrained to be positive semidefinite (PSD). 

The difference between rank-1 complete and rank-1 strictly-complete POVMs has significant consequences for QST in the presence of noise and errors. In this case, numerical optimization is required to produce an estimate of the measured quantum state. The two different types of POVMs demand different strategies for numerical optimization. Rank-1 complete POVMs necessitate algorithms that are restricted to the set all pure states. This is a nonconvex constraint and so is difficult to incorporate in numerical optimization. Rank-1 strictly-complete POVMs require optimization that is restricted to the set of quantum states, which is a convex set. Therefore, rank-1 strictly-complete POVMs are compatible with the well established methods for convex optimization while rank-1 complete POVMs are not. Moreover, we proved that for rank-$r$ strictly complete POVMs, the estimate returned by certain convex programs are robust to all sources of noise and errors. This includes preparation errors, which necessarily exist in any experiment and cause the actual state to be not exactly pure. This property makes rank-1 strictly-complete POVMs advantageous for pure-state QST.

We went on to discuss different methods to produce both rank-$r$ complete and strictly-complete POVMs in Chapter~\ref{ch:constructions}. We showed, that while rank-1 strictly-complete POVMs are inherently related to positivity, which is a difficult constraint to treat analytically, we can still construct POVMs that are provably rank-1, and more generally, rank-$r$ strictly-complete. We provided two methods for constructing strictly-complete POVMs. The first applies to a certain type of POVM, which we called element-probing (EP) POVMs. EP-POVMs allow for the direct reconstruction of density matrix elements. For these types of POVMs, we introduced tools based on the Schur complement and the Haynsworth matrix inertia to prove an EP-POVM is rank-$r$ complete or strictly-complete. These tools can also be used to construct new rank-$r$ strictly-complete POVMs, with two examples given in Appendix~\ref{app:constructions}. We also demonstrated numerically that a set of random orthonormal basis measurements form a rank-$r$ strictly-complete POVM. We applied these two methods to a simulation of QST to show that the quantum state could be efficiently and robustly estimated in the presence of sources of noise and errors. Therefore, we conclude that strictly-complete POVMs are the best choice for bounded-rank QST, due to their efficiency, robustness, and compatibility with convex optimization.

At the end of both Chapter~\ref{ch:new_IC} and Chapter~\ref{ch:constructions}, we identified how the ideas of rank-$r$ strictly-complete POVMs can be generalized to QDT and QPT.  This relation is made possible by the fact that a positivity constraint exists for both matrices that define QDT and QPT. For QDT, the POVM elements that we diagnose are constrained to be PSD matrices. For QPT, the condition that the process is completely positive (CP) is equivalent to the process matrix being PSD. Since the definition of rank-$r$ strictly-complete is only with respect to PSD matrices and not just quantum states, the same notion applies for both QDT and QPT. For QDT, the generalization is straightforward since the unknown POVM element is probed with a set of quantum states. We can translate many constructions for rank-$r$ strictly-complete POVMs in QST to a strictly-complete set of probing states for QDT, as was shown in Sec.~\ref{sec:constructions_QDT}.

It is not as straightforward to generalize the notion of strict-completeness to QPT, but in Chapter~\ref{ch:PT}, we presented such a generalization. Most quantum information protocols require unitary process, which is prior information that can be applied to QPT. Unitary processes are represented by rank-1 process matrices, so unitary QPT is analogous to pure-state QST. We defined sets of states that uniquely identify a random unitary process within the set of all unitary maps, called unitarily informationally complete (UIC) sets. We provided a few example constructions and also gave numerical evidence that these UIC sets also uniquely identify any random unitary process from within the set of all CPTP maps. In any real application of QPT, the process being measured is not exactly unitary. Therefore, we studied the problem of near-unitary QPT and considered two different types of error models that may disrupt the target unitary process. We showed that different estimators for QPT respond differently to these two types of error models, and therefore could be used to diagnose which types of errors are present.

QT is fundamentally an experimental protocol to characterize a quantum system, so any new method for QT should be tested experimentally. In Chapter~\ref{ch:experiment}, we discussed experimental tests on an ensemble of cesium atoms performed by Hector Sosa-Martinez and Nathan Lysne in the lab of Prof. Poul Jessen at the University of Arizona. Different rank-1 complete and strictly-complete POVMs were implemented, and various numerical estimation programs were compared. It was found that the POVMs with the most elements produced the lowest infidelity estimates for QT. However, some POVMs with less elements produce estimates with almost as low infidelity. The results illustrate the tradeoff between efficiency and robustness in QT. While POVMs with many elements are the most robust, and thus produce the best estimates, they require much more experimental effort. Rank-1 strictly-complete POVMs produce estimates with almost as low infidelity but are much more efficient. For QPT, the experiment demonstrates that the UIC set does produce a high fidelity estimate of the unknown unitary process that also indicates of the type of noise present. The set also allowed for the implementation of QPT for a $d = 16$ Hilbert space, which is infeasible with standard techniques. 

The experiment opens three avenues for future theoretical research in QT. First, while we have some theoretical and numerical methods to compare POVMs, the experimental results do not exactly match, as discussed in Sec.~\ref{sec:HS_analysis}. This may be due to sources of noise and errors that are unique to the experiment. Current theoretical and numerical methods for comparing POVMs do not take such differences into account. For example, in the original work by Scott~\cite{Scott2006}, it was proven that so-called ``tight'' POVMs, such as the SIC and MUB, are optimal for QST. However, this proof holds under the assumption that the experiment is only limited by finite sampling. This is not the case for the cesium spin experiment as well as most real applications of QT. It would be useful to derive methods to compare POVMs with arbitrary types of noise or errors.

Second, the experiment also confirmed that each estimator for QST and QPT perform differently, which has consequences on how we compare different methods. For example, the Tr-norm and the rank-$r$-projection algorithms produce infidelities much lower than LS and ML. However, the Tr-norm and rank-$r$-projection algorithms are biased towards pure states, and therefore may overestimate the performance of QT. This effect must be better understood in order to not make false claims of superior methods that are only due to the bias estimation. In general, it is important to study how all estimators performs in different error regimes to make sure that the estimation is reliable. If we are given prior information about the type of errors present, we may be able to choose the best suited estimator. 

Finally, the estimates produced in both QST and QPT from the experimental data exemplifies an outstanding question in QT research: what do we do with the estimates? In Chapter~\ref{ch:experiment}, we compared the infidelity to the targets states and process, but the density and process matrices in theory contain all information about the quantum states and processes. However, it is not straightforward to extract this information. Previous work has made some relations between density and process matrices to useful quantities. For example, it was shown that entanglement measures can only be calculated with full tomographic reconstructions~\cite{Carmeli2016a,Lu2016}. There have also been a proposal for QPT that relates certain elements of the process matrix to different sources of errors~\cite{Korotkov2013}. However, we lack a well defined framework for understanding both the density matrix and the process matrix. Future work, which may flush out important relations, would allow for the diagnosis of noise or error sources and make QT the useful experimental tool that is only now a promise. 

The outlook for QT as a whole is mixed. Originally, QT was only feasible for small systems (e.g. a couple of qubits). However, with the unifying techniques proposed in this dissertation, as well as related work in compressed sensing~\cite{Gross2010, Flammia2012}, QT is now possible for larger systems (e.g. 3-10 qubits). These systems are common in today's state-of-the-art experiments, so QT is currently a useful tool for experimentalists. However, with new technological advances, it is expected that soon still larger systems (e.g. $> 10$ qubits) will be more common. For these systems, even strictly-complete methods for QT, will not be feasible. This is due to the fact that most methods for QT, even the ones proposed here, scale exponentially with the number of qubits. It may be that other types of prior information, such as matrix product states~\cite{Cramer2010}, can be leveraged to make QT feasible in larger systems. In this case the notions of completeness and strict-completeness may have useful generalizations that allow for efficient and robust methods. However, it seems that QT's likely future is as one tool in the toolbox for diagnosing quantum systems. In order to build quantum information processors that demonstrate advantages over classical techniques, we will require many such tools and the fact that QT is now possible with larger systems makes it a tool of greater value.

\appendix

\chapter{Other rank-$r$ strictly-complete POVM constructions} \label{app:constructions}
In this appendix we present four rank-$r$ strictly-complete POVMs. The first three (GMB, 5PB, and PSI) were implemented in the experiment discussed in Chapter~\ref{ch:experiment}. The final POVM is a generalization of the POVM given in Eq.~\eqref{psi-complete} to be rank-$r$ strictly-complete. 

\section{Gell-Mann bases (4GMB, 5GMB, and GMB)}\label{app:GMB}
Goyeneche {\em et al.}~\cite{Goyeneche2015} proposed two sets of bases for pure-state QST, which we refer to as the 4GMB (consisting of four bases and given in Eq.~\eqref{4gmb}) and 5GMB (consisting of the 4GMB plus the computational basis). Goyeneche {\em et al.}~\cite{Goyeneche2015} proved that both these constructions are rank-1 complete by the decomposition method discussed in Sec.~\ref{sec:decomp_method}. In Sec.~\ref{sec:EP}, we showed that the 4GMB form an EP-POVM, and the same can be shown for the 5GMB. In Ref.~\cite{Goyeneche2015}, the 5GMB were proposed in order to avoid the failure set by adaptively constructing four of the bases based on the measured outcomes of the first basis. We do not consider such adaptive techniques here. Instead, we treat the 5GMB as fixed, and use the EP-POVM framework to prove the 5GMB are in fact rank-1 strictly-complete. We then show that this type of basis measurement can be extended to bounded-rank QST, and provide an algorithm to generate $4r+1$ bases that are provably rank-$r$ strictly-complete. When $r \geq d/2$, the algorithm constructs the full-IC POVM referred to as GMB, which was applied in the experiment and discussed in Chapter~\ref{ch:experiment}.

All of the constructions discussed in this section (4GMB, 5GMB, and GMB) are EP-POVMs that allow for the reconstruction of density matrix elements that make up the diagonals. For convenience, we label the upper-right diagonals $0$ to $d-1$, where the $0$th diagonal is the principal diagonal and the $(d-1)$st diagonal is the upper right element. Each diagonal, except the $0$th, has a corresponding Hermitian conjugate diagonal (its corresponding lower-left diagonal). Thus, if we measure the elements on a diagonal, we also measure the elements of its Hermitian conjugate. The computational basis corresponds to measuring the $0$th diagonal. 

We begin by considering the 5GMB construction. In Sec.~\ref{sec:EP}, we showed the 4GMB allows for reconstruction to of the elements on the first diagonals. The 5GMB additionally includes the computational basis measurement, which allows us to reconstruct of all elements on the 0th diagonal. To show that the 5GMB is rank-1 complete, we follow the general strategy outlined in Sec.~\ref{ssec:EP_comp}. First, choose the leading $3 \times 3$ principal submatrix,
\begin{equation}
M_0 = 
\begin{pmatrix}
\rho_{0,0} & \rho_{0,1} & \bm{\rho_{0,2}} \\
\rho_{1,0} & \rho_{1,1} & \rho_{1,2} \\
\bm{\rho_{2,0}} & \rho_{2,1} &\rho_{2,2}  \\
\end{pmatrix},
\end{equation}
where, hereafter, the elements in bold font are the unmeasured elements. By applying a unitary transformation, which switches the first two rows and columns, we can move $M_0$ into the block matrix form, 
\begin{equation}
M_0 \rightarrow UM_0U^\dagger=
\begin{pmatrix}
\rho_{1,1} &\rho_{1,0} & \rho_{1,2} \\
\rho_{0,1} & \rho_{0,0} & \bm{\rho_{0,2}} \\
 \rho_{2,1} & \bm{\rho_{2,0}} & \rho_{2,2}   \\
\end{pmatrix}.
\end{equation}
This matches the form in Eq.~\eqref{block_mat}, with $A = \rho_{1,1}$, $B^{\dagger} =  (\rho_{1,0}, \rho_{1,2})$ and $C$ is the bottom $2 \times 2$ submatrix. Form Eq.~\eqref{Schur_rank}, we can solve for $\rho_{0,2}$ and $\rho_{2,0}$, since $C = \rho_{1,1}^{-1} B B^{\dagger}$. The set of states with $\rho_{1,1}= 0$ corresponds to the failure set. Note that the diagonal elements of $C$, $\rho_{0,0}$ and $\rho_{2,2}$, are also measured. We repeat this procedure for the set of principal $3 \times {3}$ submatrices, $M_{i} \in \bm{M}$ for $i=0,\ldots,d-2$,
\begin{equation}
M_{i} = \begin{pmatrix}
\rho_{i,i} & \rho_{i,i+1} & \bm{\rho_{i,i+2}} \\
\rho_{i+1,i} & \rho_{i+1,i+1} & \rho_{i+1,i+2} \\
\bm{\rho_{i+2,i}} & \rho_{i+2,i+1} &\rho_{i+2,i+2}  \\
\end{pmatrix},
\end{equation}
For each $M_{i}$, the upper-right and the lower-left corners elements $\rho_{i,i+2}$ and $\rho_{i+2,i}$ are unmeasured. Using the same procedure as above, we reconstruct these elements for all values of $i$ and thereby reconstruct the 2nd diagonals. We repeat the entire procedure again choosing a similar set of $4 \times {4}$ principal submatrices and reconstruct the 3rd diagonals and so on for the rest of the diagonals until all the unknown elements of the density matrix are reconstructed. Since, we  have reconstructed all diagonal elements of the density matrix and used the assumption that $\textrm{rank}(\rho) = 1$ the 5GMB is rank-$1$ complete POVM. The first basis measures the 0th diagonal, so by Proposition~\ref{prop1} the 5GMB is also rank-1 strictly-complete.

The failure set corresponding to $\bm{M}$ is when $\rho_{i,i} = 0$ for $i = 1,\ldots,d-2$. Additionally, the 5GMB provide another set of submatrices $\bm{M}'$ to reconstruct $\rho$. This set of submatrices results from also measuring the elements $\rho_{d-1,0}$ and $\rho_{0,d-1}$, which were not used in the construction of $\bm{M}$. The failure set for $\bm{M}'$ is the same as the failure set of $\bm{M}$, but since $\bm{M'} \neq \bm{M}$, we gain additional robustness. When we consider both sets of submatrices the total failure set is $\rho_{i,i} = 0$ and $\rho_{j,j} = 0$ for $i = 1,\ldots,d-2$ and $i \neq j \pm 1$. This is the exact same set found by Goyeneche~{\em et al.}~\cite{Goyeneche2015}.

We generalize these ideas to measure a rank-$r$ state by designing $4r +1$ orthonormal bases that correspond to a rank-$r$ strictly-complete POVM. The algorithm for constructing these bases, for dimensions that are powers of two, is given in Algorithm~\ref{alg:rankr_GMB}. In principle, sets of orthonormal bases with similar properties can be designed for any dimension. Technically, the algorithm produces unique bases for $r \leq d/2$, but since $d+1$ mutually unbiased bases are full-IC, for $r \geq d/4$ one may prefer to measure the latter. The corresponding measured elements are the first $r$ diagonals of the density matrix. 

Given the first $r$ diagonals of the density matrix, we can reconstruct a state $\rho \in {\cal S}_r$ with a similar procedure as the one outlined for the five bases. First, choose the leading $(r+2) \times {(r+2)}$ principle submatrix, $M_0$. The unmeasured elements in this submatrix are $\rho_{0,r+1}$ and $\rho_{r+1,0}$. By applying a unitary transformation, we can bring $M_0$ into canonical form, and by using the rank condition from Eq.~\eqref{Schur_rank} we can solve for the unmeasured elements. We can repeat the procedure with the set of $(r+2) \times {(r+2)}$ principle submatrices $M_i \in \bm{M}$ for for $i = 0,\ldots,d-r-1$ and 
\begin{equation}
M_{i} = \begin{pmatrix}
\rho_{i,i} & \cdots & \bm{\rho_{i,i+r+1}} \\
\vdots & \ddots & \vdots  \\
\bm{\rho_{i+r+1,i}} & \cdots &  \rho_{i+r+1,i+r+1}
\end{pmatrix}.
\end{equation}
From $M_i$ we can reconstruct the elements $\rho_{i,i+r+1}$, which form the $(r+1)$st diagonal. We then repeat this procedure choosing the set of $(r+3) \times {(r+3)}$ principle submatrices to reconstruct the $(r+2)$nd diagonal and so on until all diagonals have been reconstructed. This shows the POVMs are rank-$r$ complete. By Proposition~\ref{prop1}, since we also measure the computational bases, the POVMs are also rank-$r$ strictly-complete. 

The failure set corresponds to the set of states with singular $r \times r$ principal submatrix
\begin{equation}
A_i = \begin{pmatrix}
\rho_{i+1,i+1} & \cdots & \rho_{i,i+r} \\
\vdots & \ddots & \vdots \\
\rho_{i+r,i} & \cdots & \rho_{i+r,i+r}
\end{pmatrix},
\end{equation}
for $i = 1,\ldots,d-r-1$. This procedure also has robustness to this set since, as in the case of $r = 1$, there is an additional construction $\bm{M}'$. The total failure set is then when $A_i$ is singular for $i = 0,\ldots,d-r-1$ and $A_j$ is singular for $j \neq i \pm 1$.
\clearpage

\noindent\makebox[\linewidth]{\rule{\linewidth}{1pt}}
\noindent {\bf Algorithm A.1} Construction of $4r+1$ bases in the GMB \\
\noindent\makebox[\linewidth]{\rule{\linewidth}{0.4pt}}
\begin{enumerate}\label{alg:rankr_GMB}
\item{Construction of the first basis:}
\item[]{The choice of the first basis is arbitrary, we denote it by,
\begin{equation}
\mathbbm{B}_0 = \{ |0 \rangle, | 1 \rangle, \ldots , | d-1 \rangle \}.
\end{equation}
This basis defines the representation of the density matrix. Measuring this basis corresponds to the measurement of the all the elements on the 0th diagonal of $\rho$.}
\item{Construction of the other $4r$ orthonormal bases:}
\item[]{{\bf for} $k \in [1, r]$, {\bf do}}
\begin{itemize}
\item[]{Label the elements in the $k$th diagonal of the density matrix by $\rho_{m,n}$ where $m = 0,\ldots,d - 1 - k$ and $n = m + k$.}
\item[]{For each element on the $k$th and $({d-k})$th diagonal, $\rho_{m,n}$, associate two, two-dimensional, orthonormal bases,
\begin{align} \label{2dim_bases}
\mathbbm{b}^{(m,n)}_{x}=&\Bigl\{| x_{m,n}^{\pm} \rangle = \frac{1}{\sqrt{2}} \left( | m \rangle \pm | n\rangle \right)\Bigr\}, \nonumber \\
\mathbbm{b}^{(m,n)}_{y}=&\Bigl\{| y_{m,n}^{\pm} \rangle = \frac{1}{\sqrt{2}} \left( | m \rangle \pm \ii | n \rangle \right)\Bigr\},
\end{align}
for allowed values of $m$ and $n$.}
\item[]{Arrange the matrix elements of the $k$th diagonal and $({d-k})$th diagonal into a vector with $d$ elements 
\begin{equation}
\vec{v}(k) = ( \underbrace{\rho_{0,k}, \ldots, \rho_{d-1-k,d-1}}_\text{$k$th diagonal elements},\underbrace{\rho_{0,d-i},\ldots,  \rho_{k-1,d-1}}_\text{$({d-k})$th diagonal elements})\equiv(v_1(k),\ldots,v_d(k)). 
\end{equation}}
\item[]{Find the largest integer $Z$ such that $\frac{k}{2^{Z}}$ is an integer.}
\item[]{Group the elements of $\vec{v}(k)$ into two vectors, each with $d/2$ elements, by selecting $\ell = 2^Z$ elements out of $\vec{v}(k)$ in an alternative fashion,
\begin{align}
\vec{v}^{(1)}(k)  &= ( v_1, \ldots, v_\ell, v_{2\ell+1}, \ldots, v_{3\ell}, \ldots, v_{d-2\ell+1}, \ldots, v_{d-\ell} ) \nonumber \\		&=(\rho_{0,i},\ldots,\rho_{\ell,i+\ell},\ldots),\nonumber \\
\vec{v}^{(2)}(k) &=( v_{\ell+1}, \ldots, v_{2\ell}, v_{3\ell+1}, \ldots, v_{4\ell}, \ldots, v_{d-\ell+1}, \ldots, v_{d}) \nonumber \\
	&=(\rho_{\ell+1,i+\ell+1},\ldots, \rho_{2\ell,i+2\ell},\ldots). \nonumber
\end{align}}
\item[]{{\bf for} $j=1,2$ {\bf do}}
\begin{itemize}
\item[]{Each element of $\vec{v}^{(j)}(k)$ has two corresponding bases $\mathbbm{b}^{(m,n)}_{x}$ and $\mathbbm{b}^{(m,n)}_{y}$ from Eq.~\eqref{2dim_bases}.}
\item[]{Union all the two-dimensional orthonormal $x$-type bases into one basis
\begin{equation}
\mathbbm{B}^{(k;j)}_{x}=\bigcup_{\rho_{m,n}\in\vec{v}^{(j)}(k)} \mathbbm{b}^{(m,n)}_{x}.
\end{equation}
Union all the two-dimensional orthonormal $y$-type bases into one basis
\begin{equation}
\mathbbm{B}^{(k;j)}_{y}=\bigcup_{\rho_{m,n}\in\vec{v}^{(j)}(k)} \mathbbm{b}^{(m,n)}_{y}.
\end{equation}
The two bases $\mathbbm{B}^{(k;j)}_{x}$ and $\mathbbm{B}^{(k;j)}_{y}$ are orthonormal bases for the $d$-dimensional Hilbert space. }
\end{itemize}
\item[]{{\bf end for}}
\item[]{By measuring $\mathbbm{B}^{(k;j)}_{x}$ and $\mathbbm{B}^{(k;j)}_{y}$ for $j=1,2$ (four bases in total), we measure all the elements on the $k$th and $(d-k)$th off-diagonals of the density matrix.}
\end{itemize}
\item[]{{\bf end for}}
\end{enumerate}
\noindent\makebox[\linewidth]{\rule{\linewidth}{1pt}}

\section{5PB: Construction by Carmeli} \label{app:5PB}
The five polynomial bases (5PB) was proposed by Carmeli {\em et al.}~\cite{Carmeli2016} and proven to be rank-1 strictly-complete therein. The 5PB is an extension of the four bases polynomial bases (4PB) proposed by Carmeli {\em et al.}~\cite{Carmeli2015}, which were proven to be rank-1 complete. Both constructions are based on a set of orthogonal polynomials, hence the name. We provide a summary of the construction here. Full details are given in Ref.~\cite{Carmeli2015}, and the proof of rank-1 completeness and strict-completeness can be found in Ref.~\cite{Carmeli2015} and~\cite{Carmeli2016} respectively. 

The index-0 basis is the computational basis,
\begin{equation}
\mathbbm{B}_0 = \{ \ket{0}, \dots, \ket{d-1} \},
\end{equation}
the same as for the GMB, discussed in Sec.~\ref{app:GMB}. We can generate the remaining four bases from a set of orthogonal polynomials labelled $p_n(x)$, with degree $n$. An $n$-degree polynomial has $n$ roots, labelled  by the set $\{ x_j \}$. The amplitudes for the projectors that make up the first basis correspond to the roots of a $d$-degree polynomial. We evaluate a set of $\{ p_0(x),\dots, p_{d-1}(x) \}$ polynomials at the roots of the $d$-degree polynomial such that,
\begin{equation}
\ket{\tilde{\phi}_j^{(1)}} = \left[ p_0(x_j), p_1(x_j),\dots, p_{d-1}(x_j) \right]^{\top}.
\end{equation}
By the definition of orthogonal polynomials, each vector, $\ket{\tilde{\phi}_j^{(1)}}$, is orthogonal, i.e. $\langle \tilde{\phi}_j^{(1)} | \tilde{\phi}_k^{(1)} \rangle = \delta_{j,k} \| \ket{\tilde{\phi}_j^{(1)} }\|_2$. We normalize each projector, $ \ket{\phi_j}^{(1)} =  \ket{\tilde{\phi}_j^{(1)}}/\| \ket{\tilde{\phi}_j^{(1)} }\|_2$ to get the projectors that make up the first basis,
\begin{equation}
\mathbbm{B}_1 = \{ \ket{\phi_0^{(1)}}, \dots, \ket{\phi_{d-1}^{(1)}} \}.
\end{equation}

The amplitudes for the projectors that make up the second basis correspond to the roots of a $(d-1)$-degree polynomial, which we denote by the set $\{ y_j \}$. We evaluate a set of $\{ p_0(x),\dots, p_{d-1}(x) \}$ at the roots of the $(d-1)$-degree polynomial such that,
\begin{equation} \label{4PB_2}
\ket{\phi_j^{(2)}} = \left[ p_0(y_j), p_1(y_j),\dots, p_{d-1}(y_j) \right]^{\top},
\end{equation}
for $j < d-1$, which are also orthogonal by the definition of orthogonal polynomials. This expression only applies for $j < d-1$ since a $(d-1)$-degree polynomial only has $d-1$ roots. Therefore, we supplement the $d-1$ vectors in Eq.~\eqref{4PB_2} with the final vector $\ket{\phi_{d-1}^{(2)}} = [0,\dots,0,1]^{\top}$. Then after normalizing each, the second basis is defined as
\begin{equation}
\mathbbm{B}_2 = \{ \ket{\phi_0^{(2)}}, \dots, \ket{\phi_{d-1}^{(2)}} \}.
\end{equation}
The third and fourth bases are found by shifting the amplitudes in the first and second bases by a phase $e^{{\rm i} \alpha k}$, where $\alpha$ is not a rational multiple of $\pi$, such that,
\begin{align}
\ket{\tilde{\phi}_j^{(3)} }&= \left[ p_0(x_j)e^{{\rm i} \alpha}, p_1(x_j),\dots, p_{d-1}(x_j) e^{{\rm i} \alpha (d-1)} \right]^{\top}, \nonumber \\
\ket{\tilde{\phi}_j^{(4)} }&= \left[ p_0(y_j)e^{{\rm i} \alpha}, p_1(y_j),\dots, p_{d-1}(y_j) e^{{\rm i} \alpha (d-1)} \right]^{\top},
\end{align}
and we again renormalize each and supplement the final basis with the state $\ket{\phi_{d-1}^{(4)}} = [0,\dots,0,1]^{\top}$. This gives the final bases, 
\begin{align}
\mathbbm{B}_3 &= \{ \ket{\phi_0^{(3)}}, \dots, \ket{\phi_{d-1}^{(3)}} \}, \nonumber \\
\mathbbm{B}_4 &= \{ \ket{\phi_0^{(4)}}, \dots, \ket{\phi_{d-1}^{(4)}} \}.
\end{align}
Carmeli {\em et al.}~\cite{Carmeli2015} showed the four bases, $\{\mathbbm{B}_1, \mathbbm{B}_2, \mathbbm{B}_3, \mathbbm{B}_4\}$ (4PB),  are rank-1 complete and Carmeli {\em et al.}~\cite{Carmeli2016} showed that the five bases, $\{\mathbbm{B}_0, \mathbbm{B}_1, \mathbbm{B}_2, \mathbbm{B}_3, \mathbbm{B}_4\}$ (5PB),  are rank-1 strictly-complete, each without a failure set.

\section{PSI: Construction by Flammia}
The PSI construction was also proposed by Flammia {\em et al.}~\cite{Flammia2005} and proven to be a rank-1 complete POVM by the decomposition method. The POVM consists of $3d-2$ rank-1 operators in the following form,
\begin{align}
E_0 &= a | 0 \rangle \langle 0|, \nonumber \\
E_{j, 1} &= \frac{b}{2} \left( P_{j-1,j} + \frac{2 \sqrt{2}}{3} X_{j-1,j}- \frac{1}{3}Z_{j-1,j} \right), \nonumber \\
E_{j, 2} &= \frac{b}{2} \left( P_{j-1,j} -  \frac{\sqrt{2}}{3}X_{j-1,j}+ \sqrt{\frac{2}{3}}Y_{j-1,j} - \frac{1}{3}Z_{j-1,j}\right), \nonumber \\
E_{j, 3} &= \frac{b}{2} \left( P_{j-1,j} - \frac{\sqrt{2}}{3}X_{j-1,j} -  \sqrt{\frac{2}{3}}Y_{j-1,j}  -\frac{1}{3} Z_{j-1,j}\right), 
\end{align}
for $j = 1,\dots,d-1$ and $a$ and $b$ are chosen such that $\sum_{\mu} E_{\mu} = \mathds{1}$. The operator, $X_{j,j-1} = | j \rangle \langle j-1| + |j-1 \rangle \langle j|$, is the Pauli $\sigma_x$ operator across the subspaces spanned by $\ket{j-1}$ and $\ket{j}$, similar definitions apply for $Y_{j-1,j}$ and $Z_{j-1,j}$. The operator $P_{j-1,j}$ is that projection onto the subspace. 

We can show that this POVM is rank-1 strictly-complete by considering the density matrix elements that are defined by the probability of each POVM element. For $j = 1$, the four elements $E_0,\,E_{1,1}, \, E_{1,2},$ and $E_{1,3}$ are parallel to the elements that make up the 2-dimensional SIC POVM, a.k.a. the tetrahedron. Since the SIC POVM is full-IC, these four elements define probabilities that uniquely reconstruct the matrix that spans the $\ket{j-1}$ and $\ket{j}$ subspace. Therefore,  the density matrix elements $\rho_{0,0}$, $\rho_{0,1}$, $\rho_{1,0}$, and $\rho_{1,1}$ are measured. We can combine the value of $\rho_{1,1}$ with the three POVM elements for $j=2$ to reconstruct the $\rho_{1,2}$, $\rho_{2,1}$, and $\rho_{2,2}$ density matrix elements. The procedure can be repeated to reconstruct all elements $\{\rho_{j,j}, \rho_{j-1,j}, \rho_{j,j-1} \}$. Thus, the POVM uniquely reconstructs all elements on the main diagonal (or 0th diagonal with the notation in Sec.~\ref{app:GMB}) and first off-diagonal (or 1st diagonal). Then, we apply the same method introduced as Sec.~\ref{app:GMB}, which takes principle submatrices to reconstruct the higher diagonals, to show that this POVM is rank-1 strictly-complete. For this POVM, the failure set corresponds to any $\rho_{j,j} = 0$ for $j < d-1$, which is still a set of measure zero. However, this is a ``larger'' set of measure zero since it requires that all populations are not equal to zero.

\section{Rank-$r$ Flammia}\label{app:examples_full}
Finally, we provide an additional rank-$r$ strictly-complete POVM that was not implemented in the experiment. This construction is based on the rank-1 strictly-complete POVM proposed by Flammia {\em et al.}~\cite{Flammia2005}, and given in Eq.~\eqref{psi-complete}. We construct an EP-POVM with $(2d-r)r+1$ elements and prove it is rank-$r$ strictly-complete with the methods form Sec.~\ref{sec:EP}. The POVM elements are,
\begin{align}\label{psic mixed}
&E_k=a_k\ket{k}\bra{k},\;k=0,\ldots,r-1\nonumber \\
&E_{k,n}=b_k(\mathds{1}+\ket{k}\bra{n}+\ket{n}\bra{k}),\;n=k+1,\ldots,d-1,\nonumber \\
&\widetilde{E}_{k,n}=b_k(\mathds{1}-\ii\ket{k}\bra{n}+\ii\ket{n}\bra{k}), \;n=k+1,\ldots,d-1,\nonumber \\
&E_{2d-r,r+1}=\mathds{1}-\sum_{k=0}^{r}\left[E_k +\sum_{n=1}^{d-1}(E_{k,n}+\widetilde{E}_{k,n})\right],
\end{align}
with $a_k$ and $b_k$ chosen such that $E_{(2d-r)r+1}\geq0$. The probability $p_k=\textrm{Tr}(E_k\rho)$ can be used to calculate the density matrix element $\rho_{k,k}=\braket{k | \rho | k}$, and the probabilities $p_{k,n}=\textrm{Tr}(E_{k,n}\rho)$ and $\tilde{p}_{k,n}=\textrm{Tr}(\widetilde{E}_{k,n}\rho)$ can be used to calculate the density matrix elements $\rho_{n,k}=\braket{n |\rho | k}$ and $\rho_{k,n}=\braket{k | \rho |n}$. Thus, this is an EP-POVM which reconstruct the first $r$ rows and first $r$ columns of the density matrix. 

Given the measured elements, we can write the density matrix in block form corresponding to measured and unmeasured elements,
\begin{equation} \label{block_rho_gen}
\rho= 
\begin{pmatrix}
A &  B^\dagger\\
B & C 
\end{pmatrix},
\end{equation}
where $A$ is a $r \times r$ submatrix and $A$, $B^\dagger$, and $B$ are composed of measured elements. Suppose that $A$ is nonsingular. Given that $\textrm{rank}(\rho)=r$, using the rank additivity property of Schur complement and that $\textrm{rank}(A)=r$, we obtain $\rho/A=C-BA^{-1}B^\dagger=0$. Therefore, we conclude that $C=BA^{-1}B^\dagger$.  Thus we can reconstruct the entire rank-$r$ density matrix. 

Following the arguments for the POVM in Eq.~\eqref{psi-complete}, it is straight forward to show that this POVM is in fact rank-$r$ strictly-complete. The failure set of this POVM corresponds to states for which $A$ is singular. The set is dense on a set of states of measure zero. 

The POVM of Eq.~\eqref{psic mixed} can alternatively be implemented as a series of $r-1$ POVMs, where the $k$th POVM, $k=0,\ldots,r-1$, has $2(d-k)$ elements, 
\begin{align}\label{psic mixed kth}
&E_k=a_k\ket{k}\bra{k},\nonumber \\
&E_{k,n}=b_k(\mathds{1}+\ket{k}\bra{n}+\ket{n}\bra{k}),\;n=k+1,\ldots,d-1,\nonumber \\
&\widetilde{E}_{k,n}=b_k(\mathds{1}-\ii\ket{k}\bra{n}+\ii\ket{n}\bra{k}), \;n=k+1,\ldots,d-1,\nonumber \\
&E_{2(d-k)}=\mathds{1}-\left[E_k +\sum_{n=1}^{d-1}(E_{k,n}+\widetilde{E}_{k,n})\right].
\end{align}
For this POVM, the measured elements are the same as from Eq.~\eqref{psic mixed}, and the proof of rank-$r$ strict-completeness follows accordingly.

\chapter{Quantum control with partial isometries} \label{app:control}
Quantum control is the procedure for applying external fields to a quantum system in order to create a desired quantum evolution. Quantum control is required for any QT protocol in order to prepare states, or create different POVMs. We discuss techniques for closed system control, where the evolution is unitary. 

\section{Closed system control objectives}
In closed system control, the system is evolved with unitary dynamics created by a Hamiltonian, written in standard form,
\begin{equation}
H(t) = H_0 + \sum_{j=1}^m c_j(t) H_j,
\end{equation}
where $H_0$ is referred to as the ``drift'' Hamiltonian and all $H_j$'s are referred to as the ``control'' Hamiltonians. The control Hamiltonian describes an external field that is applied to the quantum system and varied in time in order to create the desired evolution. The constants, $c_j(t)$ are called the control parameters. The corresponding evolution is found by integrating the Schr\"{o}dinger equation,
\begin{equation}
U = \textrm{exp} \left[ -{\rm i} \int_0^T dt H(t) \right],
\end{equation}
from time $t=0$ to a final time $t=T$. A system is said to be controllable if the drift and control Hamiltonians togeteher generate the Lie algebra $\mathfrak{su}(d)$; that is the linear combinations of all Hamiltonians, $\{ H_0, H_j\}$, along with all combinations from the Lie bracket, $\left[H_i, H_j \right]$, span the Hermitian operator space. When the system is controllable, there exists a set of control parameters that generate any $U \in \textrm{SU}(d)$.

Since the Hamiltonian is time dependent, we cannot analytically express the unitary at the final time for arbitrary control parameters. In order to define such an analytic expression, we consider control parameters are piecewise defined, such that,
\begin{equation}
c_j(t) = \begin{cases}
      c_{j,1}, & \text{if}\ 0 \leq t < t_1, \\
      c_{j,2}, &  \text{if}\ t_1 \leq t < t_2, \\
     		& \vdots \ \\
      c_{j,n}, &  \text{if}\ t_{n-1} \leq t < t_n = T. \\
          \end{cases}
\end{equation}
When each control parameter is piecewise defined for the same time intervals then the elements $\vec{c}_{j,k}$ make up an $m \times n$ matrix $C$, where $m$ is the number of control Hamiltonians and $n$ is the number of control steps, i.e., piecewise elements of $c(t)$. The columns of $C$ are vectors that describe a time-independent control Hamiltonian for a given time interval. For example, $\vec{c}_k$ specifies the control Hamiltonian for $t_{k-1} \leq t < t_k$. When the Hamiltonian is time-independent, the Schr\"{o}dinger equation is analytically solvable. Therefore, the total evolution is described by a series of unitary maps,
\begin{equation} \label{piecewise_U}
U(C) = U(\vec{c}_N) U(\vec{c}_{N-1}) \cdots U(\vec{c}_1).
\end{equation}
We assume that each time interval is constant, $\Delta t = t_{k} - t_{k-1} = T/n$.

Closed system quantum control can be used to accomplish partial isometries.  A one-dimensional isometry is a state-to-state map; a full $d$-dimensional isometry is a unitary map of the full Hilbert space.  Intermediately, the partial isometry maps and subspace of the Hilbert space to any other subspace of the same dimension, while preserving the inner product. The goal of state-to-state mappings is to evolve an initial state, $\ket{\psi_0}$ to a target state, $\ket{\phi}$. The final state from the controlled evolution, $\ket{\psi(T)} = U(C) \ket{\psi_0}$ . Therefore, the success of a state-to-state mapping is defined by the infidelity, or overlap, between the target and final states,
\begin{align} \label{Jpsi1}
J_1[C] &= 1 - | \langle \phi | \psi (T) \rangle |^2, \nonumber \\
	&= 1 - | \langle \phi_0 | U(C)  |\psi_0 \rangle |^2.
\end{align}
When $J_1 = 0$ then the state-to-state mapping is performed perfectly and the final state matches the target state. 
 
In unitary control, the goal is to specify the entire unitary. This is equivalent to specifying $d$ state-to-state mappings  that take the fiducial basis to any desired orthonormal basis. The success is defined by the Hilbert-Schmidt distance squared between the target unitary, $W$, and the unitary created by the control, $U(C)$,
\begin{align} \label{Jd_ReTr}
J_d[C] &= \frac{1}{2d}\| W - U(C) \|^2, \nonumber \\
	&= 1 - \frac{1}{d} \textrm{ReTr}(W^{\dagger} U(C) ),
\end{align}
since $W$ and $U(C)$ are unitaries, $\textrm{Tr}(|W|^2) = \textrm{Tr}(|U(C)|^2) = d$. The ``ReTr$(\cdot)$'' operator stands for $\textrm{Re}(\textrm{Tr}(\cdot))$. We also include a normalization factor of $\frac{1}{2d}$, such that $J_d = 0$ when $U(C) = W$, i.e. the control achieves the objective unitary map exactly. The functional $J_d[C]$ is dependent on the global phase difference between $W$ and $U(C)$ but often, the global phase is irrelevant physics.  The relevant unitaries are in the special-unitary group, $SU(d)$. Therefore, we define a functional that is not proportional to the global phase,
\begin{equation} \label{Jd_abs}
\bar{J}_d[C] =  1 - \frac{1}{d^2} \left| \textrm{Tr}(W^{\dagger} U(C) ) \right|^2,
\end{equation}
and similarly $\bar{J}_d[C] = 0$ when $W = e^{-i \theta} U(C)$ for any phase $\theta$ based on the normalization. The advantage is that this reduces the number of free parameters that must be specified by the control, thereby reducing the total time required.

State-to-state and unitary control are the two extreme cases of closed-system control. For the state-to-state control, the goal is to evolve a single state to a target state. For unitary control, the goal is equivalent to evolving a set of $d$ orthonormal states to a different set of $d$ orthonormal states. In between, we evolve $n \leq d$ orthonormal states, $\{ \ket{\psi_i} \}$ to $n$ orthonormal states target states, $\{ \ket{\phi_i} \}$. The corresponding control objectives are then,
\begin{align} \label{Jpsi}
J_n[C] &= 1 - \frac{1}{n} \textrm{Re} \left[ \sum_{i=1}^n \braket {\phi_i | U(C)| \psi_i} \right], \nonumber \\
\bar{J}_n[C] &= 1 - \frac{1}{n^2} \left| \sum_{i=1}^n \braket {\phi_i | U(C) | \psi_i} \right|^2.
\end{align}
When $n = 1$, Eq.~\eqref{Jpsi} reduces to Eq.~\eqref{Jpsi1}. If we define $\ket{\psi_i} = W \ket{\phi_i}$, then for $n = d$, the objectives in Eq.~\eqref{Jpsi} reduce to Eq.~\eqref{Jd_ReTr} and Eq.~\eqref{Jd_abs} respectively.  We refer to control objectives with $n \neq 1$ or $d$ as ``partial-isometry control.'' The total time required to implement a partial isometry roughly scales with $n^2$, since this is the number of free parameters that specify the partial isometry control task. Therefore, partial-isometry control is more efficient than unitary control and is desirable in certain applications, such as the measurements of subspaces discussed in Chapter~\ref{ch:experiment}.

We can alternatively write a partial isometry in bra-ket notation,
\begin{equation}
X_n = \sum_{i=1}^n |\phi_i \rangle \langle \psi_i |,
\end{equation}
and if $n = d$ recover a unitary matrix. This can also be expressed as a rank-$n$ projectors, $A_n = \sum_i \ket{\psi_i} \bra{\psi_i}$, that acts on full unitary
\begin{equation}
X_n = W A_n.
\end{equation}
We can also express the control objective functional in terms of the projector and unitary,
\begin{align} \label{JA}
J_n[C] &= 1 - \frac{1}{n} \textrm{ReTr} \left[ A_n W^{\dagger} U(C)  \right], \nonumber \\
\bar{J}_n[C] &= 1 - \frac{1}{n^2} \left| \textrm{Tr}\left[ A_n W^{\dagger} U(C) \right] \right|^2.
\end{align}

\section{Numerical control search}
To implement closed system control, we need to find the control parameters, $C^*$, that minimizes $J_n[C]$ or $\bar{J}_n[C]$. One way to accomplish this is through numerical optimization. There are several different choices of algorithms to determine the control parameters that minimize the objective functionals. We use a variant of the gradient ascent pulse engineering (GRAPE) algorithm, originally proposed in Ref.~\cite{Khaneja2005} and further discussed in Ref.~\cite{Machnes2011}. GRAPE starts with a set of random control parameters and evaluates the functional and the gradient of the functional. The algorithm then steps in the direction of descending\footnote{The original proposal stepped in ascending direction but we look to {\em minimize} our control objective so the step is in the descending direction.} direction by some amount and recalculates the objective functional and the gradient. It continues this process until a measure of the gradient is smaller than some threshold. This point then corresponds to a local minima in the functional. If the functional was convex then this local minima would be guaranteed to be the global minima. However, none of the objective functionals discussed above are convex. In Refs.~\cite{Rabitz2005,Ho2009,Dominy2011}, it was shown that while the functionals are not convex, i.e., there is not a single global minima, they do have a favorable landscape for gradient-based algorithms. Instead of having a single global minimum the functionals introduced in the previous section have many global minimum but all give the same value of the functional. Therefore, any time the algorithm stops, with the gradient equal to zero, then the corresponding control parameters produce one of the many global minimum. However, this does mean that there are many (in fact infinite) different control parameters that achieve the same control objective.

In order to use the gradient descent methods we need to know the gradient of the objective functional. This was originally derived in Ref.~\cite{Ho2009} for the partial isometry objective. We present a brief outline here only for $\bar{J}_n[C]$ objective in Eq.~\eqref{JA}, but the derivation is similar for $J_n[C]$. The gradient with respect to the control parameter $c_{j,k}$ is,
\begin{equation} \label{grad_barJ}
\frac{\partial J_n[C]}{\partial c_{j,k}} =  - \frac{2 t}{n^2} \textrm{Tr} \left[ A_n W^{\dagger} \frac{\partial U(C)}{\partial c_{j,k}}  \right],
\end{equation}
where $t = \textrm{Tr}\left[ A_n A_n^{\dagger} W^{\dagger}  U(C) \right]$. The partial derivative of the unitary can be found by expanding in terms of Eq.~\eqref{piecewise_U},
\begin{equation} \label{dUdc}
\frac{\partial U(C)}{\partial c_{j,k}} = U(\vec{c}_N) \cdots U(\vec{c}_{k+1}) \frac{\partial U(\vec{c}_k)}{\partial c_{j,k}} U(\vec{c}_{k-1}) \cdots U(\vec{c}_1).
\end{equation}
The partial derivative of the unitary for the $k$th control parameter was solved in Refs.~\cite{Rabitz2005,Machnes2011} by expanding in the eigenbasis of $U(\vec{c}_k) = V \Lambda V^{\dagger}$, with eigenvalues $\{ \lambda_{\alpha} \}$ and corresponding eigenvectors $\{ | \lambda_{\alpha} \rangle \}$,
\begin{equation}
\frac{\partial U(\vec{c}_k)}{\partial c_{j,k}} = V D_{j,k} V^{\dagger}
\end{equation}
where $D_{j,k}$ is a $d \times d$ matrix with elements in the eigenbasis of $U(\vec{c}_k)$,
\begin{equation} \label{D}
\langle \lambda_{\alpha} | D_{j,k} | \lambda_{\beta} \rangle = \begin{cases}
\Delta t \langle \lambda_{\alpha} | H_{j} | \lambda_{\beta} \rangle  e^{-i \Delta t \lambda_{\alpha}} & \text{if}\ \lambda_{\alpha} = \lambda_{\beta}, \\
i \Delta t \langle \lambda_{\alpha} | H_{j} | \lambda_{\beta} \rangle  \frac{e^{-i \Delta t \lambda_{\alpha} }- e^{-i \Delta t \lambda_{\beta}}}{\Delta t(\lambda_{\alpha} - \lambda_{\beta})} & \text{if}\ \lambda_{\alpha} \neq \lambda_{\beta}.
\end{cases}
\end{equation}
We combine Eq.~\eqref{dUdc}-\eqref{D} into Eq.~\eqref{grad_barJ} to write the general form of the gradient,
\begin{equation} \label{grad_barJ}
\frac{\partial J_n[C]}{\partial c_{j,k}} =  - \frac{2 t}{n^2} \textrm{Tr} \left[ A_n W^{\dagger}  U(C) D'_{j,k}\right].
\end{equation}
where $D'_{j,k} = U^{\dagger}(\vec{c}_1) \cdots U^{\dagger}(\vec{c}_k) V D_{j,k} V^{\dagger} U(\vec{c}_{k-1}) \cdots U(\vec{c}_1)$. This expression can also be used to show that there are no local minimum under a few assumptions, which was done in Ref.~\cite{Ho2009}.

With the analytic form of the gradient of $J_n[C]$ and $\bar{J}_n[C]$, and the assumption that there exist no local minimum, we can use gradient based algorithm to efficiently find a global minimum of either functional. We apply  MATLAB's fminunc routine which uses the BFGS quasi-Newton technique with variables $\{ c_{j,k} \}$. The algorithm calculates the function value and gradient at a given point and then numerically finds the hessian in order to calculate how large a step to take in the direction of the gradient. It then repeats this iteration until the maximum value in the gradient is below a pre-specified threshold.

\bibliographystyle{utphys}
\bibliography{thebibliography_edit2}{}

\end{document}